\documentclass[%
 reprint,
 amsmath,amssymb,
 aps,
]{revtex4-2}

\usepackage{graphicx}
\usepackage{dcolumn}
\usepackage{braket}
\usepackage{bm}
\usepackage{dsfont}
\usepackage{mathrsfs}
\usepackage{xcolor}

\usepackage{hyperref}

\usepackage{tikz}

\newcommand{\beginsupplement}{%
	\setcounter{table}{0}
	\renewcommand{\thetable}{S\arabic{table}}%
	\setcounter{figure}{0}
	\renewcommand{\thefigure}{S\arabic{figure}}%
	\renewcommand{\theequation}{S\arabic{equation}}
}

\begin{document}

\title{Non-equilibrium topological order from Floquet anyon proliferation in the toric code}

\author{Francesco Petiziol}
\affiliation{%
Technische Universität Berlin, Institut für Physik und Astronomie, Hardenbergstra\ss e 36, 10623 Berlin, Germany}%

\date{\today}

\begin{abstract}
We analyze the dynamics of Kitaev's toric code in Floquet quenches where the drive coherently creates and propagates fermionic $f$ anyons. We show that the long-time state departs from pure-state topological order but retains long-range entanglement, thereby realizing a Floquet analog of intrinsic mixed-state topological order. As a key result enabling this analysis, we derive decompositions of the entanglement entropy and logarithmic negativity that separate the ground-state value, including topological terms, from a contribution given by the driven quasiparticles. For a drive inducing non-interacting $f$ dynamics, these decompositions allow us to leverage numerical methods for free fermions and to link the emergence of the proposed long-time ordered state to the equilibration of $f$ anyons to a periodic Gibbs ensemble. For drives inducing Floquet eigenstate thermalization of $f$, our results suggest a scenario where Floquet heating of the $f$ sector stabilizes a genuine non-equilibrium topological phase, in contrast with the usual view of heating as an obstacle to order in periodically driven systems.  
\end{abstract}

\maketitle

Investigating how topologically ordered (TO) systems behave out of equilibrium, such as in response to external driving or non-thermal environments, is of fundamental importance both for their realizations in quantum simulators~\cite{Satzinger2021, Semeghini2021, Leonard2023, Iqbal2024, Evered2025, Gammon-Smith2026, Sun2023, Petiziol2024toric} and for the search for novel collective phenomena~\cite{Po2017, Jin2023, Petiziol2024anyons, Poilblanc2024, wangIntrinsicMixedStateTopological2025, sohalNoisyApproachIntrinsically2025, Will2025, Will2026}. 
As an example, notions of intrinsic TO in mixed states (dubbed imTO) have recently been proposed in systems subject to tailored decoherence channels proliferating anyons~\cite{wangIntrinsicMixedStateTopological2025, sohalNoisyApproachIntrinsically2025}. 
While decoherence and thermal noise generically disrupt pure-state TO~\cite{castelnovoEntanglementTopologicalEntropy2007, Castelnovo2007a, Nussinov2008, Alicki2009, Hastings2011, Brown2016, Lu2020, Fan2024}, imTO states are characterized by the persistence of long-range entanglement, which is possible when the proliferated anyons possess non-trivial self-statistics (e.g., fermions).

We show that a phenomenology akin to imTO can emerge in the long-time state of TO systems driven by a coherent time-periodic modulation. We analyze Floquet quenches in the toric code~\cite{Kitaev2003}, where the modulation resonantly creates pairs of fermionic quasiparticle excitations ($f$), while not disturbing other quasiparticles ($e$ or $m$), analogously to imTO decoherence models~\cite{wangIntrinsicMixedStateTopological2025}. 
The drive considered couples neighboring spins and induces non-interacting $f$ dynamics: On the one hand, the restriction to free dynamics allows us to investigate numerically systems large enough to probe long-time equilibration, by leveraging generalized Jordan-Wigner mappings~\cite{Chen2022}. On the other, it provides an interpretation of the dynamics of TO through quench theory in free-fermion systems~\cite{Russomanno2012, Lazarides2014a, Yamashika2024, Calabrese2005}. Our findings indicate that, for drives inducing non-integrable $f$ dynamics, a genuine non-equilibrium TO phase may be stabilized by Floquet heating, despite the latter being commonly seen as preventing order out of equilibrium~\cite{Moessner2017}.  

The above analysis is supported by an additional key result of this work. We find decompositions of the entanglement (and R\'enyi) entropy and of the logarithmic negativity~\cite{Negativity1998, Vidal2002, Plenio2005} of the system, for partitions of interest to detect topological entanglement, into the sum of two contributions: One is given by their value for the toric-code ground state, which includes the subleading topological term~\cite{Hamma2005, Kitaev2006a, Levin2006, Lee2013, Castelnovo2013}; the other is a contribution from a fermionic state describing the proliferated $f$ quasiparticles only, obtained via the generalized Jordan-Wigner mapping. These decompositions are non-trivial due to the non-locality of the spin-fermion mapping. They crucially allow us to link the entanglement entropy of the driven spin state to that of the free fermion problem, which can be computed efficiently. Moreover, they trace the departure from pure-state TO induced by the drive back to the equilibration of the $f$ particles to an infinite-temperature-like (periodic Gibbs) ensemble. In turn, the decomposition of the logarithmic negativity shows that its value in the long-time ensemble obeys an area law with a non-zero topological entanglement negativity, which signals the presence of long-range entanglement and thus of a non-equilibrium counterpart of TO. 

\begin{figure}
\centering
\includegraphics{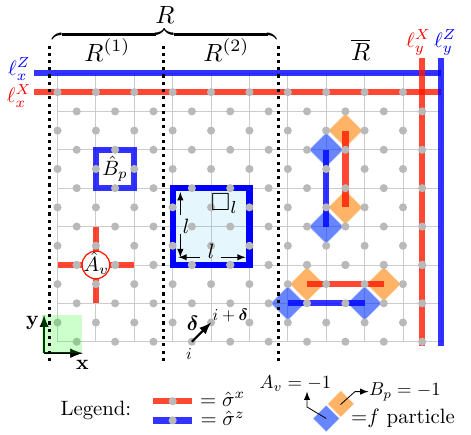}
\caption{Sketch of the toric-code system, of the vertex ($\hat{A}_v$) and plaquette ($\hat B_p$) operators, and of the paths $\ell_{x}^{X}$, $\ell_{x}^{Z}$, $\ell_{y}^{X}$ and  $\ell_{y}^{Z}$ defining the Wilson loops $\hat{W}^{ZX}_x$ and $\hat{W}^{ZX}_y$. The $f$ quasiparticles sit at the ends of strings of $\hat{\sigma}_i^z\hat{\sigma}_{i+\bm{\delta}}^x$ operators (the vector $\bm{\delta}$ is indicated). Entanglement in the evolved state is analyzed for the non-contractible regions $R=R^{(1)}\cup R^{(2)}$ and $\bar{R}$. The expectation values of contractible Wilson-loop operators supported on squares $\square_l$ probe the departure from pure-state TO.}
\label{fig:1}
\end{figure}

\paragraph*{The model.} We consider the toric-code model~\cite{Kitaev2003} under time-periodic driving, describing two-level systems (spins) on the bonds of a square $L_x\times L_y $ lattice on a torus. The Hamiltonian is $\hat{H}_{\rm tc} + \hat{H}_d(t)$ with
\begin{equation}
\hat{H}_{\rm tc} = -\frac{g}{2} \sum_v \hat A_v -\frac{g}{2}\sum_p \hat B_p, 
\end{equation}
and $\hat H_d(t)$ describing the Floquet drive, to be specified below. The vertex operator $\hat A_v=\prod_{i\in v} \hat \sigma_i^x$ acts on spins on the bonds stemming from the lattice vertex $v$, while $\hat B_p=\prod_{i \in p} \hat \sigma_i^z$ involves spins surrounding a plaquette $p$ [Fig.~\ref{fig:1}]. The Hamiltonian $\hat{H}_{\rm tc}$ is exactly solvable, as all operators $\hat A_v$ and $\hat B_p$ commute. The eigenstates can be labelled by the expectation values of $ \hat A_v$ and $ \hat B_p$, with the ground state(s) satisfying $ A_v = B_p=+1$. Due to the global parity constraint $\prod_v \hat{A}_v=\prod_p \hat B_p = \mathds{1}$, not all vertex and plaquette operators are independent. A complete set of observables can then be obtained by including two Wilson-loop operators which trace non-contractible loops around the torus, commute with each other and with all $\hat A_v$ and $\hat B_p$ (and, thus, also with $\hat{H}_{\rm tc}$). We will work with $\hat{W}^{ZX}_x = \prod_{i\in \ell^{Z}_x} \hat \sigma_i^z \prod_{j\in \ell^X_x} \hat \sigma_j^x$ and $\hat W^{ZX}_y= \prod_{i\in \ell^{Z}_y} \hat \sigma_i^z \prod_{j\in \ell^X_y} \hat \sigma_j^x$, with the loops $\ell^{X}_{x/y}$ and $\ell^Z_{x/y}$ depicted in Fig.~\ref{fig:1}. The $\pm 1$ eigenvalues of $\hat{W}^{ZX}_x$ and $\hat{W}_y^{ZX}$ label four superselection sectors and thus four degenerate ground states $\ket{g_{\pm1, \pm1}^{ZX}}$. The excitations of $\hat H_{\rm tc}$ realize an Abelian anyon theory, including three non-trivial anyon types --- $e$, $m$, and $f$--- and the vacuum. The anyons $e$ and $m$ correspond to a vertex ($A_v=-1$) and a plaquette ($B_p=-1$) defect, respectively, and have bosonic self-statistics but non-trivial mutual statistics: braiding $e$ around $m$, or vice versa, yields the accumulation of a $\pi$ phase. The third particle, $f=e\times m$, will be the focus of this work, as it couples to the drive: it has fermionic self-statistics and corresponds to a joint violation of two adjacent vertex and plaquette operators, $ A_v =  B_{v+\bm{\delta}} =-1$, where $\bm{\delta}=(1/2, 1/2)$ in units of the lattice constant, as shown in Fig.~\ref{fig:1}. Fermion pairs are created and moved by the action of strings of $\hat\sigma^z_i \hat\sigma^x_{i+\bm{\delta}}$ operators. Such operators commute with $\hat{W}^{ZX}_x$ and $\hat{W}^{ZX}_y$, thus preserving the superselection sector defined by them.

The drive Hamiltonian $\hat H_d(t)$ couples to $f$ quasiparticles only, without creating $e$ or $m$, and has the form
\begin{equation} \label{eq:drive}
\hat H_d(t) = -\Omega (t)\sum_{i} \hat \sigma^z_i \hat \sigma^x_{i+\bm{\delta}},
\end{equation}
where $\Omega(t) = \Omega \cos(\omega t)$ is a weak oscillating field, $\Omega\ll \omega$. The driving frequency $\omega$ is set on resonance with the fermion-pair creation gap, $\omega=4g$, with the goal of deliberately forcing the creation of fermionic quasiparticles. For the different case of off-resonant dressing, $\omega \ne 4 g$, and a circular-like modulation, a drive similar to \eqref{eq:drive} has been shown to Floquet-engineer a topological band structure for $f$, leading to the emergence of Majorana zero modes and thus an effectively non-Abelian model~\cite{Petiziol2024anyons}. Static quenches in the toric code, without a focus on $f$, have been studied, e.g., in Refs.~\cite{Tsomokos2009, Rahmani2010, Halasz2013}.

The impact of the drive can be conveniently analyzed by applying a generalized Jordan-Wigner mapping~\cite{Chen2022}, illustrated in the Supplemental Material~\cite{SM}: In the absence of $e$ particles in the system, the driven spin Hamiltonian $\hat H_{\rm tc} + \hat H_d(t)$ maps to the non-interacting fermion Hamiltonian
\begin{multline} \label{eq:fermion_driven_H}
\hat H_f(t) = \mu_f \sum_p \hat f_p^\dag \hat f_p \\
- \Omega(t)\sum_{\bm{r}=\bm{\mathrm{x}},-\bm{\mathrm{y}}}\sum_p(\hat f_{p+\bm{r}}^\dag \hat f_{p} +   \hat f_{p+\bm{r}}^\dag\hat f_p^\dag+ \mathrm{H.c.}),
\end{multline}
with $\mu_f=2 g$. The operators $\hat f_p$ ($\hat f_p^\dag$) annihilate (create) an $f$ particle on plaquette $p$ and their vacuum represents the toric-code ground state $\ket{g_{1,1}^{ZX}}$. The drive thus modulates hopping and pairing terms for $f$ in time. 
 In the following, we analyze the dynamics of $f$ in Floquet quenches starting from $\ket{g_{1,1}^{ZX}}$, by time-propagating the fermionic model in Eq.~\eqref{eq:fermion_driven_H}. The latter being quadratic provides a striking advantage for both the analytical and numerical treatment. For concreteness, we focus on the sector $W^{ZX}_{x}=W_{y}^{ZX}=+1$, which corresponds to adopting periodic boundary conditions in Eq.~\eqref{eq:fermion_driven_H} \footnote{The cases $W^{ZX}_{x}=-1$ and $W_{y}^{ZX}=-1$ correspond to anti-periodic boundary conditions instead~\cite{Chen2022, Petiziol2024anyons}.}. 

\paragraph*{Entanglement decompositions.} To investigate the fate of TO in response to the drive, we would like to characterize entanglement in the driven system. 
However, due to the strong non-locality of the generalized Jordan-Wigner mapping leading to Eq.~\eqref{eq:fermion_driven_H}, the question arises of whether and how the entanglement structure of the fermionic system, which we can time-evolve efficiently, relates to that of the underlying physical spins. This becomes evident, for instance, when noticing that the (topological) entanglement of the spin ground state is hidden in the vacuum of the fermion system. To proceed, we first develop analytical tools addressing this problem.

We will be interested in the entanglement entropy $\mathcal{S}_{R}=-\mathrm{tr}\hat{\rho}_R \log \hat{\rho}_R$ between sub-regions $R$ and $\bar{R}$ as in Fig.~\ref{fig:1} and in the logarithmic negativity $\mathcal{E}=\log \Vert \hat{\rho}_R^{T_2}\Vert_1$ between $R^{(1)}$ and $R^{(2)}$, where $\hat{\rho}_R$ is the reduced density operator of the spin system in $R$, $(\cdot)^{T_2}$ denotes the (bosonic, here and hereafter) partial transpose with respect to $R^{(2)}$ and $\Vert \hat{\rho} \Vert_1=\mathrm{tr}\sqrt{\hat{\rho}^\dag \hat{\rho}}$ is the trace norm~\cite{Negativity1998, Vidal2002, Plenio2005}. The sub-regions considered are non-contractible along the vertical direction. For this geometry, $\mathcal{E}$ exhibits an area law and a negative subleading contribution, dubbed topological entanglement negativity, which diagnoses long-range entanglement~\cite{Lu2020, Fan2024}, for both pure TO states~\cite{Lee2013, Castelnovo2013} and imTO states~\cite{wangIntrinsicMixedStateTopological2025}. The entropy $\mathcal{S}_R$ behaves analogously for pure-state TO~\cite{Hamma2005, Kitaev2006a, Levin2006}. However, for mixed states, entropy-based diagnostics become sensitive to classical correlations and can no longer faithfully detect long-range entanglement within $R$~\cite{castelnovoEntanglementTopologicalEntropy2007, Lu2020, Fan2024, wangIntrinsicMixedStateTopological2025}. Since the driven dynamics makes the system locally indistinguishable from a mixed state, we use $\mathcal{S}_R$ to monitor the departure from pure-state TO and rely on $\mathcal{E}$ to probe topological entanglement in the long-time state. 

Our first result, derived in \cite{SM}, is that the moments of $\hat{\rho}_R$ factorize as 
$\mathrm{tr}(\hat{\rho}_R^n) = \mathrm{tr}[(\hat{\rho}_R^{(g)})^n] \mathrm{tr}[(\hat{\rho}_R^{(f)})^n]$
 for the entanglement cuts considered. Here, $\hat\rho_R^{(g)}=\mathrm{tr}_{\bar{R}}\ket{g_{1,1}^{ZX}}\!\bra{g_{1,1}^{ZX}}$ and $\hat\rho_R^{(f)}$ denote the reduced density operators of the toric-code ground state and of the fermionic system governed by Eq.~\eqref{eq:fermion_driven_H}, respectively. The fermion sites belonging to $R$ include all plaquettes that lie entirely inside $R$ and those crossed by the cut between $R^{(2)}$ and $\bar{R}$ in Fig.~\ref{fig:1}. 
The factorization of the moments leads, via the replica limit $\mathcal{S}_R=-\lim_{n\to 1}\partial_n \mathrm{tr} \hat{\rho}_R^n $, to the decomposition
\begin{eqnarray}
\mathcal{S}_R  = (2L_y-1) \log(2) + \mathcal{S}_R^{(f)}. 
\label{eq:vNentr}
\end{eqnarray}
The entropy thus decomposes into the ground-state contribution, which includes the topological term $-\log(2)$, and the entropy of the fermionic state describing the $f$ quasiparticles, $\mathcal{S}^{(f)}_R=-\mathrm{tr} \hat{\rho}^{(f)}_R \log \hat{\rho}^{(f)}_R$.
Equation~\eqref{eq:vNentr} provides a very useful tool, as it allows us to investigate the entanglement entropy of the spin system in terms of the non-interacting fermion dynamics given by Eq.~\eqref{eq:fermion_driven_H}, which can efficiently be computed~\cite{peschelCalculationReducedDensity2003, Vidal2003, peschelReducedDensityMatrices2009}. The factorization of the moments for all $n$ also provides information on the entanglement spectrum of $\hat{\rho}_R$~\cite{Yao2010}, as it implies that the non-zero eigenvalues of $\hat{\rho}_R$ are pairwise products of the non-zero eigenvalues of $\hat{\rho}^{(g)}_R$, which all equal $2^{-(2L_y-1)}$~\cite{SM}, and of $\hat{\rho}^{(f)}_R$. Hence, the spectrum of $\hat{\rho}^{(f)}_R$ determines that of $\hat{\rho}_R$, with each level acquiring an additional $2^{2L_y-1}$-fold degeneracy from $\hat{\rho}^{(g)}_R$~\cite{Yao2010}. 
General decompositions of the von Neumann entropy into contributions from the TO ground state and from anyonic entanglement, conceptually similar to Eq.~\eqref{eq:vNentr}, have been predicted for generic anyon theories~\cite{Bonderson2017} and found for Kitaev's honeycomb model in Majorana representation~\cite{Yao2010}. However, results for other concrete models, such as the one considered here, have been lacking, to our knowledge.

A decomposition separating the anyon-vacuum contribution from that of $\hat{\rho}_R^{(f)}$ is also found for $\mathcal{E}$,
\begin{equation}
  \mathcal{E}  = (L_y-1)\log(2) + \widetilde{\mathcal{E}}^{(f)} 
 \label{eq:logneg}.
\end{equation} 
The first addend is the logarithmic negativity of the toric-code ground state. The $f$ contribution $\widetilde{\mathcal{E}}^{(f)}$, instead, is not directly related to definitions of negativity for fermionic systems~\cite{Eisler2015, Shapourian2017, Shapourian2019a, Shapourian2019}. It is given by $\widetilde{\mathcal{E}}^{(f)} = \log(\Vert \hat{\Sigma}_+\Vert_1 + \Vert \hat{\Sigma}_-\Vert_1)$, where $\hat{\Sigma}_\pm$ describe the components of $(\hat{\rho}_R^{(f)})^{T_2}$ for which the parity in $R^{(2)}$ remains unchanged under partial transpose ($+$) and those for which it flips ($-$), $\hat{\Sigma}_\pm = [(\hat{\rho}_R^{(f)})^{T_2} \pm \hat{\mathcal{P}}_2  (\hat{\rho}_R^{(f)})^{T_2} \hat{\mathcal{P}}_2]/2$ with $\hat{\mathcal{P}}_2=(-1)^{\sum_{p\in R^{(2)}} \hat{f}_p^\dag \hat{f}_p}$~\cite{SM}. Unlike $\mathcal{S}^{(f)}_R$, $\widetilde{\mathcal{E}}^{(f)} $ cannot, to our knowledge, be computed efficiently even for Gaussian $\hat{\rho}^{(f)}_R$, since $ \hat{\Sigma}_\pm$ are non-Gaussian and standard free-fermion techniques do not apply. 
Nonetheless, it will allow us to draw conclusions about the existence of long-range entanglement in the long-time steady-state of the driven system. 

\paragraph*{Floquet dynamics and anyon proliferation.} We now discuss the $f$-anyon dynamics induced by the weak resonant drive, leveraging the fermionic representation of Eq.~\eqref{eq:fermion_driven_H} and the entropy decomposition~\eqref{eq:vNentr}. We shall see that, although the integrability of $\hat{H}_f(t)$ prevents genuine Floquet eigenstate thermalization~\cite{Lazarides2014, Moessner2017}, $\hat H_f(t)$ still drives the system from the vacuum to a locally maximally mixed state at long times. 

The Hamiltonian \eqref{eq:fermion_driven_H} in quasimomentum representation, $\hat H_f(t) =(1/2) \sum_{\bm{k}} (\hat{f}_{\bm{k}}^\dag, \hat{f}_{-\bm{k}}) {\mathcal{H}}_{\bm{k}}(t)(\hat{f}_{\bm{k}}, \hat{f}_{-\bm{k}}^\dag)^T$, defines a collection of non-interacting pseudospin-1/2 problems, each governed by ${\mathcal{H}}_{\bm{k}}(t) = \bm{B}_{\bm{k}}(t)\cdot {\bm{\sigma}}$, where ${\bm{\sigma}}$ is the vector of Pauli matrices. The effective magnetic field $\bm{B}_{\bm{k}}(t)=[0, \Delta_{\bm{k}}(t),  \varepsilon_{\bm{k}}(t)]$ has components $\varepsilon_{\bm{k}}(t) =\mu_f -2\Omega(t)[\cos(k_x) +\cos(k_y)]$ and $\Delta_{\bm{k}}(t) = -2\Omega(t)[\sin(k_x) -\sin(k_y)]$. 
To leading order in a high-frequency expansion for $\omega\gg \Omega$, corresponding to a rotating-wave approximation, the dynamics is well described by the effective Floquet Hamiltonian ${\mathcal{H}}^{(F)}_{\bm{k}}=B^{(F)}_{\bm{k},y}  \sigma_y $, with $B^{(F)}_{\bm{k}, y}=-\Omega[\sin(k_x) -\sin(k_y)]$ (see~\cite{SM}). The drive forces the precession of the $\bm{k}$-pseudospins about the $y$ axis of the Bloch sphere at an inhomogeneous Rabi frequency $2B^{(F)}_{\bm{k}, y}$. This corresponds to oscillations between the vacuum $\ket{0}$ and states $\hat{f}^\dag_{\bm{k}} \hat{f}^\dag_{-\bm{k}}\ket{0}$ hosting a pair of $f$ particles with opposite quasimomenta. Accordingly, the two-point correlations oscillate as $\langle \hat{f}_{\bm{k}}^\dag \hat{f}_{\bm{k}}\rangle(n\tau)=\sin^2(B^{(F)}_{\bm{k},y} n \tau)$ and $\langle \hat{f}_{\bm{k}}^\dag \hat{f}^\dag_{-\bm{k}}\rangle(n\tau)=-(1/2)\sin(2 B^{(F)}_{\bm{k}, y} n \tau)$ at stroboscopic times, $t=n \tau$ with integer $n$ and $\tau=2\pi/\omega$. 
The modes at quasimomenta where $B^{(F)}_{\bm{k}, y}$ vanishes, i.e. $k_y=\pi-k_x$ or $ k_y=k_x$ $(\mathrm{mod}\ 2 \pi)$, remain unoccupied. We denote this set as $K_0$, having cardinality $|K_0|\le 2 \min(L_x, L_y)$. 

The above high-frequency analysis also allows us to investigate the long-time dynamics, where local observables of periodically driven non-interacting fermions as in Eq.~\eqref{eq:fermion_driven_H} are expected to be well captured by a periodic Gibbs ensemble (PGE), $\hat{ \rho}^{(f)}_{\rm PGE}=\exp(-\sum_j\zeta_j \hat{m}_j^\dag \hat{m}_j)/Z$~\cite{Russomanno2012, Lazarides2014a}. Here, $\hat{m}_j^\dag$ and $\hat{m}_j$ are operators creating and annihilating, respectively, $f$ quasiparticles in the $j$th single-particle Floquet mode and $Z$ is a normalization constant. The coefficients $\zeta_j$ are fixed by the initial occupations of the corresponding Floquet mode, through $\bra{0}\hat{m}_j^\dag \hat{m}_j \ket{0} =1/[1+\exp(\zeta_j)]$ for our initial vacuum state. Since all observables, and thus the reduced density matrix, are fully determined by the two-point correlations via Wick's theorem, the PGE is also expected to capture faithfully the system's entanglement, for non-disordered systems~\cite{Ziraldo2013, Ziraldo2012}. By diagonalizing ${\mathcal{H}}^{(F)}_{\bm{k}}$ we obtain the Floquet modes (in high-frequency approximation) and determine $\hat{\rho}^{(f)}_{\rm PGE} = (1/2^{|\overline{K}_0|})\prod_{\bm{k}\in K_0} (1-\hat{n}_{\bm{k}})$~\cite{SM}, with $|\overline{K}_0|$ denoting the number of modes not in $K_0$. The PGE is, thus, fully mixed except for the contribution of the undriven modes, $\bm{k}\in K_0$. It possesses average occupations $\langle \hat{f}_{\bm{k}}^\dag \hat f_{\bm{k}} \rangle=1/2$ for $\bm{k}\not\in K_0$ and $\langle \hat{f}_{\bm{k}}^\dag \hat f_{\bm{k}} \rangle=0$ for $\bm{k}\in K_0$, with all pairing amplitudes $\langle\hat{f}_{\bm{k}}^\dag \hat f_{-\bm{k}}^\dag \rangle$ vanishing. In real space, this leads to a trivial correlation matrix up to corrections decaying with the system size, $\langle \hat{f}_{p}^\dag \hat{f}_{p'} \rangle = \delta_{pp'}/2 - (1/2L_xL_y)\sum_{\bm{k}\in K_0} \exp[-i \bm{k}\cdot (\bm{r}_p-\bm{r}_{p'})] = \delta_{pp'}/2 +\mathcal{O}(1/\max(L_x, L_y))$ and $\langle \hat{f}_p^\dag \hat f_{p'}^\dag \rangle = 0$. These results depict the following picture: As the drive coherently modulates the creation of $f$ pairs spreading throughout the system, the asynchronous precession of the $\bm{k}$-pseudospins leads to a dephasing of local real-space observables, making the $f$ sector featureless (infinite-temperature-like) at long times, as described by the PGE. As the $f$ system equilibrates to the PGE reaching maximal entropy, the system then departs from pure-state TO~\cite{wangIntrinsicMixedStateTopological2025, Castelnovo2007a}. 
The time required to equilibrate over macroscopic scales, at fixed driving amplitude, can be expected to scale with the linear size of the system~\cite{Bravyi2006}, such that the ground state topological entanglement may still be detected, at finite times, over length scales outside of the Lieb-Robinson light cone of quasiparticle spreading~\cite{Hastings2010,Chen2010a}.

\begin{figure}
\includegraphics[width=\linewidth]{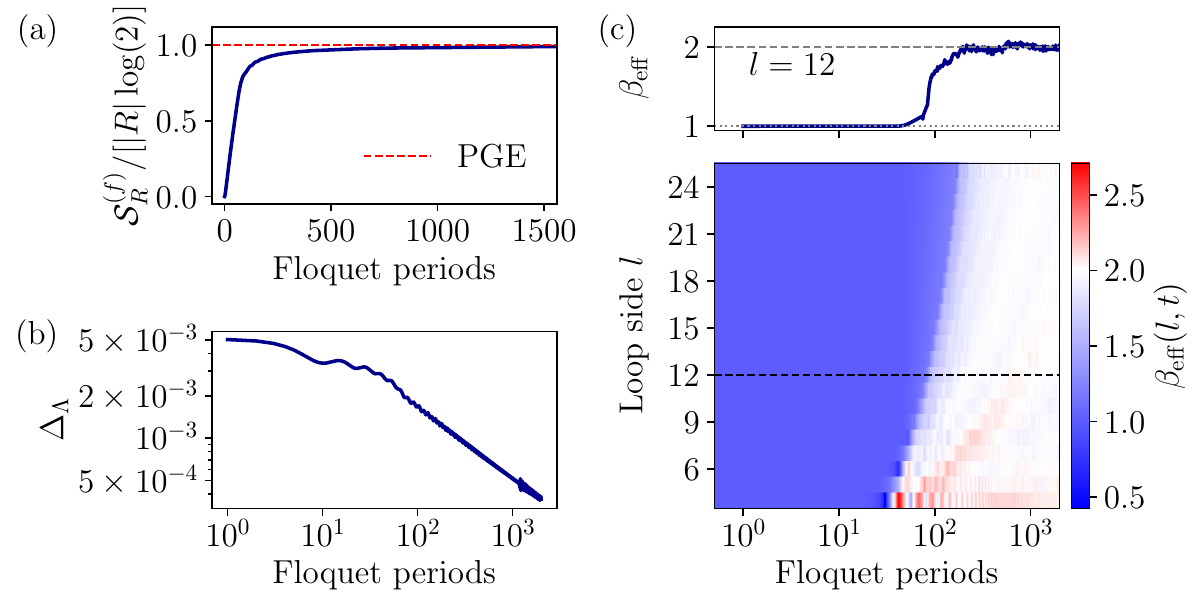}
\caption{(a) Equilibration of the fermion entropy per site $\mathcal{S}_R^{(f)}/[|R|\log(2)]$ to the PGE value, for $R$ a strip of width $w_R=8$ sites. (b) Convergence of the correlation matrix to the PGE, quantified by $\Delta_\Lambda=\Vert \Lambda(t)-\Lambda^{\rm {PGE}}\Vert_2/2|R|$~\cite{Sen2016b}, with $\Vert O \Vert_2=\sqrt{\mathrm{tr}O^\dag O}$. (c) Crossover from a perimeter ($\beta_{\rm eff}=1$) to area-law ($\beta_{\rm eff}=2$) behaviour of contractible Wilson $Z$-loops, quantified by the effective exponent $\beta_{\rm eff}$ (see text). A horizontal cut at $l=12$ is reported in the upper panel. In all panels, $L_x=L_y=300$ and $\Omega=\mu_f/50$ were used. }
\label{fig2}
\end{figure}

We verify the above analytical predictions for the stroboscopic Floquet dynamics in Fig.~\ref{fig2}, through exact numerical time propagation of $\hat H_f(t)$ and determination of the PGE from the numerically exact Floquet modes~\cite{SM}. The entropy $\mathcal{S}_R^{(f)}$ for the $R$-$\bar{R}$ bipartition of Fig.~\ref{fig:1} grows linearly in time and converges to the PGE value [Fig.~\ref{fig2}(a)], reaching the maximum value of $\log(2)$ per site, in agreement with the high-frequency estimate. Concerning the transient behaviour, we note that the bipartition considered preserves translational invariance in the $y$ direction, yielding a decoupled one-dimensional problem for each value of the transverse quasimomentum $k_y$. For the latter, Calabrese-Cardy's quasiparticle picture explains the linear entropy growth in terms of the ballistic spreading of entangled pairs of quasiparticles across the sub-region boundary~\cite{Yamashika2024, Calabrese2005}, with eventual saturation to the PGE entropy. 
To further confirm that the PGE captures the entanglement structure of the system's state at long times, the convergence of the two-point correlation matrix ${\Lambda}_{ij}=\langle \hat{F}_i^\dag \hat{F}_j \rangle$, for $\bm{\hat{F}}=(\bm{\hat{f}}, \bm{\hat{f}}^\dag)$, to the PGE prediction in $R$ is exemplified in Fig.~\ref{fig2}(b).

The departure from pure-state TO as the $f$ particles equilibrate to the PGE can be observed also in the expectation values of contractible Wilson loops $W^Z_{\square}(t) = \langle \prod_{p\in \mathrm{Int}(\square)} \hat{B}_p\rangle(t)$, where we will consider square $l\times l$ loops $\square_l$ of area $\mathscr{A}_{l}$ and perimeter $\mathscr{P}_{l}$. The quantities $W^Z_{\square}(t)$ are expected to scale approximately as $ W^{Z}_{\square} \sim\exp(\chi \mathscr{A} + \chi' \mathscr{P})$ for constants $\chi, \chi'$~\cite{Fradkin1978, Kogut1979}, and to exhibit a perimeter law ($\chi=0$) in the TO phase and an area law ($\chi\ne 0$) when the quasiparticles are confined or thermal.
We track the crossover between these two regimes by monitoring the effective exponent $\beta_{\rm eff}(l, t)=\partial \log(-\log|W^Z_{\square_l}(t)|)/\partial \log l$ over time, computed via numerical gradient. Away from the crossover region, $\beta_{\rm eff}$ approximates the leading exponent of $-\log|W_{\square_l}^Z(t)|$ up to corrections decaying with $l$~\cite{SM}.
A crossover from $\beta_{\rm eff}\approx 1$ to $\approx 2$ is indeed found in $\beta_{\rm eff}(l, t)$, as reported in Fig.~\ref{fig2}(c), clearly signaling the emergence of a non-zero area-law term as Floquet equilibration to the PGE takes place within larger and larger regions.  

\textit{Long-range entanglement.} We established that the long-time state of the system is not TO in the pure-state sense, as it approaches a state with maximal entropy in the $f$ sector on macroscopic scales. We next show that, nevertheless, the system still exhibits long-range entanglement and can thus be thought of as possessing a non-equilibrium notion of TO, akin to imTO~\cite{wangIntrinsicMixedStateTopological2025}. Indeed, analyzing the logarithmic negativity $\mathcal{E}$ between adjacent noncontractible sub-regions $R^{(1)}$ and $R^{(2)}$ as in Fig.~\ref{fig:1}, we find that $\mathcal{E}$ equilibrates to the ground-state (and area-law) value for large systems and thus retains its topological entanglement negativity. Unlike the topological entanglement entropy, the latter has been shown to reliably detect long-range entanglement in toric-code mixed states for the sub-region geometry considered~\cite{Castelnovo2013, Lee2013, Lu2020, Fan2024, wangIntrinsicMixedStateTopological2025}. 
The above behaviour of $\mathcal{E}$ can easily be determined through the negativity decomposition of Eq.~\eqref{eq:logneg} and the high-frequency analysis of the PGE. For a fully mixed ensemble, the $f$-contribution $\widetilde{\mathcal{E}}^{(f)}$ would vanish, since it would yield $\Vert 
\hat{\Sigma}_+ \Vert_1 + \Vert 
\hat{\Sigma}_- \Vert_1=1$. The PGE of our driven system involves corrections to this picture due to the undriven modes $\bm{k}\in K_0$; however, these can be shown to vanish for large $L_x=L_y=L$, as they can be bounded by $\Vert\hat{\Sigma}_{+}^{\rm PGE} \Vert_1+\Vert\hat{\Sigma}_{-}^{\rm PGE} \Vert_1 \le 1 + 2 \sqrt{\exp(4 w_R^2/L)-1}$ where $w_R$ is the fixed width of $R$~\cite{SM}. 
All in all, we conclude that the $f$ contribution $\widetilde{\mathcal{E}}^{(f)}$ in Eq.~\eqref{eq:logneg} vanishes for growing system size at high frequency, such that the logarithmic (and topological) negativity of the long-time state converges to that of the ground state. 

Our analysis built on the integrability of the $f$ dynamics to both facilitate the analytical treatment and allow us to simulate systems large enough to probe local equilibration to the PGE. In turn, integrability constrains the dynamics through the conservation of the initial occupation of the single-particle Floquet modes~\cite{Lazarides2014a, Moessner2017}, such that the achievement of an infinite-temperature-like state at long times depends on the choice of initial state (here, the $f$ vacuum) and is not a universal feature.  
However, our results indicate that the non-equilibrium form of TO discussed here should emerge universally for generic $f$ drives that induce non-integrable dynamics. For the latter, the system is expected to reach infinite temperature in the $f$ sector, for generic initial states and parameter values~\cite{Lazarides2014, Moessner2017} and, since Eqs.~\eqref{eq:vNentr} and \eqref{eq:logneg} are not restricted to Gaussian fermionic states, $\mathcal{E}$ would converge to the ground-state value also in this case. This scenario could then represent a genuine non-equilibrium TO phase which is enabled by Floquet heating. 

\textit{Outlook.} We have investigated the emergence of a non-equilibrium form of TO in the long time dynamics of the toric code under a coherent Floquet drive coupling to its fermionic quasiparticles ($f$), sharing similarities with notions of intrinsic mixed-state TO~\cite{wangIntrinsicMixedStateTopological2025, sohalNoisyApproachIntrinsically2025}. %
In the case where the drive heats all anyon sectors, thus fully destroying TO at long times, an open question is whether the ordered states discussed here may still exist prethermally via fractionalized prethermalization~\cite{Jin2023}. Our main findings include decompositions of the entropy and of the logarithmic negativity of the toric code, which explicitly decouple the contribution given by the TO ground state from that given by the driven $f$ dynamics. Generalizing such decompositions for concrete models beyond the scenario considered here would be a strong asset for the investigation of TO out of  equilibrium.

This work has been funded by the Deutsche Forschungsgemeinschaft (DFG, German Research Foundation) through the Emmy Noether Programme – project number 555842149. The author is grateful to Alexander Hahn for insightful discussions about anyon entanglement. Input from Claude Opus 5.5 has been used to improve the performance of the final code for numerical simulations and to streamline the derivation in Eqs.~(S162)-(S167) in the Supplemental Material~\cite{SM}.

 \let\oldaddcontentsline\addcontentsline
\renewcommand{\addcontentsline}[3]{}
\bibliography{zot_library_aps}
\let\addcontentsline\oldaddcontentsline

\cleardoublepage
 \let\oldaddcontentsline\addcontentsline
\renewcommand{\addcontentsline}[3]{}

\let\addcontentsline\oldaddcontentsline

\beginsupplement
\setcounter{page}{1}
\onecolumngrid
\begin{center}
{\bf \large Supplemental Material to} \\
\vspace{0.2cm}
{\bf \large Non-equilibrium topological order from Floquet anyon proliferation in the toric code} \\
\vspace{0.4cm}

{ Francesco Petiziol }\\

{\itshape
Technische Universität Berlin, Institut für Physik und Astronomie, Hardenbergstra{\ss}e 36, Berlin 10623, Germany}
\end{center}

\vspace{.5cm}

\onecolumngrid

\tableofcontents

\section{Toric-code ground states and topological entanglement}
\label{sec:recap}

In this section, we recapitulate relevant properties of the toric-code model and analyze the entanglement structure of the ground state that corresponds to the $f$-particle vacuum and which is the initial state in the Floquet quenches studied in the main text. We will analyze, in particular, R\'enyi and von Neumann entropies, and the logarithmic negativity. 

\subsection{Ground states and $f$ vacuum}

To discuss the structure of the toric-code $f$-vacuum ground state, we start by considering the ground states in the sector defined by the non-contractible Wilson-loop operators  
\begin{equation}
\hat W^Z_x = \prod_{i\in \ell^Z_x} \hat{\sigma}_i^z, \quad \hat W^Z_y = \prod_{i\in \ell^Z_y} \hat{\sigma}_i^z,
\end{equation}
where $\ell^Z_x$ and $\ell^Z_y$ are non-contractible loops on the lattice wrapping around the $x$ and $y$ direction, respectively, as depicted in Fig.~\ref{fig:1}. The ground state in the sector $W^Z_x =  W^Z_y= +1$ can explicitly be constructed as 
\begin{equation} \label{eq:gs11}
\ket{g^Z_{1,1}} = \frac{1}{\mathcal{N}} \prod_v (1 + \hat{A}_{v}) \ket{\Uparrow},
\end{equation}
with respect to the reference product state $\ket{\Uparrow}=\bigotimes_i \ket{\uparrow_i}$, where $\hat \sigma^z \ket{\uparrow} = \ket{\uparrow}$ and where $\mathcal{N}$ is a normalization constant. Other ground states can be obtained from $\ket{g_{1,1}^Z}$ by acting with the operators 
\begin{equation}
\hat W^X_x = \prod_{i\in \ell^X_x} \hat{\sigma}_i^x, \quad \hat W^X_y = \prod_{i\in \ell^X_y} \hat{\sigma}_i^x,
\end{equation}
defined on non-contractible loops $\ell_x^X$ and $\ell_y^X$ on the dual lattice [Fig.~\ref{fig:1}]. These operators anti-commute with $\hat{W}_y^Z$ and $\hat{W}_x^Z$, respectively, while still commuting with one another, with the Hamiltonian and with the vertex and plaquette stabilizers. The other three degenerate ground states are given, for $n,n'\in\{0, 1\}$, by
\begin{equation}
 \ket{g^Z_{1-2n, 1-2n'}} =(\hat{W}^X_x)^{n'} (\hat{W}^X_y)^n \ket{g^Z_{1,1}}.
 \label{eq:gsZothers}
\end{equation}

The $f$-particle vacuum, given the convention used in the main text to define $f$ (defined through the vector $\bm{\delta}$), is found by using a different complete set of stabilizers (which still commute with the Hamiltonian) to label the toric-code eigenstates. This includes $\hat B_p$ and $\hat{A}_{v_p} \hat B_p$, which still have eigenvalue $+1$ in the ground state, and the Wilson loops $\hat W^{ZX}_{x}$ and $\hat W^{ZX}_y$ defined in the main text,
\begin{subequations} \label{eq:wilsonZX}
\begin{align}
& \hat W^{ZX}_x = \hat{W}_x^Z\hat{W}_x^X = \prod_{i\in \ell^{Z}_x} \hat \sigma_i^z \prod_{j\in \ell^X_x} \hat \sigma_j^x\ , \\
& \hat W^{ZX}_y=\hat{W}_y^Z\hat{W}_y^X= \prod_{i\in \ell^{Z}_y} \hat \sigma_i^z \prod_{j\in \ell^X_y} \hat \sigma_j^x\ .
\end{align}
\end{subequations}
The Wilson-loop operators $\hat W^{ZX}_{x}$ and $\hat W^{ZX}_y$ commute also with the operators $\hat{\sigma}_i^z \hat{\sigma}_{i+\bm{\delta}}^x$ that create and move $f$ pairs, such that the $f$ motion preserves the superselection sector defined by $W^{ZX}_{x}$ and $W^{ZX}_y$. 
A flipped eigenvalue of the plaquette stabilizers, ${B}_p=-1$, induced by the action of strings of $\hat{\sigma}_i^z \hat{\sigma}_{i+\bm{\delta}}^x$ operators on the ground state, is interpreted as the presence of an $f$ particle at plaquette $p$.  A flipped eigenvalue of $\hat{A}_{v} \hat B_{v+\bm{\delta}}$, induced by the action of strings of $\hat{\sigma}_i^z$, corresponds to the presence of a $e$ particle on the vertex $v$ instead. Note that the $ZX$ strings $\hat{\sigma}_i^z \hat{\sigma}_{i+\bm{\delta}}^x$ commute with the $\hat{A}_{v} \hat B_{v+\bm{\delta}}$ stabilizers and, thus, cannot create independent $e$ pairs. 

The ground state in the sector labelled by $ W^{ZX}_x= W^{ZX}_y=+1$, which is the initial $f$-particle vacuum considered in the main text, is then obtained as
\begin{subequations}
\label{eq:g11ZX}
\begin{align}
\ket{g^{ZX}_{1,1}} = & \frac{1}{2}(1+\hat{W}^{ZX}_x)(1+\hat{W}^{ZX}_y)\ket{g_{1,1}^Z} ,\\
 = & \frac{1}{2}(1+ \hat{W}^{X}_x + \hat{W}_y^X - \hat{W}_x^X\hat{W}^{X}_y)\ket{g_{1,1}^Z} , \label{eq:gZX2}\\
 = & \frac{1}{2}\left(\ket{g^{Z}_{1,1}} + \ket{g^{Z}_{-1,1}} +\ket{g^{Z}_{1,-1}} -\ket{g^{Z}_{-1,-1}}\right),
 \label{eq:gZX3}
\end{align}
\end{subequations}
in terms of the ground states $\ket{g_{\cdot ,\cdot}^Z}$ of Eqs.~\eqref{eq:gs11} and \eqref{eq:gsZothers}.

\subsection{Entanglement entropy}
\label{sec:schmidt}

\begin{figure}[t]
\centering
\includegraphics{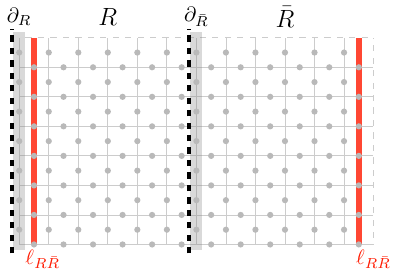}
\caption{Bipartition of the system into non-contractible regions $R$ and $\bar{R}$. The vertices belonging to the boundary regions $\partial_R$ and $\partial_{\bar{R}}$ are indicated in gray shading. The non-contractible operator $\ell_{R\bar{R}}=\prod_{v\in\partial_R}\hat{A}_v$ is depicted in red colour.}
\label{fig:toriccodesketch}
\end{figure}

We analyze the entanglement entropy of the $f$-particle ground state $\ket{g_{1,1}^{ZX}}$ of Eqs.~\eqref{eq:g11ZX} for a bipartition of the system into regions $R$ and $\bar{R}$ as in Fig.~\ref{fig:1}. The regions $R$ and $\bar{R}$ are connected, adjacent and non-contractible along the $y$ direction. Moreover, with the choice of unit cell depicted in Fig.~\ref{fig:1}, they both contain an integer number of unit cells. While the following discussion can be adapted to any choice of entanglement cuts, the bipartition treated here will specifically be relevant for later sections.

We first review the Schmidt decomposition of $\ket{g_{1,1}^{Z}}$, starting from Eq.~\eqref{eq:gs11}. The $\hat A_v$ stabilizers and their products form configurations of closed ``$X$-loops'' $\hat\ell$ (loops of $\hat{\sigma}^x$ operators), such that Eq.~\eqref{eq:gs11} can also be interpreted as a loop expansion, 
\begin{equation}
\label{eq:loopexpan}
\ket{g_{1,1}^{Z}} = \frac{1}{\sqrt{|\mathsf{X}|}}\sum_{\hat \ell\in \mathsf{X}}\hat \ell\ket{\Uparrow}.
\end{equation}
Here, we indicated with $\mathsf{X}$ the group generated by the $\hat A_v$ operators, which has $|\mathsf{X}|=2^{{N_v}-1}$ elements if $N_v$ is the total number of vertices (the ``$-1$'' at the exponent stems from the global even-parity constraint, $\prod_v\hat{A}_v=\mathds{1}$). Moreover, we indicate with $\mathsf{X}_R$ and $\mathsf{X}_{\bar{R}}$ the subgroups of loops that act only on spins in $R$ and $\bar{R}$, respectively. Their cardinality is $|\mathsf{X}_R|=2^{N_{vR}}$ and $|\mathsf{X}_{\bar{R}}|=2^{N_{v\bar{R}}}$, where $N_{vR}$ and $N_{v\bar{R}}$ denote the number of vertex stabilizers fully contained in region $R$ and $\bar{R}$, respectively. The remaining stabilizers straddle the cuts, thus acting on both $R$ and $\bar{R}$ spins, and belong to the set $\mathsf{X}_{R \bar{R}}=\mathsf{X}/(\mathsf{X}_R\cdot  \mathsf{X}_{\bar{R}})$. 
Due to the global parity constraint, the set $\mathsf{X}_{R \bar{R}}$ contains $|\mathsf{X}_{R \bar{R}}|=2^{2 L_{y}-1}$ elements, as $2 L_y$ is the number of vertices intersected by the cuts. In the following, for the set of loops $\mathsf{X}_{\mathscr{R}}$ in a sub-region $\mathscr{R}$, we will indicate with $\ket{\mathscr{R}}$ the equal-weight superposition of loop configurations in $\mathscr{R}$,
\begin{equation}\label{eq:chilambda}
\ket{\mathscr{R}}=  \frac{1}{\sqrt{|\mathsf{X}_{\mathscr{R}}|}} \sum_{\hat\ell\in \mathsf{X}_{\mathscr{R}}} \hat\ell \ket{\Uparrow}_{\mathscr{R}}.
\end{equation}
We can then rewrite Eq.~\eqref{eq:loopexpan} as
\begin{equation}
\label{eq:g11zstart}
\ket{g_{1,1}^{Z}} = \frac{1}{\sqrt{|\mathsf{X}_{R \bar{R}}|}}\sum_{\hat\ell \in \mathsf{X}_{R \bar{R}}}\hat\ell \ket{R}\ket{\bar{R}}.
\end{equation}
It is now convenient to select an independent set of stabilizers generating $\mathsf{X}_{R \bar{R}}$ as follows. We include all stabilizers $\hat{A}_v$ along the cuts, namely such that $v$ belongs to the boundary regions $\partial_R$ and $\partial_{\bar{R}}$ in Fig.~\ref{fig:toriccodesketch}, but we exclude one vertex operator in each of these two regions. The selected stabilizers generate the sub-group $\overline{\mathsf{X}}_{R \bar{R}}$, having cardinality $|\overline{\mathsf{X}}_{R \bar{R}}|=2^{2L_y-2}$. We then complete the set by including the loop $\hat{{\ell}}_{R \bar{R}} = \prod_{v \in \partial_{R}} \hat{A}_v$, that is, the product of all $\hat{A}_v$ with $v$ in region $\partial_R$ [Fig.~\ref{fig:toriccodesketch}]. The sum over loops in $\mathsf{X}_{R \bar{R}}$ in Eq.~\eqref{eq:g11zstart} can then be rewritten like
\begin{equation}
\label{eq:decomptildeg}
\ket{g_{1,1}^{Z}} = \frac{(1 + \hat{\ell}_{R \bar{R}})}{\sqrt{2}} \frac{1}{\sqrt{2^{2L_y-2}}}\sum_{\hat\ell \in \overline{\mathsf{X}}_{R \bar{R}}}  \hat \ell\ket{R}\ket{\bar{R}} .
\end{equation} 

Consider the spins belonging to the shaded region $\partial_{R \bar{R}}=\partial_R\cup \partial_{\bar{R}}$ depicted in Fig.~\ref{fig:toriccodesketch}. Only loops $\hat\ell \in \overline{\mathsf{X}}_{R \bar{R}}$ can act non-trivially on that region and the action of each loop results in a different state $\ket{\lambda}_{\partial_{R \bar{R}}}$ of the boundary spins with an even number of spin flips at each boundary (as the loop must intersect those spins an even number of times; note here that $\hat{{\ell}}_{R \bar{R}}$ winds around the torus along the vertical direction and acts trivially inside $\partial_R$). We express this correspondence by labeling the loops in $\overline{\mathsf{X}}_{R \bar{R}}$ as $\hat \ell_\lambda$, where $\lambda$ enumerates the states $\ket{\lambda}_{\partial_{R \bar{R}}} = \hat\ell_\lambda^{\partial_{R\bar{R}}}\ket{\Uparrow}_{\partial_{R \bar{R}}}$. Here, $\hat\ell_\lambda^{\partial_{R\bar{R}}}$ denotes the restriction of $\hat\ell_\lambda$ to the $\partial_{R \bar{R}}$ region and, similarly, we will use a superscript $\mathscr{R}\in \{ R, \bar{R}\}$ to denote the restrictions of $\hat\ell_\lambda$ to region $\mathscr{R}$, i.e. $\hat\ell_\lambda^{\mathscr{R}}$. Since different loops $\hat \ell_\lambda$ act differently on spins near the cuts, it also holds that $\bra{R}\hat\ell_\lambda^R\hat\ell_{\lambda'}^R\ket{R}=\bra{\bar{R}} \hat\ell^{\bar{R}}_{\lambda}\hat\ell^{\bar{R}}_{\lambda'}\ket{\bar{R}}=\delta_{\lambda\lambda'} $.
Therefore, defining the states 
\begin{equation}\label{eq:chitildelambda}
\ket{R_{\lambda}} =  \hat\ell_{\lambda}^R\ket{R},\quad 
\ket{\bar{R}_{\lambda}} =  \hat\ell_{\lambda}^{\bar{R}}\ket{\bar{R}},
\end{equation}
Eq.~\eqref{eq:decomptildeg} can be rewritten like 
\begin{equation}
\label{eq:schmidtZ}
\ket{g_{1,1}^{Z}} =\frac{(1 + \hat{\ell}_{R \bar{R}})}{\sqrt{2}}\frac{1}{\sqrt{2^{2L_y-2}}}\sum_{\lambda=1}^{2^{2L_y-2}}\ket{R_{\lambda} \bar{R}_{\lambda}}.
\end{equation}
The states $\ket{R_{\lambda}}$ and $\ket{\bar{R}_{\lambda}}$ of Eq.~\eqref{eq:chitildelambda} are equal-weight superpositions of all loops in $R$ and $\bar{R}$, respectively, that are compatible with (terminate at the boundary with) a given state $\ket{\lambda}_{\partial_{ R\bar{R}}}$ of the boundary spins.

We now proceed to evaluate $\ket{g_{1,1}^{ZX}}$ from Eqs.~\eqref{eq:gZX2} and \eqref{eq:schmidtZ}. We first note that, by decomposing $\hat{\ell}_{R \bar{R}}=\hat{\ell}_{R \bar{R}}^{R}\otimes \hat{\ell}_{R \bar{R}}^{\bar{R}}$ into its $R$ and $\bar{R}$ part, the following relation holds,
\begin{equation}
\label{eq:WygA}
\hat{W}_y^X \hat{{\ell}}_{R \bar{R}} = \hat{\ell}^{R}_{R \bar{R}},
\end{equation}
or, equivalently, $\hat{W}_y^X  = \hat{\ell}^{\bar{R}}_{R \bar{R}}$. Hence, the product $\hat{W}_y^X \hat{\ell}_{R \bar{R}}$ acts trivially in $\bar{R}$. 
From Eq.~\eqref{eq:gZX2}, using Eq.~\eqref{eq:WygA} and defining the states
\begin{align}
& \ket{R_{\lambda,s}}= \frac{1}{\sqrt{2}}\left[1+(-1)^s \hat{\ell}^{R}_{R \bar{R}}\right]\ket{R_{\lambda}},\\
& \ket{\bar{R}_{\lambda,s}}=  \frac{1}{\sqrt{2}} \left[1+(-1)^s\hat W_y^X\right]\ket{\bar{R}_{\lambda}},
\end{align}
one finds that $\ket{g_{1,1}^{ZX}}$ can be written in Schmidt form as
\begin{eqnarray}
\ket{g_{1,1}^{ZX}}&& = \frac{1}{\sqrt{2^{2L_y-1}}}\sum_{\lambda=1}^{2^{2L_y-2}} \sum_{s\in \{ 0,1\} } (W_x^X)^s \ket{R_{\lambda,s} \bar{R}_{\lambda,s}}\\
&&= \frac{1}{\sqrt{2^{2 L_y-1}}}\sum_{s\in\{0, 1 \}}\sum_{\lambda=1}^{2^{2L_y-2}} \ket{s_{\lambda}^R s_\lambda^{\bar{R}}}.
\label{eq:schmidtZX}
\end{eqnarray}
Here, we defined the states
\begin{eqnarray}
&& \ket{0_\lambda^{R}} =  \ket{R_{\lambda,0} },\quad \ket{0_\lambda^{\bar{R}}} =  \ket{\bar{R}_{\lambda,0} },\quad  \ket{1_\lambda^{R}} = \hat{W}_x^{X,R} \ket{R_{\lambda,1} }, \quad \ket{1_\lambda^{\bar{R}}} = \hat{W}_x^{X,\bar{R}} \ket{\bar{R}_{\lambda,1} }.
\end{eqnarray}
Since the Schmidt decomposition of Eq.~\eqref{eq:schmidtZX} involves $2^{2L_y -1}$ non-zero Schmidt coefficients with equal value $1/\sqrt{2^{2L_y-1}}$, the von Neumann entropy of the reduced state can easily be computed as 
\begin{align}
\mathcal{S}^{(g)} = - \mathrm{tr} \hat{\rho}_R^{(g)} \log \hat{\rho}_R^{(g)} & = (2 L_y -1)\log 2 ,
\label{eq:SAgroundstateZ}
\end{align}
where we indicated with $\hat{\rho}_R^{(g)}=\mathrm{tr}_{\bar{R}} \ket{g_{1,1}^{ZX}}\!\bra{g_{1,1}^{ZX}}$ the reduced density matrix of the ground state in $R$. The subleading contribution $-\log(2)$ in Eq.~\eqref{eq:SAgroundstateZ} is the topological entanglement entropy~\cite{Hamma2005, Kitaev2006a, Levin2006}, signaling the presence of long-range entanglement. From Eq.~\eqref{eq:schmidtZX}, it further follows that the moments of $\hat{\rho}_R^{(g)}$ have value
\begin{equation}
\label{eq:groundstatemoments}
\mathrm{tr} \left(\hat{\rho}_R^{(g)}\right)^n = 2^{-(n-1)(2L_y-1)} .
\end{equation}

\subsection{Logarithmic negativity}
\label{sec:logneggs}

We next compute the logarithmic negativity between two sub-regions $R^{(1)}$ and $R^{(2)}$ for the $\ket{g_{1,1}^{ZX}}$ ground state, considering the tripartition of the system as in Fig.~1. We will recover results known for other toric-code ground states~\cite{Castelnovo2013, Lee2013}. In particular, for the adjacent non-contractible regions considered, the logarithmic negativity acquires a sub-leading term reflecting long-range entanglement akin to the topological entanglement entropy, known as the topological entanglement negativity. The logarithmic negativity $\mathcal{E}$ is defined as 
\begin{equation}
\mathcal{E} = \log \Vert \hat{\rho}_R^{T_{2}} \Vert_1,
\end{equation}
where $\Vert \hat\rho\Vert_1=\mathrm{tr}\sqrt{\hat\rho^\dag \hat\rho}$ is the trace norm, and is based on the partial transpose $\hat{\rho}_R^{T_{2}}$ of the reduced density matrix with respect to region $R^{(2)}$. It can be obtained via the replica method by considering the analytic continuation of the even series of moments of the partial transpose (R\'enyi negativities)~\cite{Calabrese2012},
\begin{equation} \label{eq:replicanegativity}
\mathcal{E} = \lim_{2n\to 1}\log \mathrm{tr}\left[(\hat{\rho}_R^{T_2})^{2n}\right].
\end{equation}

To determine the partial transpose of $\hat{\rho}_R^{(g)}$ with respect to a subregion $R^{(2)}$, we first derive a convenient decomposition of $\ket{g_{1,1}^{ZX}}$ for a tripartition into regions $R^{(1)}$, $R^{(2)}$ and $\bar{R}$. This can be obtained from the loop expansion of Eq.~\eqref{eq:chilambda} proceeding in the same way as done in Sec.~\ref{sec:schmidt}. We start by expressing $\ket{g_{1,1}^Z}$ as
\begin{eqnarray}
\label{eq:g11Zneg}
\ket{g_{1,1}^{Z}} = \frac{1}{\sqrt{|\mathsf{X}_{R^{(1)}R^{(2)} \bar{R}}|}}\sum_{\hat{\ell}\in\mathsf{X}_{R^{(1)} R^{(2)} \bar{R}}} \hat{\ell} \ket{R^{(1)}R^{(2)} \bar{R} },
\end{eqnarray}
where we considered the groups $\mathsf{X}_{R^{(1)}}$ and $\mathsf{X}_{R^{(2)}}$ of loops fully belonging to $R^{(1)}$ and $R^{(2)}$, respectively, and the set of possible boundary loops $ \mathsf{X}_{R^{(1)}R^{(2)} \bar{R}}\equiv\mathsf{X}/(\mathsf{X}_{R^{(1)}} \mathsf{X}_{R^{(2)}} \mathsf{X}_{\bar{R}})$. 
Since the boundary regions are non-contractible, the product of all $\hat{A}_v$ stabilizers intersected by one vertical cut acts trivially on the boundary spins and it is convenient to address this as done in Eqs.~\eqref{eq:chitildelambda} and \eqref{eq:schmidtZ}. For $\mathscr{R}, \mathscr{R}'\in\{R^{(1)}, R^{(2)}, \bar{R}\}$, we define $\overline{\mathsf{X}}_{\mathscr{R} \mathscr{R}'}$ as the group generated by products of the stabilizers $\hat{A}_v\in\partial_{\mathscr{R}\mathscr{R}'}$ acting at the boundary $\partial_{\mathscr{R}\mathscr{R}'}$ between two regions $\mathscr{R}$ and $\mathscr{R}'$, with one such $\hat{A}_v$ excluded. We further denote with $\hat{\ell}_{\mathscr{R}\mathscr{R}'}$ the non-contractible operators resulting from the product of all stabilizers along $\partial_{\mathscr{R}\mathscr{R}'}$, $\hat{\ell}_{\mathscr{R}\mathscr{R}'}=\prod_{v \in \partial_{\mathscr{R}\mathscr{R}'}}\hat{A}_v$. Only two such operators need be considered (out of three vertical cuts), due to the global parity constraint: the product of all stabilizers across all $\partial_{\mathscr{R}\mathscr{R}'}$ is an element of $\mathsf{X}_{R^{(1)}}\cdot \mathsf{X}_{R^{(2)}}\cdot \mathsf{X}_{\bar{R}}$ and not of $\mathsf{X}_{R^{(1)}R^{(2)} \bar{R}}$. The groups $\overline{\mathsf{X}}_{\mathscr{R} \mathscr{R}'}$ contain $2^{L_{y}-1}$ elements (note that, within individual sub-regions, there is no additional global parity constraint). As each loop $\hat{\ell}\in \overline{ \mathsf{X}}_{\mathscr{R}\mathscr{R}'}$ acts on a different subset of boundary spins in $\partial_{\mathscr{R}\mathscr{R}'}$ and is thus in one-to-one correspondence with a state of those spins, we can label it as $\hat{\ell}_\lambda$ with $\lambda$ identifying the corresponding spin state, as done in Sec.~\ref{sec:schmidt}. We can then rewrite the summation in Eq.~\eqref{eq:g11Zneg}  as follows,
\begin{eqnarray}
\ket{g_{1,1}^{Z}} =&& \frac{(1 + \hat{\ell}_{R^{(1)}R^{(2)}})}{\sqrt{2}}\frac{(1 + \hat{\ell}_{R^{(1)} \bar{R}})}{\sqrt{2}}\frac{1}{\sqrt{2^{3(L_y-1)}}}\sum_{ \lambda_1,\lambda_2, \lambda_3=1 }^{2^{L_y-1}}\ket{R^{(1)}_{\lambda_1\lambda_2}R^{(2)}_{\lambda_2\lambda_3}\bar{R}_{\lambda_3\lambda_1}}, 
\end{eqnarray}
where we extracted the noncontractible loops $\hat{\ell}_{R^{(1)}R^{(2)}}$ and $\hat{\ell}_{R^{(1)} \bar{R}}$ from the sum and defined the states
\begin{equation}
\ket{R_{\lambda\lambda'}} = \hat{\ell}_\lambda^R \hat{\ell}_{\lambda'}^R \ket{R}.
\end{equation}
To determine the decomposition of $\ket{g_{1,1}^{ZX}}$ from $\ket{g_{1,1}^{Z}}$ and Eq.~\eqref{eq:gZX2}, we first note, as done in Sec.~\ref{sec:schmidt}, that the product
\begin{equation}
 \hat{W}_y^X \hat{\ell}_{R^{(1)} \bar{R}} = \hat{\ell}_{R^{(1)} \bar{R}}^{R^{(1)}}
\end{equation}
acts purely in $R^{(1)}$, that is, $ \hat{W}_y^X= \hat{\ell}_{R^{(1)} \bar{R}}^{\bar{R}}$. Moreover, we also note that, as $\hat{\ell}_{R^{(1)}R^{(2)}}^{R^{(1)}}$ and $\hat{\ell}_{R^{(1)}\bar{R}}^{R^{(1)}}$ differ by multiplication by the closed loop $\prod_{v\in R^{(1)}}\hat{A}_v \in \mathsf{X}_{R^{(1)}}$ fully contained in $R^{(1)}$,  and since $\ket{R^{(1)}}$ is a condensate of closed loops in $R^{(1)}$, it holds that 
\begin{equation}
\hat{\ell}_{R^{(1)}R^{(2)}}^{R^{(1)}} \ket{R^{(1)}} =   \hat{\ell}_{R^{(1)}\bar{R}}^{R^{(1)}} \prod_{v\in R^{(1)}}\hat{A}_v \ket{R^{(1)}}  =  \hat{\ell}_{R^{(1)} \bar{R}}^{R^{(1)}} \ket{R^{(1)}} .
\end{equation}
Then, by introducing the orthogonal states
\begin{eqnarray}
&& \ket{R^{(1)}_{\lambda\lambda',s} }= \frac{1}{\sqrt{2}} \left[1+(-1)^s \hat{\ell}_{R^{(1)}\bar{R}}^{R^{(1)}}\right]\ket{R^{(1)}_{\lambda\lambda'}},\nonumber \\
&& \ket{R^{(2)}_{\lambda\lambda',s} }= \frac{1}{\sqrt{2}} \left[1+(-1)^s \hat{\ell}_{R^{(1)}R^{(2)}}^{R^{(2)}}\right] \ket{R^{(2)}_{\lambda\lambda'}},\nonumber  \\
&& \ket{\bar{R}_{\lambda \lambda',s} }= \frac{1}{\sqrt{2}} \left[1+(-1)^s \hat{\ell}_{R^{(1)}\bar{R}}^{\bar{R}}\right] \ket{\bar{R}_{\lambda \lambda'}},
\end{eqnarray}
and
\begin{align}
& \ket{0_{\lambda\lambda'}^{(1)}} = \ket{R^{(1)}_{\lambda\lambda',0} },  && \ket{0_{\lambda\lambda'}^{(2)}} = \ket{R^{(2)}_{\lambda\lambda',0}}, && \ket{0_{\lambda\lambda'}^{\bar{R}}} = \ket{\bar{R}_{\lambda\lambda',0} },\nonumber \\
& \ket{1_{\lambda\lambda'}^{(1)}} = \hat{W}_x^{X,R^{(1)}}\ket{R^{(1)}_{\lambda\lambda',1} },  && \ket{1_{\lambda\lambda'}^{(2)}} =  \hat{W}_x^{X,R^{(2)}}\ket{R^{(2)}_{\lambda \lambda',1}}, && \ket{1_{\lambda \lambda'}^{\bar{R}}} =\hat{W}_x^{X,\bar{R}}  \ket{\bar{R}_{\lambda \lambda',1} } ,
\end{align}
which fulfil $\langle s^{\mathscr{R}}_{\lambda\lambda'}| {s}^{\prime \mathscr{R}}_{\mu\mu'}\rangle=\delta_{ss'}\delta_{\lambda\mu}\delta_{\lambda'\mu'}$ for $s,s'\in\{0, 1\}$ and region $\mathscr{R}$, a few manipulations lead to the tripartite decomposition
\begin{align}
\label{eq:schmidttripartition}
\ket{g_{1,1}^{ZX}} =& \frac{1}{\sqrt{2^{3 L_y-2}}} \sum_{\lambda_1, \lambda_2, \lambda_3=1}^{2^{L_y-1}} \sum_{s\in\{0,1\}} \ket{s_{\lambda_1\lambda_2}^{(1)} s_{\lambda_2\lambda_3}^{(2)} s_{\lambda_3\lambda_1}^{\bar{R}}} .
\end{align}

We can also construct an alternative decomposition in the form
\begin{equation} 
\ket{g_{1,1}^{ZX}} = \frac{1}{\sqrt{2^{3 L_y-1}}} \sum_{\substack{w,\bar{w},\bar{\bar{w}}=\pm, \\ : w \bar{w}\bar{\bar{w}}=1} } \sum_{\lambda_1, \lambda_2, \lambda_3=1}^{2^{L_y-1}}  \ket{w_{\lambda_1\lambda_2}^{(1)} \bar{w}_{\lambda_2\lambda_3}^{(2)} \bar{\bar{w}}_{\lambda_3\lambda_1}^{\bar{R}}} ,
\end{equation}
where the states $\ket{w_{\lambda \lambda'}^{\mathscr{R}}}$ in each sub-region $\mathscr{R}$ are also eigenstates of the horizontal non-contractible Wilson loop operator restricted to that sub-region, 
\begin{equation}
\hat W_x^{ZX, \mathscr{R}} \ket{w_{\lambda \lambda'}^{\mathscr{R}}} = w \ket{w_{\lambda \lambda'}^{\mathscr{R}}},
\end{equation}
with $w=\pm$. The states $\ket{w_{\lambda \lambda'}^{\mathscr{R}}}$ are defined as
\begin{equation}
\ket{w_{\lambda \lambda'}^{\mathscr{R}}} =\frac{1}{\sqrt{2}}\left( \ket{0_{\lambda\lambda'}^{\mathscr{R}}}+ w \ket{1_{\lambda\lambda'}^{\mathscr{R}}} \right).
\end{equation}
%
%
The reduced density matrix $\hat{\rho}_R^{(g)}$ in $R$ can then equivalently be written in the two forms
\begin{align}
\hat{\rho}_R^{(g)}& = \frac{1}{2^{3L_y-2}} \sum_{\substack{s\in \{0, 1\}}} \sum_{\bm{\lambda}} \ket{s_{\lambda_1\lambda_2}^{(1)},s_{\lambda_2\lambda_3}^{(2)}}\bra{s_{\lambda_1\lambda_4}^{ (1)},s_{\lambda_4\lambda_3}^{(2)}},  \nonumber \\
& = \frac{1}{2^{3L_y-1}}  \sum_{
\substack{v \bar{v}, w,\bar{w}=\pm,\\ : w\bar{w}=v \bar{v}} } \sum_{\bm{\lambda}}  \ket{w_{\lambda_1\lambda_2}^{(1)} \bar{w}_{\lambda_2\lambda_3}^{(2)}}  \bra{v_{\lambda_1 \lambda_4}^{(1)} \bar{v}_{\lambda_4 \lambda_3}^{(2)}},
\label{eq:decotrip}
 \end{align}
 where we used the short-hand notation $\sum_{\bm{\lambda}}=\sum_{\lambda_1, \lambda_2, \lambda_3, \lambda_4}  $. 
We derived both representations, as it will be convenient for some of the calculations presented in later sections. 
To compute the logarithmic negativity between $R^{(1)}$ and $R^{(2)}$, we compute the partial transpose of the reduced density matrix in $R$,
\begin{align}
\label{eq:rhoAgtransposeeta}
\left(\hat{\rho}_R^{(g)}\right)^{T_{2}}& = \frac{1}{2^{3L_y-2}} \sum_{s\in \{0, 1\}} \sum_{\bm{\lambda}} \ket{s_{\lambda_1\lambda_2}^{(1)},s_{\lambda_4\lambda_3}^{ (2)}}\bra{s_{\lambda_1\lambda_4}^{ (1)},s_{\lambda_2 \lambda_3}^{(2)}}, \\
& = \frac{1}{2^{3L_y-1}} \sum_{
\substack{v, \bar{v}, w,\bar{w}=\pm,\\ : w\bar{w}=v \bar{v}} } \sum_{\bm{\lambda}}  \ket{w_{\lambda_1\lambda_2}^{(1)} \bar{v}_{\lambda_4 \lambda_3}^{(2)}}  \bra{v_{\lambda_1 \lambda_4}^{(1)}  \bar{w}_{\lambda_2\lambda_3}^{(2)}}.
\label{eq:rhoAgtransposew}
 \end{align}
Following the replica approach, we determine the even series of moments of the partial transpose, starting from $n=2$. Using the representation of Eq.~\eqref{eq:rhoAgtransposeeta}, one finds
\begin{align}
& \left[\left(\hat{\rho}_R^{(g)}\right)^{T_{2}}\right]^2  =\frac{1}{(2^{3L_y-2})^2} 
\sum_{s \in \{0, 1\}} \sum_{\bm{\lambda}} \ket{s_{\lambda_1\lambda_2}^{(1)},s_{\lambda_4\lambda_3}^{ (2)}}\! \bra{s_{\lambda_1\lambda_2}^{(1)},s_{\lambda_4\lambda_3}^{(2)}}.
\end{align}
As $[(\hat{\rho}_R^{(g)})^{T_{2}}]^2 $ is diagonal, the negativity is immediately obtained as
\begin{eqnarray}
\label{eq:negativitygs}
\mathcal{E}^{(g)}= &&\log \mathrm{tr} \sqrt{\left[\left(\hat{\rho}_R^{(g)}\right)^{ T_{2}}\right]^2} =   
  (L_y-1) \log(2).
\end{eqnarray}
The negativity thus exhibits an area law term proportional to the length of the non-contractible vertical cut and a sub-leading contribution of topological origin, as known for other toric-code ground states~\cite{Castelnovo2013, Lee2013}.

\section{Generalized Jordan-Wigner mapping}
\label{sec:spinfermionmapping}

In this section, we discuss how to construct the fermionic representation of the toric code used in the main text, which maps the Hilbert space of spins to a many-fermion system describing the emergent $f$ quasiparticles. In this procedure, we adapt the approach of Ref.~\cite{Chen2022}, especially in view of facilitating the entanglement analysis presented in Secs.~\ref{sec:entropydecomposition} and \ref{sec:decompnegativity} for the entanglement cuts chosen.

As discussed in Sec.~\ref{sec:recap}, pairs of fermionic quasiparticles can be created from the toric-code ground state through the action of operators $\hat{\sigma}_i^z \hat{\sigma}_{i+\bm{\delta}}^x$. Moreover, the stabilizers that detect the presence of $e$ and $f$ particles are $\hat{A}_{v} \hat{B}_{v+\bm{\delta}}$ and $\hat{B}_p$, respectively. These stabilizers are first mapped to two flavours of pseudospins 1/2 living on the vertices and plaquettes, respectively,
\begin{equation}
\hat{A}_{v} \hat{B}_{v+\bm{\delta}} \leftrightarrow \hat{Z}_{v}^e, \qquad \hat{B}_p \leftrightarrow \hat{Z}_p^f,
\end{equation}
with spin up (e.g., $  Z_v^e = +1$) corresponding to the absence of particles and spin down ($  Z_v^e = -1$) to the presence of a particle on that vertex or plaquette. Since we are studying the behaviour of $f$ particles, we assume that no vertex defects $e$ are present in the lattice, such that $ Z_v^e =+1$ for all $v$. The noncontractible Wilson loop operators $\hat{W}_{x}^{ZX}$ and $\hat{W}_{y}^{ZX}$, defined in Eq.~\eqref{eq:wilsonZX}, are also mapped to pseudospins
\begin{equation}
\hat{W}_x^{ZX} \leftrightarrow \hat{Z}^{W}_{x} , \quad \hat{W}_y^{ZX} \leftrightarrow \hat{Z}^{W}_{y} ,
\end{equation}
whose eigenvalues encode the topological superselection sector.

\begin{figure}
\centering
\includegraphics{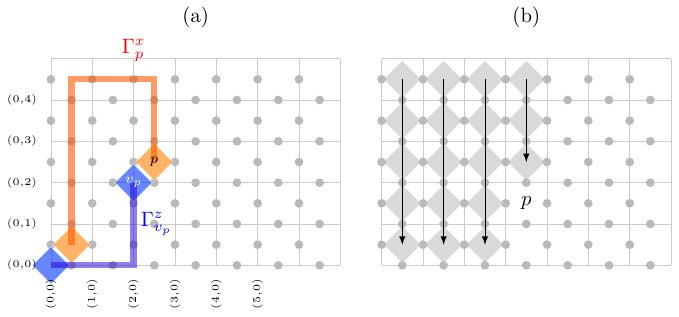}
\caption{(a) Convention for the $Z$ and $X$ strings defining $\hat{X}_1^f\hat{X}_p^f$ in Eq.~\eqref{eq:XfXf}. (b) Path used to define the Jordan-Wigner mapping in Eqs.~\eqref{eq:JWmapping} and \eqref{eq:JWstring}.  }
\label{fig:JWmapping}
\end{figure}

We next define operators that anti-commute with $\hat{Z}_p^f$, associated with the creation of $f$ particles. To this end, we attribute coordinates $(x, y)$ to the lattice vertices as in Fig.~\ref{fig:JWmapping}(a) and denote with $(x_p, y_p)$ the coordinates of the vertex $v_p$ such that $v_p+\bm{\delta}=p$. Since physical states admit only an even number of $f$ particles, operators $\hat{X}_p^f$ that anti-commute with only one $\hat{Z}_p^f$ have no non-zero matrix elements between physical states. However, one can define 
\begin{align}
\label{eq:XfXf}
\hat{X}_1^f \hat{X}_p^f \leftrightarrow \prod_{i \in \Gamma^z_{v_p}} \hat{\sigma}_i^z \prod_{j \in \Gamma^x_p} \hat{\sigma}_{j}^x .
\end{align}
The paths $\Gamma_{v_p}^z$ and $\Gamma_p^x$ in Eq.~\eqref{eq:XfXf} are chosen as follows and depicted in Fig.~\ref{fig:JWmapping}(a). The path $\Gamma_{v_p}^z$ emanates from the vertex at coordinate (0,0) and reaches $v_p$ by (i) extending horizontally until $(x_p,0)$ and (ii) ascending vertically until $(x_p, y_p)$. The path $\Gamma_p^x$ emanates from the plaquette associated with $(0,0)$, ascends vertically along the dual lattice until the top-most plaquette, continues horizontally and finally descends vertically until $p$. The creation of a particle at plaquette $p$ is thus conventionally paired with the creation of a particle at the bottom-left of the lattice. 

\noindent Proceeding with the definition of Eq.~\eqref{eq:XfXf}, the desired anti-commutation and commutation relations are fulfilled, 
\begin{align}
& \{ \hat{X}_1^f \hat{X}_p^f, \hat{Z}_p^f \} = 0, \nonumber \\
&[\hat{X}_1^f \hat{X}_p^f, \hat{Z}_{p'}^f ] = 0, \quad \mbox{for } p'\ne 1,p, \nonumber \\
& [\hat{X}_1^f \hat{X}_p^f, \hat{X}_1^f \hat{X}_{p'}^f] = 0.
\end{align}
We introduce fermionic operators via the Jordan-Wigner mapping
\begin{equation} 
\label{eq:JWmapping}
\hat{Z}_p^f \leftrightarrow -i \hat\gamma_{2p-1} \hat\gamma_{2p} , \qquad  \hat{\mathscr{T}}_{0,{p}} \hat{X}_p^f  \leftrightarrow \hat\gamma_{2p},
\end{equation}
with
\begin{equation}
\label{eq:JWstring}
\hat{\mathscr{T}}_{p,{p'}} = \prod_{p<p''<{p'}} \hat{Z}_{p''}^f,
\end{equation}
where $\hat\gamma_p$ are Majorana operators, $\{ \hat\gamma_p , \hat{\gamma}_{p'}\} = 2\delta_{pp'}$, 
from which one can construct complex fermionic operators $\hat{f}_p$ and $\hat{f}_p^\dag$ as
\begin{equation}
\hat\gamma_{2p-1} = \hat{f}_p + \hat{f}_p^\dag,  \qquad \hat\gamma_{2p} = -i (\hat f_p - \hat f_p^\dag).
\end{equation}
The ordering of the plaquettes $p$ defining the Jordan-Wigner string $\hat{\mathscr{T}}_{p{p'}}$ in Eq.~\eqref{eq:JWstring} is chosen as in Fig.~\ref{fig:JWmapping}(b), starting from the top left plaquette (throughout the paper, we are assuming that the left edge in the figures corresponds to the boundary of the region $R^{(1)}$, as in Fig.~1). 
The vacuum of the fermion operators in a given superselection sector, $\ket{0}$, corresponds to the toric code ground state in the spin picture,
\begin{equation}
\ket{g_{1,1}^{ZX}} \leftrightarrow \ket{0} .
\end{equation} 

\textit{Mapping of the Hamiltonian.} The pseudospin operators that create $f$ pairs on neighbouring plaquettes map to hopping and pairing operators of fermions, as expected: for a site $i$ on a vertical bond to the left of the plaquette $p$, one finds (with a slight abuse of notation on the indices)

\begin{equation}
\begin{tikzpicture}[scale=.9]
\draw[gray!40, thin] (0,0) grid (2,2);
\foreach \x in {0.5, 1.5}
  \foreach \y in {0,1}
    \fill[gray!55] (\x,\y) circle (2pt);
\foreach \x in {0, 1}
  \foreach \y in {0.5,1.5}
    \fill[gray!55] (\x,\y) circle (2pt);

\node[blue, left] at (0,.5) {$\hat\sigma_i^z$};
\draw[line width=3pt, blue!80!purple, opacity=.5] (0,0)--(0,1);
\node[above, red] at (0.5,1) {$\hat\sigma_{i+\bm{\delta}}^x$};
\draw[line width=3pt, orange!80!red, opacity=.5] (0,1)--(1,1);

\node[] at (0.5,.5) {\scriptsize $p$};
\node[] at (0.5,1.7) {\scriptsize $p+\mathbf{y}$};
\node[] at (1.5,.5) {\scriptsize $p+\mathbf{x}$};
\node at (5, -.25) {
$\begin{aligned}
& \leftrightarrow \hat X^f_p \hat X^f_{p+\mathbf{y}},\\
& \leftrightarrow i \hat\gamma_{2(p+\mathbf{y})-1} \hat\gamma_{2p},\\
& = (\hat f_{p+\mathbf{y}} + \hat f_{p+\mathbf{y}}^\dag)(\hat f_p - \hat f_p^\dag),
\end{aligned}$
};
\end{tikzpicture}
\label{eq:vertXX}
\end{equation}
which amounts to hopping and pairing along the vertical direction. 
For site $i$ on the horizontal bond below plaquette $p$ 

\begin{equation}
\begin{tikzpicture}[scale=.9]
\draw[gray!40, thin] (0,0) grid (2,2);
\foreach \x in {0.5, 1.5}
  \foreach \y in {0,1}
    \fill[gray!55] (\x,\y) circle (2pt);
\foreach \x in {0, 1}
  \foreach \y in {0.5,1.5}
    \fill[gray!55] (\x,\y) circle (2pt);

\node[blue, below] at (.5,0) {$\hat\sigma_i^z$};
\draw[line width=3pt, blue!80!purple, opacity=.5] (0,0)--(1,0);
\node[left, red] at (1,.5) {$\hat\sigma_{i+\bm{\delta}}^x$};
\draw[line width=3pt, orange!80!red, opacity=.5] (1,0)--(1,1);

\node[] at (0.5,1.7) {\scriptsize $p+\mathbf{y}$};
\node[] at (1.5,.5) {\scriptsize $p+\mathbf{x}$};
\node at (6, -.25) {
$\begin{aligned}
& \leftrightarrow  -\hat{Z}_p^f \hat{\mathscr{T}_{0,p}} \hat{X}^f_p \hat{\mathscr{T}}_{0,p+\mathbf{x}} \hat X^f_{p+\mathbf{x}} \prod_{v\ \mathrm{ above }\ v_{p+\mathbf{x}}} \hat Z_v^e,\\
& \leftrightarrow i \hat\gamma_{2p-1} \hat\gamma_{2(p+\mathbf{x})} \prod_{v\ \mathrm{ above }\ v_{p+\mathbf{x}}} \hat Z_v^e ,\\
& = (\hat{f}_p+\hat{f}_p^\dag)(\hat{f}_{p+\mathbf{x}}-\hat{f}_{p+\mathbf{x}}^\dag) \prod_{v\ \mathrm{ above }\ v_{p+\mathbf{x}}} \hat Z_v^e.
\end{aligned}$
};
\end{tikzpicture}
\label{eq:horizXX}
\end{equation}
The product of $e$-particle parities $\hat{Z}_v^e$ appearing in Eq.~\eqref{eq:horizXX} reproduces in an explicit manner the mutual statistics between $f$ and $e$: an $f$ particle hopping around a loop will acquire a phase $0$ or $\pi$ depending on the parity of the number of $e$ particles inside that loop. As we are assuming that no $e$ particles are present in the system, this term becomes irrelevant in the following. All in all, the driven toric-code Hamiltonian studied in the main text maps to
\begin{eqnarray}
-\frac{g}{2}\sum_v\hat{A}_v&&- \frac{g}{2}\sum_p \hat{B}_p -\Omega(t)\sum_i \hat{\sigma}_i^z \hat{\sigma}_{i+\bm{\delta}}^{x} \nonumber \\ && \longleftrightarrow  g\sum_p ( \hat{Z}_{v_p}^e + 1)\hat{f}_p^\dag \hat{f}_p -\frac{g}{2}\sum_v \hat{Z}_v^e  - \Omega(t)\sum_{\mathbf{r}=\mathbf{x},- \mathbf{y}} \sum_p \hat{\mathscr{Z}}^e_{p, p+\mathbf{r}} \left(\hat{f}_{p+\mathbf{r}}^\dag \hat{f}_p^\dag + \hat{f}_{p+\mathbf{r}}^\dag \hat{f}_p + \mathrm{H.c.}\right),
\end{eqnarray}
where we defined the string of $e$ parities
\begin{equation}
 \hat{\mathscr{Z}}^e_{p, p+\mathbf{x}} = \prod_{v\ \mathrm{above}\ v_{p+\mathbf{x}}} \hat{Z}_v^e, \qquad \hat{\mathscr{Z}}^e_{p, p-\mathbf{y}}=1, 
\end{equation}
and reduces to Eq.~(3) in the absence of $e$ particles in the system (i.e., $Z_v^e=+1$ for all $v$) and defining the chemical potential $\mu_f=2g$ for $f$ particles. 
Note that the mapping in Eq.~\eqref{eq:vertXX} and \eqref{eq:horizXX} applies to $\hat\sigma^z_i \hat\sigma_{i+\bm{\delta}}^x$ terms that do not straddle the system's boundary: for the latter, the mapping also depends on the Wilson-loop pseudospins $\hat{Z}^{W}_x$ and $\hat{Z}^{W}_y$, such that the choice of superselection sector translates into a choice of boundary conditions (either periodic or anti-periodic) for the fermion hopping and pairing, see Refs.~\cite{Chen2022, Petiziol2024anyons}. The choice $W^{ZX}_x=W^{ZX}_y=1$ corresponds to periodic boundary conditions in both directions. 

\textit{Mapping of the contractible Wilson-loop operators.} In the main text, the fate of pure-state topological order in the long-time Floquet equilibration is analyzed by probing the decay of the expectation values of closed Wilson operators
\begin{equation}
\hat W^Z_{ \square} =\prod_{p\in \mathrm{Int}(\square)} \hat{B}_p,
\end{equation}
on contractible square $Z$-loops $\square$. 
Indeed, the time-periodic forcing disrupts the $Z$-loop structure enforced by the ground state constraints $B_p=1$ (while it preserves the loop structure associated with $e$ particles, namely, loops given by products of $\hat{A}_{p-\bm{\delta}}\hat{B}_{p}$ stabilizers). As the stabilizers $\hat B_p$ map to Majorana fermion parities in the generalized-Jordan-Wigner mapping, $\hat B_p \leftrightarrow -i \gamma_{2p-1}\gamma_{2p}$ [Eq.~\eqref{eq:JWmapping}], $\hat W_{\square}^Z$ has the fermionic representation
\begin{equation}
\label{eq:wilsontoparity}
\hat W^Z_{\square} \quad  \leftrightarrow \quad \prod_{p\in \mathrm{Int}(\square) } (-i\hat\gamma_{2p-1}\hat\gamma_{2p}).
\end{equation}
By Wick's theorem, the expectation values $W^Z_{\square}$ can be expressed as the Pfaffian of the Majorana covariance matrix $\Gamma_{ij} =(i/2) \langle [\hat\gamma_i, \hat \gamma_j ]\rangle$ restricted to the modes inside $\square$, $|W^Z_{\square}| = |\mathrm{Pf} (\Gamma_\square)|$~\cite{Surace2021}. To determine the perimeter-to-area law crossover in Fig.~2 in numerical simulations, we rather computed directly $-\log | W^Z_{\square}| = -(1/2) \log |\det \Gamma_\square |$, which can be computed more efficiently with NumPy routines~\cite{Harris2020}. 

\begin{figure}
\centering
\includegraphics{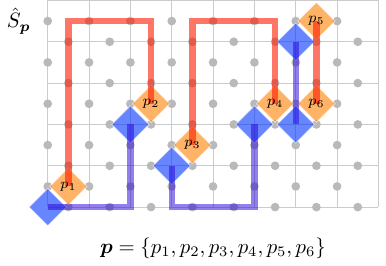}
\caption{Example of a $ZX$-string configuration $\hat{S}_{\bm p}$ corresponding to an $f$ particle configuration $\bm{p}$, as defined in Eq.~\eqref{eq:Sstring} [up to the $(-1)^{M_{\bm{p}}}$ factor].}
\label{fig:Sp}
\end{figure}

\textit{Mapping of fermionic Fock states.} We finally determine the mapping for a Fock state $\ket{\bm{p}}$ of an even number $2M_{\bm{p}}$ of fermions in a fixed superselection sector, with $\bm{p}=\{p_1, \dots p_{2M_{\bm{p}}}\}$ denoting the occupied plaquettes,
\begin{equation}
\label{eq:ketp}
\ket{\bm{p}} = \hat{f}_{p_1}^\dag \dots \hat{f}_{p_{2M_{\bm{p}}}}^\dag \ket{0},
\end{equation}
and $p_1<p_2<\dots<p_{2M_{\bm{p}}}$ respecting the plaquette ordering defined above [Fig.~\ref{fig:JWmapping}(b)]. From Eq.~\eqref{eq:JWmapping}, an operator creating a particle pair, $\hat{f}_p^\dag \hat{f}_{p'}^\dag$ with $p'>p$, thus corresponds, in the pseudo-spin picture, to
\begin{align}
\hat{f}_p^\dag \hat{f}_{p'}^\dag & =  \frac{1}{4}[\hat\gamma_{2p-1} \hat\gamma_{2p'-1}-i \hat\gamma_{2p-1} \hat\gamma_{2p'} - i \hat\gamma_{2p}\hat \gamma_{2p'-1} -\hat \gamma_{2p}\hat \gamma_{2p'}] , \nonumber\\
& \leftrightarrow - \hat{X}_p^f \hat{\mathscr{T}}_{p,{p'}} \hat{X}_{p'}^f \frac{(1+\hat{Z}_p^f)}{2}\frac{(1+ \hat{Z}_{p'}^f)}{2}.
\end{align}
Recalling that $\hat{Z}_p^f$ acts trivially on the ground state for all $p$, the state $\ket{\bm{p}}$ has spin representation  
\begin{equation}
\label{eq:ptoS}
\ket{\bm{p}} \leftrightarrow \hat{{S}}_{\bm{p}} \ket{g^{ZX}_{1,1}},
\end{equation}
where $\hat{{S}}_{\bm{p}}$ denotes the configuration of $ZX$-strings that produces the fermion configuration $\bm{p}$. In pseudospin representation, $\hat{{S}}_{\bm{p}}$ reads as
\begin{equation}
\label{eq:Sstring}
\hat{ {S}}_{\bm{p}} \leftrightarrow (-1)^{M_{\bm{p}}}\hat{X}_{p_1}^f \hat{X}_{p_2}^f \dots \hat{X}_{p_{2M_{\bm{p}}-1}}^f \hat{X}_{p_{2M_{\bm{p}}}}^f
\end{equation}
and an example in spin representation is shown in Fig.~\ref{fig:Sp}.

For a generic state $\ket{\psi}$ in the Hilbert space of $f$ excitations created from the ground state $\ket{g_{1,1}^{ZX}}$, we thus obtain the dual spin-fermion representation
\begin{equation}
\label{eq:psif}
\ket{\psi} = \sum_{\bm{p}}c_{\bm{p}} \hat{{S}}_{\bm{p}} \ket{g_{1,1}^{ZX}}\quad   \leftrightarrow \quad  \sum_{\bm{p}}c_{\bm{p}} \ket{\bm{p}}  .
\end{equation}
In Secs.~\ref{sec:entropydecomposition} and \ref{sec:decompnegativity}, we will discuss how entanglement properties in the two (spin vs fermion) representations depend on one another.

\section{Entropy decomposition} 
\label{sec:entropydecomposition}

In this section, we use the generalized Jordan-Wigner mapping of Sec.~\ref{sec:spinfermionmapping} to show that the entanglement entropy of a generic toric-code state that contains fermion quasiparticles, for certain bipartitions, can be decomposed into two contributions as in Eq.~(4) of the main text: One is given by the entanglement entropy of the ground state and the other by the entanglement entropy of a fermionic state describing the $f$ quasiparticles. 

We consider a generic $\ket{\psi}$ of the form of Eq.~\eqref{eq:psif} and a bipartition of the spin system into subsystems $R$ and $\bar{R}$ of the type depicted in Fig.~1. The regions $R$ and $\bar{R}$ are non-contractible along the vertical direction and, for the choice of unit cell shown in Fig.~1, each region contains an integer number of unit cells. Considering $\hat\sigma_i^z \hat\sigma^x_{i+\bm{\delta}}$ terms that can create anyon pairs or induce anyon hopping, the entanglement cuts can intersect only $\hat\sigma_i^z \hat\sigma^x_{i+\bm{\delta}}$ terms that induce horizontal tunnelling and pairing across the cut (namely, for site $i$ on a horizontal bond), but not terms that induce vertical fermion motion (for site $i$ on a vertical bond). As we shall see below, this makes such cuts particular in the light of the convention to define the fermions adopted here, as it provides a criterion for translating the bipartition for spins into a bipartition of fermionic sites. 

Let $G$ be the set of all even-parity fermion configurations $\bm{p}$ in the system, having cardinality $|G|=2^{N_p-1}$, if $N_p$ is the total number of plaquettes. 
We further denote with $G_R^{(\mathrm{e})}$ and $G_R^{(\mathrm{o})}$ the configurations in $R$ with an even (`e') and odd (`o') number of anyons, respectively (and the same for $G_{\bar{R}}$). If $N_{R}$ ($N_{\bar{R}}$) is the number of plaquette operators $\hat B_p$ that can be violated by a $ZX$-string acting fully in $R$ ($\bar{R}$), then $|G_R^{\rm (e)}| = 2^{N_R-1}$ and $|G_{\bar{R}}^{(\rm e)}| = 2^{N_{\bar{R}}-1}$. The set $G_R^{(\rm e)}$ ($G_{\bar{R}}^{(\rm e)}$) can also be thought of as the set of $f$ configurations that can be created by a $ZX$-string acting solely in $R$ ($\bar{R}$), where the $ZX$-string convention defined in Eq.~\eqref{eq:XfXf} is understood (other conventions can be converted into the one we use through multiplication by stabilizers).

Before proceeding forward, we point out that the above partition of $f$ configurations $\bm{p}$ between $R$ and $\bar{R}$ is unambiguous because there are no configurations that can be generated both by acting solely in $R$ and solely in $\bar{R}$. This property is specific to the entanglement cut chosen, in the light of our convention to define the fermions, i.e., how the plaquette and vertex defects defining an $f$ particle are positioned with respect to one another (which, in turn, is motivated by the orientation of the vector $\bm{\delta}$ that defines the driving Hamiltonian). These observations are relevant to clarify that the decomposition of the entropy and logarithmic negativity, presented in the following and used in the main text, does not apply without modification to arbitrary choices of the cut, that are different from those analyzed in this work.

\subsection{Entropy of the fermionic state} 
\label{sec:entang_fermions}

Let us first analyse the generic state $\ket{\psi}=\sum_{\bm{p}}c_{\bm{p}}\ket{\bm{p}}$ of Eq.~\eqref{eq:psif} in fermionic representation. The fermionic density operator $\hat{\rho}^{(f)} = \ket{\psi}\!\bra{\psi}$ has even fermion parity, hence, it can be decomposed by resolving the parity in each sub-region (even-even or odd-odd in $R$-$\bar{R}$) as
\begin{equation}
\hat{\rho}^{(f)} = \sum_{\substack{\bm{p}_R, \bm{p}'_R \in G_R^{(\mathrm{e})} \\\bm{p}_{\bar{R}}  , \bm{p}'_{\bar{R}} \in G_{\bar{R}}^{(\mathrm{e})}}} \rho^{(f)}_{\bm{p}_R \bm{p}_{\bar{R}},\bm{p}'_R \bm{p}'_{\bar{R}} } \ket{\bm{p}_R \bm{p}_{\bar{R}}} \bra{\bm{p}'_R \bm{p}'_{\bar{R}}} + \sum_{\substack{\tilde{\bm{p}}_R, \tilde{\bm{p}}'_R \in G_R^{(\mathrm{o})} \\ \tilde{\bm{p}}_{\bar{R}} , \tilde{\bm{p}}'_{\bar{R}} \in G_{\bar{R}}^{(\mathrm{o})}}} \rho^{(f)}_{\tilde{\bm{p}}_R \tilde{\bm{p}}_{\bar{R}},\tilde{\bm{p}}'_R \tilde{\bm{p}}'_{\bar{R}} } \ket{\tilde{\bm{p}}_R \tilde{\bm{p}}_{\bar{R}}} \bra{\tilde{\bm{p}}'_R \tilde{\bm{p}}'_{\bar{R}}},
\end{equation}
where we denoted with a tilde, $\tilde{\bm{p}}$, configurations with an odd number of particles in $R$ and $\bar{R}$, $\tilde{\bm{p}}_R \in G_R^{(\mathrm{o})}$ and  $\tilde{\bm{p}}_{\bar{R}} \in G_{\bar{R}}^{(\mathrm{o})}$. For a pure $\ket{\psi}$ as in Eq.~\eqref{eq:psif}, the density matrix has the form $\rho^{(f)}_{\bm{p}_R \bm{p}_{\bar{R}},\bm{p}'_R \bm{p}'_{\bar{R}} } = c_{\bm{p}_R\bm{p}_{\bar{R}}} c^*_{\bm{p}'_R \bm{p}'_{\bar{R}}}.$ 
The reduced density matrix, tracing out the modes in region $\bar{R}$~\cite{Vidal2021}, is
\begin{equation}
\label{eq:rhoAfermions}
 \hat{ \rho}^{(f)}_R = \sum_{\bm{p}_R, \bm{p}'_R \in G_R^{(\mathrm{e})}}\rho^{(f, \mathrm{e})}_{\bm{p}_R, \bm{p}'_R} \ket{\bm{p}_R}\bra{\bm{p}'_R} +
   \sum_{\tilde{\bm{p}}_R, \tilde{\bm{p}}'_R \in G_R^{(\mathrm{o})}}  \rho^{(f, \mathrm{o})}_{\tilde{\bm{p}}_R, \tilde{\bm{p}}'_R} \ket{\tilde{\bm{p}}_R}\bra{\tilde{\bm{p}}'_R},
\end{equation}
with
\begin{align}
\label{eq:rhofe}
\rho^{(f, \mathrm{e})}_{\bm{p}_R, \bm{p}'_R} = \sum_{\bm{p}_{\bar{R}} \in G_{\bar{R}}^{(\mathrm{e})}} \rho^{(f)}_{\bm{p}_R \bm{p}_{\bar{R}},\bm{p}'_R \bm{p}_{\bar{R}} }, \\
\label{eq:rhofo}
\rho^{(f, \mathrm{o})}_{\tilde{\bm{p}}_R, \tilde{\bm{p}}'_R} = \sum_{\tilde{\bm{p}}_{\bar{R}} \in G_{\bar{R}}^{(\mathrm{o})}}\rho^{(f)}_{\tilde{\bm{p}}_R \tilde{\bm{p}}_{\bar{R}},\tilde{\bm{p}}'_R \tilde{\bm{p}}_{\bar{R}} }.
\end{align}
Due to the global even parity, the reduced density matrix has thus a block-diagonal structure in the $\{\ket{\bm{p}_R}, \ket{\tilde{\bm{p}}_R}\}$ basis, 
\begin{equation}
\label{eq:blockdiagf}
\rho_R^{(f)} = \rho^{(f, \mathrm{e})}_R \oplus \rho^{(f, \mathrm{o})}_R,
\end{equation}
 with one block corresponding to configurations with an even number of particles in $R$ and in $\bar{R}$ and the other to configurations with an odd number of particles in $R$ and in $\bar{R}$. Accordingly, the R\'enyi entropies 
  \begin{equation}
 \mathcal{R}_n^{(f)} = \frac{1}{1-n} \log \mathrm{tr} [(\hat{\rho}_R^{(f)})^n]
 \end{equation}
 have the generic form 
\begin{equation}
\label{eq:renyi_fermions}
 \mathcal{R}_n^{(f)}= \frac{1}{1-n} \log\left( \mathrm{tr} [(\rho_R^{(f, \mathrm{e})})^n]+  \mathrm{tr} [(\rho_R^{(f, \mathrm{o})})^n]\right).
 \end{equation} 
This result will be helpful to relate the entanglement entropies of the spin system to that of the fermionic quasiparticles in the next section, Sec.~\ref{sec:redspin}.

\subsection{Reduced density matrix of the spin system}
\label{sec:redspin}

We next compute the reduced density matrix in the $R$ region for the spin system. By virtue of the mapping~\eqref{eq:ptoS}, the state $\ket{\psi}$ of Eq.~\eqref{eq:psif} has the spin representation
\begin{equation}
\label{eq:startpsi}
\ket{\psi} = \sum_{\bm{p}} c_{\bm{p}} \hat{{S}}_{\bm{p}}\ket{g_{1,1}^{ZX}},
\end{equation}
with the toric-code ground state $\ket{g_{1,1}^{ZX}}$ chosen to be in the superselection sector $W_x^{ZX}=  W_y^{ZX} =+1$. Expanding $\ket{g^{ZX}_{1,1}}$ in its Schmidt decomposition, Eq.~\eqref{eq:schmidtZX},
\begin{equation}
\ket{g^{ZX}_{1,1}} = \frac{1}{\sqrt{2^{2 L_y-1}}}\sum_{s\in\{0,1 \}}\sum_{\lambda=1}^{2^{2L_y-2}} \ket{s_{\lambda}^R s_\lambda^{\bar{R}}},
\end{equation}
the reduced density matrix in $R$ can be written in the form
\begin{equation} \label{eq:rho_reduced}
\hat{\rho}_R =\frac{1}{2^{2 L_y-1}}\sum_{s,s'} \sum_{\bm{p},\bm{p'}}\sum_{\lambda,\lambda'} c_{\bm{p}} c_{\bm{p}'}^* \hat{{S}}_{\bm{p}}^R \ket{s_{\lambda}^R}\bra{s_{\lambda'}^{\prime R}} \hat{{S}}_{\bm{p}'}^R \bra{s_{\lambda'}^{\prime \bar{R}}}\hat{{S}}_{\bm{p}'}^{\bar{R}} \hat{{S}}_{\bm{p}}^{\bar{R}} \ket{s_{\lambda}^{\bar{R}}},
\end{equation}
where we decomposed the string configuration $\hat{{S}}_{\bm{p}}$ into their $R$ and $\bar{R}$ parts,
\begin{equation}
\hat{{S}}_{\bm{p}} =\hat{{S}}_{\bm{p}}^R \otimes \hat{{S}}_{\bm{p}}^{\bar{R}}.
\end{equation}
Note that the truncated string configurations $\hat{{S}}_{\tilde{\bm{p}}}^R$ and $\hat{{S}}_{\tilde{\bm{p}}}^{\bar{R}}$ for odd-odd configurations $\tilde{\bm{p}}$ do not correspond to a valid (fully formed) string configuration in their respective regions, namely, $\hat{S}_{\tilde{\bm{p}}}^R\not\in G_R^{\rm (e)}$ and $\hat{S}_{\tilde{\bm{p}}}^{\bar{R}}\not\in G_{\bar{R}}^{\rm (e)}$.
 
The matrix element $\bra{s_{\lambda'}^{\prime \bar{R}}}\hat{{S}}_{\bm{p'}}^{\bar{R}} \hat{{S}}_{\bm{p}}^{\bar{R}} \ket{s_{\lambda}^{\bar{R}}}$ in Eq.~\eqref{eq:rho_reduced} can be evaluated by first noting that:
\begin{itemize}
\item[(i)] if $\bm{p}$ and $\bm{p}'$ have a different configuration of anyons in $\bar{R}$, then this matrix element vanishes. Indeed, in this case $\hat{{S}}_{\bm{p}}^{\bar{R}}$ and $\hat{{S}}_{\bm{p'}}^{\bar{R}}$ involve $ZX$ strings with different open ends inside the region $\bar{R}$ that act on loop condensates $\ket{s_\lambda^{\bar{R}}}$ and $\ket{s_{\lambda'}^{\prime \bar{R}}}$ (with loops potentially terminating at the boundary). Here and in the following, we always assume that the vertical strips $R$ and $\bar{R}$ are larger than one unit cell, such that the shortest $ZX$ string $\hat{\sigma}_i^z\hat{\sigma}_{i+\bm{\delta}}^x$ is not spanning their full width. 
\item[(ii)] if $\hat{{S}}_{\bm{p}}$ and $\hat{{S}}_{\bm{p}'}$ have the same $\bar{R}$ configuration, then their $\bar{R}$ part must coincide, $\hat{{S}}_{\bm{p}}^{\bar{R}}=\hat{{S}}_{\bm{p}'}^{\bar{R}}$, as it follows from the construction of the string configurations $\hat{{S}}_{\bm{p}}$ in the spin-fermion mapping of Sec.~\ref{sec:spinfermionmapping}. 
\end{itemize}
The matrix element then yields
\begin{equation}
\label{eq:deltas}
\bra{s_{\lambda'}^{\prime \bar{R}}} \hat{{S}}_{\bm{p'}}^{\bar{R}} \hat{{S}}_{\bm{p}}^{\bar{R}} \ket{s_{\lambda}^{\bar{R}}} = \delta_{ss'}\delta_{\lambda\lambda'}\delta_{\hat{{S}}_{\bm{p}}^{\bar{R}}, \hat{{S}}_{\bm{p}'}^{\bar{R}}},
\end{equation}
with $\delta_{\hat{{S}}_{\bm{p}}^{\bar{R}}, \hat{{S}}_{\bm{p}'}^{\bar{R}}}$ denoting the matching condition of $\hat{{S}}_{\bm{p}}^{\bar{R}}$ and $\hat{{S}}_{\bm{p}'}^{\bar{R}}$ in $\bar{R}$.

To evaluate how $\delta_{\hat{{S}}_{\bm{p}}^{\bar{R}}, \hat{{S}}_{\bm{p}'}^{\bar{R}}}$ affects the sums in Eq.~\eqref{eq:rho_reduced}, we further distinguish the two cases in which the configurations are either even or odd in $\bar{R}$, and denote odd configurations with a tilde in the following, $\tilde{\bm{p}}$. Accordingly, we decompose the sums over anyon configurations in Eq.~\eqref{eq:rho_reduced} as
\begin{align}
\label{eq:rhoredzero}
\hat \rho_R = & \hat \rho_R^{(\mathrm{e})}  +\hat \rho_R^{(\mathrm{o})} \\
= &  \sum_{\bm{p}, \bm{p'}}^{\rm even} c_{\bm{p}} c_{\bm{p}'}^* \hat{{S}}_{\bm{p}}^R \hat{\rho}_R^{(g)} \hat{{S}}_{\bm{p}'}^R \delta_{\hat{{S}}_{\bm{p}}^{\bar{R}}, \hat{{S}}_{\bm{p}'}^{\bar{R}}} +  \sum_{\tilde{\bm{p}}, \tilde{\bm{p}}'}^{\rm odd} c_{\tilde{\bm{p}}} c_{\tilde{\bm{p}}'}^* \hat{{S}}_{\tilde{\bm{p}}}^R \hat{\rho}_R^{(g)}  \hat{{S}}_{\tilde{\bm{p}}'}^R \delta_{\hat{{S}}_{\tilde{\bm{p}}}^{\bar{R}},\hat{ {S}}_{\tilde{\bm{p}}'}^{\bar{R}}}, 
\label{eq:rhoredzerob}
\end{align}
where $\hat{\rho}_R^{(g)}$ is the reduced density matrix of the ground state,
\begin{equation}
\hat{\rho}_R^{(g)} = \frac{1}{2^{2 L_y-1}}\sum_{s\in \{0,1\}} \sum_{\lambda=1}^{2^{2L_y-2}} \ket{s_{\lambda}^R}\!\bra{s_{\lambda}^R}.
\end{equation}
Further progress for  $\hat \rho_R^{(\mathrm{e})}$ and $\hat \rho_R^{(\mathrm{o})}$ can be made as follows:

(i) \textbf{Concerning $\hat \rho_R^{(\mathrm{e})}$}: For even anyon configurations in $R$ and $\bar{R}$, the related string configurations $\hat{{S}}_{\bm{p}}$ involved do not straddle the cuts and their $R$ and $\bar{R}$ parts define independent (up to global parity) string configurations fully contained in $R$ and $\bar{R}$. Namely, we can write $\hat{{S}}_{\bm{p}}^R=\hat{{S}}_{\bm{p}_R}$ and $\hat{{S}}_{\bm{p}}^{\bar{R}}=\hat{{S}}_{\bm{p}_{\bar{R}}}$, where $\bm{p}_R \in G_R^{(\mathrm{e})}$ and $\bm{p}_{\bar{R}} \in G_{\bar{R}}^{(\mathrm{e})}$ are even particle configurations in $R$ and in $\bar{R}$ that compose $\bm{p}$, $\bm{p}=\bm{p}_R\cup  \bm{p}_{\bar{R}}$. The sums over even configurations in the first term of Eq.~\eqref{eq:rhoredzerob} can then be rewritten as $\sum_{\bm{p}}^{\rm even} = \sum_{\bm{p}_R\in G_R^{(\mathrm{e})}} \sum_{\bm{p}_{\bar{R}}\in G_{\bar{R}}^{(\mathrm{e})}} $, leading to
\begin{align}
\hat\rho_R^{ (\mathrm{e})} =& \sum_{\substack{\bm{p}_R,\bm{p'}_R \in G_R^{(\mathrm{e})}\\ \bm{p}_{\bar{R}}, \bm{p}'_{\bar{R}}\in G_{\bar{R}}^{(\mathrm{e})}} } c_{\bm{p}_R, \bm{p}_{\bar{R}}} c_{\bm{p}'_R, \bm{p}'_{\bar{R}}}^* \hat{{S}}_{\bm{p}_R} \hat\rho_R^{(g)} \hat{{S}}_{\bm{p}'_R} \delta_{\bm{p}_{\bar{R}}, \bm{p}'_{\bar{R}}},\\
= &  \sum_{\bm{p}_R,\bm{p'}_R\in G_R^{(\mathrm{e})}} \rho^{(f, \mathrm{e})}_{\bm{p}_R\bm{p}'_R} \hat{{S}}_{\bm{p}_R} \hat\rho_R^{(g)} \hat{{S}}_{\bm{p}'_R}.
\label{eq:rhoredone}
\end{align}
The ``even part'' of the density matrix,  $\rho^{(f, \mathrm{e})}_{\bm{p}_R, \bm{p}'_R }$, has the same form defined in Eq.~\eqref{eq:rhofe} for the purely fermionic state, namely,
\begin{equation}
 \rho^{(f, \mathrm{e})}_{\bm{p}_R, \bm{p}'_R } = \sum_{\bm{p}_{\bar{R}}\in G_{\bar{R}}^{(\mathrm{e})}} c_{\bm{p}_R, \bm{p}_{\bar{R}}} c_{\bm{p}'_R, \bm{p}_{\bar{R}}}^*.
\end{equation}

\begin{figure}[b]
\centering
\includegraphics{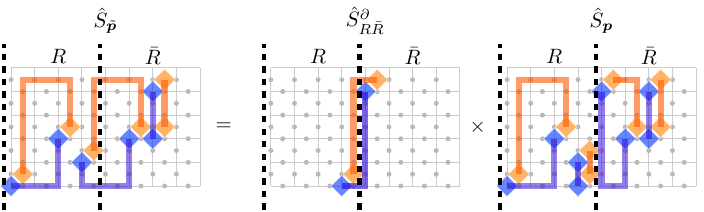}
\vspace{.5cm}
\includegraphics{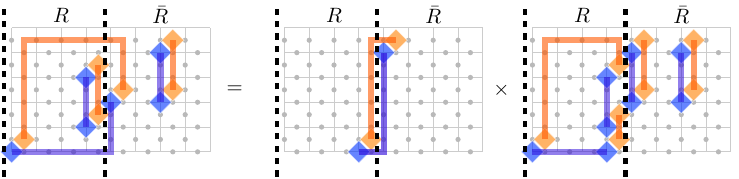}
\caption{Decomposition of a string $\hat{S}_{\tilde{\bm{p}}}$ with an odd number of anyons in both $R$ and $\bar{R}$ into the product of a boundary string $\hat{S}_{R\bar{R}}^\partial$ and a string $\hat{S}_{\bm{p}}$ with an even number of anyons in both $R$ and $\bar{R}$, according to Eq.~\eqref{eq:decompbound} for two examples.}
\label{fig:handlecut}
\end{figure}

(ii) \textbf{Concerning $\hat \rho_R^{(\mathrm{o})}$}: For odd anyon configurations in $R$ and $\bar{R}$, the related string configurations $\hat{{S}}_{\tilde{\bm{p}}}$ must straddle one cut. It is useful at this point to establish a one-to-one correspondence between each odd configuration $\tilde{\bm{p}}$ and an even configuration $\bm{p}$. We do so by decomposing $\hat{{S}}_{\tilde{\bm{p}}}$ as
\begin{align}
\label{eq:decompbound}
\hat{{S}}_{\tilde{\bm{p}}} & = (-1)^{\theta_{\tilde{\bm{p}}}}
 \hat{{S}}_{\bm{p}_R} \hat{S}^\partial_{R \bar{R}} \hat{{S}}_{\bm{p}_{\bar{R}}},
 \end{align}
where $\hat{{S}}_{\bm{p}_R} $ and $\hat{{S}}_{\bm{p}_{\bar{R}}}$ describe even configurations $\bm{p}_R\in G_R^{(\mathrm{e})}$ and $\bm{p}_{\bar{R}}\in G_{\bar{R}}^{(\mathrm{e})}$, which uniquely define the even configuration $\bm{p}=\bm{p}_R \cup \bm{p}_{\bar{R}}$ associated with $\tilde{\bm{p}}$. The configuration $\hat{S}^\partial_{R\bar{R}}$ is a two-anyon string crossing the cut which would glue $\hat{{S}}_{\bm{p}_R} $ and $\hat{{S}}_{\bm{p}_{\bar{R}}}$ back into $\hat{{S}}_{\tilde{\bm{p}}}$, graphically represented as 
\begin{equation}
\label{eq:Sboundary}
\includegraphics{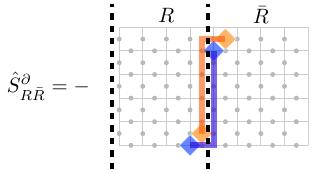}.
\end{equation}
The pre-factor $(-1)^{\theta_{\tilde{\bm{p}}}}$ arises due to the factor $(-1)^{M_{\bm{p}}}$ in the definition of the $\hat{{S}}_{\bm{p}}$ of Eq.~\eqref{eq:Sstring} and is given by $\theta_{\tilde{\bm{p}}} = n^\partial_{R}(\tilde{\bm{p}}) + n^\partial_{\bar{R}}(\tilde{\bm{p}})$, where $n^\partial_{R}(\tilde{\bm{p}})$ and $n^\partial_{\bar{R}}(\tilde{\bm{p}})$ denote the number of $f$ anyons occupying the plaquette at the end of the $\hat{S}_{R\bar{R}}^\partial$ string in $R$ and $\bar{R}$, respectively. Hence, the factor $(-1)^{\theta_{\tilde{\bm{p}}}}$ is equal to $+1$ if $\tilde{\bm{p}}$ has no $f$ anyons on the end-plaquettes of $\hat{S}_{R\bar{R}}^\partial$ or if it contains both, while it is $-1$ otherwise.  
 The glueing string $\hat{S}_{R\bar{R}}^\partial$ is the same for all $\hat{{S}}_{\tilde{\bm{p}}}$ and, as indicated in Eq.~\eqref{eq:Sboundary}, it is defined including a minus sign. The decomposition for two examples is sketched in Fig.~\ref{fig:handlecut}. 
 From Eq.~\eqref{eq:decompbound}, we can rewrite Eq.~\eqref{eq:deltas} as
\begin{align}
\bra{s_{\lambda'}^{\prime \bar{R}}} \hat{{S}}_{\tilde{\bm{p}}'}^{\bar{R}}\hat{{S}}_{\tilde{\bm{p}}}^{\bar{R}} \ket{s_{\lambda}^{\bar{R}}} & = \bra{s_{\lambda'}^{\prime \bar{R}}}  \hat{{S}}_{\bm{p}'_{\bar{R}}}   \hat{S}^{\partial,\bar{R}}_{R \bar{R}} \hat{S}^{\partial,\bar{R}}_{R \bar{R}}\hat{{S}}_{\bm{p}_{\bar{R}}}  \ket{s_{\lambda}^{\bar{R}}} \nonumber \\
& = \delta_{ss'}\delta_{\lambda\lambda'}\delta_{\tilde{\bm{p}}_{\bar{R}} \tilde{\bm{p}}'_{\bar{R}}},
\label{eq:deltas2}
\end{align}
where $\hat{S}^{\partial,\mathscr{R}}_{R \bar{R}} $ is the restriction of $\hat{S}^{\partial}_{R \bar{R}} $ to the region $\mathscr{R}$. In Eq.~\eqref{eq:deltas2}, $\delta_{\tilde{\bm{p}}_{\bar{R}}, \tilde{\bm{p}}'_{\bar{R}}}$ descends from $\delta_{\bm{p}_{\bar{R}}, \bm{p}'_{\bar{R}}}$ and the one-to-one correspondence between even and odd configurations. Since $n^\partial_{\bar R}(\tilde{\bm{p}})$ depends only on $\tilde{\bm p}_{\bar R}$, the matching condition makes the factors $(-1)^{n^\partial_{\bar R}(\tilde{\bm p})+n^\partial_{\bar R}(\tilde{\bm p}')}$ equal to one and the odd component of the reduced density matrix is thus obtained as
\begin{equation}
\label{eq:rhoodd}
 \hat{\rho}_R^{(\mathrm{o})}=\sum_{\substack{\tilde{\bm{p}}_R, \tilde{\bm{p}}'_R\in G_R^{(\mathrm{o})} \\ \tilde{\bm{p}}_R\leftrightarrow \bm{p}_R, \tilde{\bm{p}}_R'\leftrightarrow\bm{p}_R' } } (-1)^{n^{\partial}_{R}(\tilde{\bm{p}}_R)+n^{\partial}_{R}(\tilde{\bm{p}}'_R)} {\rho}_{\tilde{\bm{p}}_R\tilde{\bm{p}}'_R}^{(f, \mathrm{o})}
\hat{{S}}_{\bm{p}_R} \hat{S}^{\partial,R}_{R \bar{R}} \hat{\rho}_R^{(g)} \hat{S}^{\partial,R}_{R \bar{R}}  \hat{{S}}_{\bm{p}'_R}, 
\end{equation}
where $\tilde{\bm{p}}_{R}\leftrightarrow \bm{p}_{R}$ indicates the correspondence between even and odd (sub-)configurations and where ${\rho}_{\tilde{\bm{p}}_R\tilde{\bm{p}}'_R}^{(f, \mathrm{o})}$ has the same form as Eq.~\eqref{eq:rhofo} for the purely fermionic state,
\begin{equation}
\label{eq:rhooddfin}
 {\rho}_{\tilde{\bm{p}}_R\tilde{\bm{p}}'_R}^{(f, \mathrm{o})}=\sum_{\tilde{\bm{p}}_{\bar{R}} \in G_{\bar{R}}^{(\mathrm{o})}} c_{\tilde{\bm{p}}_R\tilde{\bm{p}}_{\bar{R}}} c_{\tilde{\bm{p}}'_R\tilde{\bm{p}}_{\bar{R}}}^*.
\end{equation}
Collecting the results in Eqs.~\eqref{eq:rhoredzero}, \eqref{eq:rhoredone} and \eqref{eq:rhoodd}, the reduced density matrix in $R$ reads as
\begin{equation}
\label{eq:rhoAfinal}
\hat{\rho}_R =   \sum_{\bm{p}_R,\bm{p'}_R\in G_R^{(\mathrm{e})}} \rho^{(f, \mathrm{e})}_{\bm{p}_R\bm{p}'_R} \hat{{S}}_{\bm{p}_R} \hat\rho_R^{(g)} \hat{{S}}_{\bm{p}'_R} 
+
\sum_{\substack{\tilde{\bm{p}}_R, \tilde{\bm{p}}'_R\in G_R^{(\mathrm{o})} \\ \tilde{\bm{p}}_R\leftrightarrow \bm{p}_R, \tilde{\bm{p}}_R'\leftrightarrow\bm{p}_R' } }
 (-1)^{n^{\partial}_{R}(\tilde{\bm{p}}_R)+n^{\partial}_{R}(\tilde{\bm{p}}'_R)}
 {\rho}_{\tilde{\bm{p}}_R\tilde{\bm{p}}'_R}^{(f, \mathrm{o})}
\hat{{S}}_{\bm{p}_R}  \hat{S}^{\partial,R}_{R \bar{R}}  \hat{\rho}_R^{(g)} \hat{S}^{\partial,R}_{R \bar{R}} \hat{{S}}_{\bm{p}'_R}. 
\end{equation}
By noting that
\begin{equation}
\bra{s_\lambda^R} \hat{S}_{\bm{q}_R} \hat{S}_{\bm{p}_R} \hat{S}^{\partial,R}_{R \bar{R}}  \ket{\bar{s}_\mu^R} = 0,
\end{equation}
as $ \hat{S}^{\partial,R}_{R \bar{R}} $ acts on boundary spins and has a zero matrix element with respect to the states $\ket{s_\lambda^R}$, one finds that $\mathrm{tr}[\hat{\rho}_R^{(\rm e)} \hat{\rho}_R^{(\rm o)} ]=0$, i.e., $\hat{\rho}_R^{(\rm e)}$ and $\hat{\rho}_R^{(\rm o)}$ identify disjoint matrix blocks of $\hat{\rho}_R$. The reduced density matrix has thus a block-diagonal structure in a basis labelled by the parity of the configurations $\bm{p}_R$. This structure descends from the global even-parity symmetry (parity superselection rule), akin to the case of a purely fermionic system, Eq.~\eqref{eq:blockdiagf}. 

\subsection{R\'enyi and von Neumann entropies}
Given the reduced density matrix of Eq.~\eqref{eq:rhoAfinal}, we compute the R\'enyi entropies,
\begin{equation}
\mathcal{R}_n= \frac{1}{1-n}\log \mathrm{tr}[\hat\rho_R^n].
\end{equation}
We first determine the moments $\mathrm{tr}[\hat\rho_R^n]$ of the reduced density matrix by means of the following identities,
\begin{subequations}\label{eq:lemmas}
\begin{align}
& \bra{s_{\lambda'}^{\prime R}} \hat{{S}}_{\bm{p}'_R} \hat{{S}}_{\bm{p}_R} \ket{s_{\lambda}^R} = 
\bra{s_{\lambda'}^{\prime R}} \hat{S}^{\partial,R}_{R \bar{R}} \hat{{S}}_{{\bm{p}}'_R} \hat{{S}}_{{\bm{p}}_R} \hat{S}^{\partial,R}_{R \bar{R}}\ket{s_{\lambda}^R} 
=  \delta_{ss'}\delta_{\lambda, \lambda'}\delta_{{\bm{p}}_R, {\bm{p}}'_R} ,
\\
& \bra{s_{\lambda'}^{\prime R}} \hat{{S}}_{\bm{p}'_R} \hat{{S}}_{{\bm{p}}_R} \hat{S}^{\partial,R}_{R \bar{R}} \ket{s_{\lambda}^R} = 0,
\end{align} 
\end{subequations}
obtaining
\begin{equation}
\label{eq:renyi01}
\mathrm{tr}\left( \hat{\rho}_R^n \right) = 2^{-(n-1)(2L_y -1)}\left(\mathrm{tr}[(\rho_R^{(f, \mathrm{e})})^n] + \mathrm{tr}[(\rho_R^{(f, \mathrm{o})})^n]\right).
\end{equation}
We recognize that the term in parentheses is the $n$th moment of the reduced density matrix in fermionic representation, $\hat{\rho}_R^{(f)}$, discussed in Sec.~\ref{sec:entang_fermions}, see Eq.~\eqref{eq:renyi_fermions}, while the prefactor is the $n$th moment of the ground-state reduced density matrix, Eq.~\eqref{eq:groundstatemoments}. Note also that the factor $ (-1)^{n^{\partial}_{R}(\tilde{\bm{p}}_R)+n^{\partial}_{R}(\tilde{\bm{p}}'_R)}$ in Eq.~\eqref{eq:rhoAfinal} disappears when computing the trace. Equation \eqref{eq:renyi01} can then be written like
\begin{equation}
\mathrm{tr}\left[( \hat{\rho}_R)^n \right] = \mathrm{tr}\left[\left(\hat{\rho}^{(g)}_R\right)^n\right] \mathrm{tr}\left[\left(\hat{\rho}^{(f)}_R\right)^n\right],
\end{equation}
namely, it decomposes into the product of the moment of the toric code ground state and that of the fermionic state describing the $f$ quasiparticles.

In the replica limit $\mathcal{S}_R=-\lim_{n\to 1}\partial_n \mathrm{tr}[\hat{\rho}_R^n]$, the von Neumann entropy is found as
\begin{equation}
\label{eq:decentropy}
\mathcal{S}_R = (2 L_y-1)\log(2) + \mathcal{ S}^{(f)}_R,
\end{equation}
with $\mathcal{ S}^{(f)}_R = -\mathrm{tr}\hat\rho^{(f)}_R\log (\hat\rho^{(f)}_R)$. Hence, the R\'enyi and von Neumann entropies decompose as the sum of the pure ground state contribution and a contribution given by the entropy of the state describing the emergent $f$ quasiparticles. The decomposition \eqref{eq:decentropy} has been verified numerically against the calculation in the full spin picture for a bipartition into two regions $R$ and $\bar{R}$ of equal size of the form considered here (Fig.~1) for a $L_x=4$, $L_y=3$ torus.

\section{Decomposition of the logarithmic negativity}
\label{sec:decompnegativity}

In this section, we derive a decomposition of the logarithmic negativity into a ground state contribution and an $f$-quasiparticle contribution, analogous to the decomposition of the entropies analysed in Sec.~\ref{sec:entropydecomposition}. A notable difference, compared to the result for the entropy, is that the $f$ contribution will not directly relate to known definitions of negativity for fermionic systems~\cite{Eisler2015, Shapourian2017, Shapourian2019a} and methods to compute it efficiently for Gaussian fermion states are not known. Specifically, we analyse the negativity between two adjacent non-contractible strips $R^{(1)}$ and $R^{(2)}$, whose union forms the sub-region $R$ as depicted in the main text (Fig.~1). 

\subsection{Fermions}
\label{eq:fermionnegativity}

Consider the reduced density matrix in fermionic representation, reported here from Eq.~\eqref{eq:rhoAfermions},
\begin{equation}
 \hat{ \rho}^{(f)}_R = \sum_{\bm{p}_R, \bm{p}'_R \in G_R^{(\mathrm{e})}}\rho^{(f, \mathrm{e})}_{\bm{p}_R, \bm{p}'_R} \ket{\bm{p}_R}\bra{\bm{p}'_R} +
   \sum_{\tilde{\bm{p}}_R, \tilde{\bm{p}}'_R \in G_R^{(\mathrm{o})}}  \rho^{(f, \mathrm{o})}_{\tilde{\bm{p}}_R, \tilde{\bm{p}}'_R} \ket{\tilde{\bm{p}}_R}\bra{\tilde{\bm{p}}'_R}.
\end{equation}
By decomposing the configurations $\bm{p}_R$ into their parts $\bm{p}_1$ and $\bm{p}_2$ in $R^{(1)}$ and $R^{(2)}$, respectively, consistently with the parity constraints, we obtain
\begin{align}
 \hat{ \rho}^{(f)}_R = & \left(\sum_{\substack{\bm{p}_{1}\in G_{1}^{(\mathrm{e})}, \\ \bm{p}_{2} \in G_{2}^{(\mathrm{e})}}} + \sum_{\substack{\bm{p}_{1}\in G_{1}^{(\mathrm{o})}, \\ \bm{p}_{2} \in G_{2}^{(\mathrm{o})}}} \right) \left(\sum_{\substack{\bm{p}'_{1}\in G_{1}^{(\mathrm{e})}, \\ \bm{p}'_{2} \in G_{2}^{(\mathrm{e})}}} + \sum_{\substack{\bm{p}_{1}'\in G_{1}^{(\mathrm{o})}, \\ \bm{p}'_{2} \in G_{2}^{(\mathrm{o})}}} \right) 
  \rho^{(f, \mathrm{e})}_{\bm{p}_{1}\bm{p}_{2}, \bm{p}'_{1} \bm{p}'_{2}} \ket{\bm{p}_{1}\bm{p}_{2}}\bra{\bm{p}'_{1}\bm{p}'_{2}} \nonumber \\
&+  \left(\sum_{\substack{\tilde{\bm{p}}_{1}\in G_{1}^{(\mathrm{e})}, \\ \tilde{\bm{p}}_{2} \in G_{2}^{(\mathrm{o})}}} + \sum_{\substack{\tilde{\bm{p}}_{1}\in G_{1}^{(\mathrm{o})}, \\ \tilde{\bm{p}}_{2} \in G_{2}^{(\mathrm{e})}}} \right) \left(\sum_{\substack{\tilde{\bm{p}}'_{1}\in G_{1}^{(\mathrm{e})}, \\ \tilde{\bm{p}}'_{2} \in G_{2}^{(\mathrm{o})}}} + \sum_{\substack{\tilde{\bm{p}}'_{1}\in G_{1}^{(\mathrm{o})}, \\\tilde{ \bm{p}}'_{2} \in G_{2}^{(\mathrm{e})}}} \right) 
  \rho^{(f, \mathrm{o})}_{\tilde{\bm{p}}_{1} \tilde{\bm{p}}_{2}, \tilde{\bm{p}}'_{1} \tilde{\bm{p}}'_{2}} \ket{\tilde{\bm{p}}_{1}\tilde{\bm{p}}_{2}}\bra{\tilde{\bm{p}}'_{1} \tilde{\bm{p}}'_{2}}.
    \label{eq:fpa1}
\end{align}
In Eq.~\eqref{eq:fpa1}, $G_i^{(\rm e)}$ and $G_i^{(\rm o)}$ denote the set of $f$ configurations which are even or odd in region $R^{(i)}$, respectively. Note that the way we constructed the generalized Jordan-Wigner mapping fixes an ordering of the fermionic operators $\hat{f}_p^\dag$ which lists operators acting in $R^{(1)}$ first, i.e. ``to the left'', such that the decomposition $\ket{\bm{p}_R}=\ket{\bm{p}_1\bm{p}_2}$ does not require any re-ordering of fermionic operators. Schematically, $\hat{\rho}_R^{(f)}$ has thus the block matrix structure
\begin{equation}
\label{eq:matreprhoAf}
{\rho}^{(f)}_R =\left( \begin{array}{cc|cc}
\rho^{(f,\mathrm{e})}_{\rm ee,ee} & \rho^{(f,\mathrm{e})}_{\rm ee,oo} &  & \\
\rho^{(f,\mathrm{e})}_{\rm oo,ee} & \rho^{(f,\mathrm{e})}_{\rm oo,oo} &  & \\
\hline
 && \rho^{(f,\mathrm{o})}_{\rm eo,eo} & \rho^{(f,\mathrm{o})}_{\rm eo,oe}    \\
& & \rho^{(f,\mathrm{o})}_{\rm oe,eo} & \rho^{(f,\mathrm{o})}_{\rm oe,oe} 
\end{array}
\right),
\end{equation}
where the subscripts in $\rho^{(f)}_{ab,cd}$, with $a,b,c,d\in\{ \mathrm{e}, \mathrm{o}\}$, denote the parity in $R^{(1)}$ and $R^{(2)}$ of the states corresponding to that matrix block.
We then consider the \textit{bosonic} partial transpose~\cite{Shapourian2019a, Eisler2015},
\begin{align}
\left( \hat{ \rho}^{(f)}_R\right)^{T_2} = & \left(\sum_{\substack{\bm{p}_{1}\in G_{1}^{(\mathrm{e})}, \\ \bm{p}_{2} \in G_{2}^{(\mathrm{e})}}} + \sum_{\substack{\bm{p}_{1}\in G_{1}^{(\mathrm{o})}, \\ \bm{p}_{2} \in G_{2}^{(\mathrm{o})}}} \right) \left(\sum_{\substack{\bm{p}'_{1}\in G_{1}^{(\mathrm{e})}, \\ \bm{p}'_{2} \in G_{2}^{(\mathrm{e})}}} + \sum_{\substack{\bm{p}_{1}'\in G_{1}^{(\mathrm{o})}, \\ \bm{p}'_{2} \in G_{2}^{(\mathrm{o})}}} \right) 
  \rho^{(f, \mathrm{e})}_{\bm{p}_{1}\bm{p}_{2}, \bm{p}'_{1} \bm{p}'_{2}} 
  \ket{\bm{p}_{1}\bm{p}'_{2}}\bra{\bm{p}'_{1}\bm{p}_{2}} + \nonumber \\
&  \left(\sum_{\substack{\tilde{\bm{p}}_{1}\in G_{1}^{(\mathrm{e})}, \\ \tilde{\bm{p}}_{2} \in G_{2}^{(\mathrm{o})}}} + \sum_{\substack{\tilde{\bm{p}}_{1}\in G_{1}^{(\mathrm{o})}, \\ \tilde{ \bm{p}}_{2} \in G_{2}^{(\mathrm{e})}}} \right) \left(\sum_{\substack{\tilde{\bm{p}}'_{1}\in G_{1}^{(\mathrm{e})}, \\ \tilde{ \bm{p}}'_{2} \in G_{2}^{(\mathrm{o})}}} + \sum_{\substack{\tilde{\bm{p}}'_{1}\in G_{1}^{(\mathrm{o})}, \\ \tilde{\bm{p}}'_{2} \in G_{2}^{(\mathrm{e})}}} \right) 
    \rho^{(f, \mathrm{o})}_{\tilde{\bm{p}}_{1}\tilde{\bm{p}}_{2} , \tilde{\bm{p}}'_{1}\tilde{\bm{p}}'_{2}}
     \ket{\tilde{\bm{p}}_{1}\tilde{\bm{p}}'_{2}}\bra{\tilde{\bm{p}}'_{1}\tilde{\bm{p}}_{2}}.
\end{align}
Its schematic block-matrix representation, compared to Eq.~\eqref{eq:matreprhoAf}, has the structure
\begin{equation}
\label{eq:rhot2fermion}
\left[ {\rho}^{(f)}_R \right]^{T_2} =\left( \begin{array}{cc|cc}
\left[\rho^{(f,\mathrm{e})}_{\rm ee,ee} \right]^{T_2} & \left[ \rho^{(f,\mathrm{o})}_{\rm eo,oe} \right]^{T_2}&  & \\
\left[ \rho^{(f,\mathrm{o})}_{\rm oe,eo} \right]^{T_2} &\left[ \rho^{(f,\mathrm{e})}_{\rm oo,oo}  \right]^{T_2}&  & \\
\hline
 && \left[ \rho^{(f,\mathrm{o})}_{\rm eo,eo}  \right]^{T_2}&\left[ \rho^{(f,\mathrm{e})}_{\rm ee,oo} \right]^{T_2}   \\
& & \left[\rho^{(f,\mathrm{e})}_{\rm oo,ee} \right]^{T_2} &\left[  \rho^{(f,\mathrm{o})}_{\rm oe,oe}  \right]^{T_2}
\end{array}
\right).
\end{equation}
Having in mind these matrix structures will be helpful for comparison with the results for the spin system in the next section. 

\subsection{Negativity of the spin system}

To analyze the partial transpose and the negativity in spin representation, it is first convenient to decompose the string operators $\hat{S}_{\bm{p}}$ further. We showed in Sec.~\ref{sec:redspin} [Eq.~\eqref{eq:decompbound}] how a $ZX$-string $\hat{S}_{\tilde{\bm{p}}}$ that crosses a bipartition can be decomposed into the product of $ZX$-strings that do not cross the partition, which define valid $f$ configurations in each sub-region, and boundary operators $\hat{S}_{RR'}^\partial$. This approach can be generalized to multiple vertical cuts partitioning the system into more than two sub-regions, producing boundary operators for each cut. For example, for the partition into three non-contractible regions $R^{(1)}$, $R^{(2)}$ and $\bar{R}$ we are interested in, a string $\hat{{S}}_{\tilde{\bm{p}}}$ describing an $f$ configuration $\tilde{\bm{p}}$ containing an odd number of quasiparticles in $R^{(1)}$ and $\bar{R}$ and an even number of particles in $R^{(2)}$ can be decomposed as
\begin{eqnarray}
\hat{{S}}_{\tilde{\bm{p}}} &&= \hat{{S}}_{\tilde{\bm{p}}}^{R^{(1)}}\hat{{S}}_{\tilde{\bm{p}}}^{R^{(2)}}\hat{{S}}_{\tilde{\bm{p}}}^{\bar{R}} ,\nonumber \\
&& = 
  (-1)^{\theta_{\tilde{\bm{p}}}}
  \hat{{S}}_{\bm{p}_{1}} \hat{S}_{R^{(1)}R^{(2)}}^\partial  \hat{{S}}_{\bm{p}_{2}}\hat{S}_{R^{(2)} \bar{R}}^\partial \hat{{S}}_{\bm{p}_{\bar{R}}},
\label{eq:decompo1}
\end{eqnarray}
where $\bm{p}=\bm{p}_{1}\cup \bm{p}_{2} \cup \bm{p}_{\bar{R}} $ with $\bm{p}_{\mathscr{R}} \in G_{\mathscr{R}}^{(\mathrm{e})}$ is the configuration that is even in all sub-regions associated with $\tilde{\bm{p}}$. The phase factor $\theta_{\tilde{\bm{p}}}$ is given by
\begin{equation}
\theta_{\tilde{\bm{p}}} =  {n_{1}(\tilde{\bm{p}}) + n_{2L}(\tilde{\bm{p}})+{n}_{2R}(\tilde{\bm{p}}) + {n}_{\bar{R}}(\tilde{\bm{p}})}.
\end{equation}
The coefficients $n_1(\tilde{\bm{p}})$ and $n_{\bar{R}}(\tilde{\bm{p}})$ denote the $f$ occupation of the end-plaquettes of $\hat{S}_{R^{(1)} R^{(2)}}^\partial$ in $R^{(1)}$ and of $\hat{S}_{R^{(2)} \bar{R}}^\partial$ in $\bar{R}$, respectively, while $n_{2L}(\tilde{\bm{p}})$ and $n_{2R}(\tilde{\bm{p}})$ denote the occupation of the end-plaquettes of $\hat{S}_{R^{(1)} R^{(2)}}^\partial$ and of $\hat{S}_{R^{(2)} \bar{R}}^\partial$ in $R^{(2)}$, respectively. If only one gluing string is needed, the phase factor $\Theta$ contains only the occupations of its two end plaquettes.

\begin{figure}[t]
\begin{center}
\includegraphics{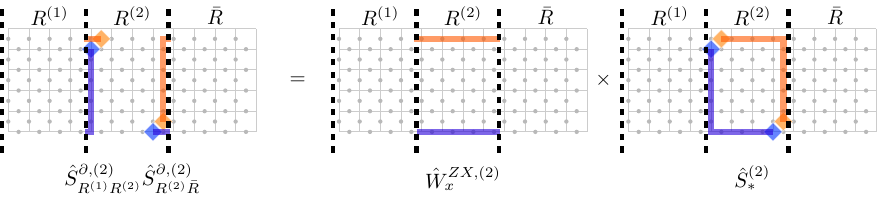}
\end{center}
\caption{ Decomposition of the product $\hat{S}_{R^{(1)} R^{(2)}}^{\partial, (2)}\hat{S}_{R^{(2)} \bar{R}}^{\partial, (2)}$ into the product of a two-anyon string $\hat{S}_*^{(2)}$ fully contained in $R^{(2)}$ and the restriction $\hat{W}_x^{ZX, (2)}$ of the horizontal Wilson-loop operator to $R^{(2)}$.}
\label{fig:Sstar}
\end{figure}

Resolving the parity of the number of fermions in each sub-region and decomposing the related $ZX$ strings accordingly, steps analogous to Sec.~\ref{sec:redspin} lead to the reduced density matrix in $R$ in the form 
\begin{equation}
\hat{\rho}_R = \sum_{i,j\in\{\mathrm{e, o}\}} \hat{\rho}^{\rm (e)}_{{ii, jj}} + \sum_{\substack{i,j, k ,l\in\{\mathrm{e, o}\}\\ i\ne j, k\ne l}} \hat{\rho}^{\rm (o)}_{{ij, k l}} ,
\end{equation}
with the subscripts $(ij,kl)$ indicating the fermion parity in $R^{(1)}$ and $R^{(2)}$ of the related fermion configuration $(\bm{p}_1 \bm{p}_2, \bm{p}_1'\bm{p}_2')$, while the superscript labels the global parity in $R$. Using the following shortcut notation for the sums,
\begin{equation}
\sum_{\rm ee, oo} = \sum_{\substack{\bm{p}_{1} \in G_{1}^{(\mathrm{e})},  \bm{p}_{2} \in G_{2}^{(\mathrm{e})} \\ \tilde{\bm{p}}'_{1} \in G_{1}^{(\mathrm{o})},  \tilde{\bm{p}}'_{2} \in G_{2}^{(\mathrm{o})} \\ (\tilde{\bm{p}}'_{1} \leftrightarrow \bm{p}'_{1} ,\tilde{\bm{p}}'_{2} \leftrightarrow \bm{p}'_{2})}}
\end{equation}
and likewise for the other parity cases. As in Sec.~\ref{sec:redspin}, the tilde denotes the actual $f$ configuration whenever it differs from the even configuration defining the strings. This is the case for the odd parts and, for configurations of type ``oe'', also for the even $R^{(2)}$ part $\tilde{\bm{p}}_2$, which differs from $\bm{p}_2$ by the occupations of both end plaquettes in $R^{(2)}$, since both gluing strings end there. The individual terms explicitly read as
\begin{eqnarray}
&& \hat{\rho}_{\rm ee, ee}^{\rm (e)}  =\sum_{\rm ee, ee}\rho_{\bm{p}_{1}\bm{p}_{2}, \bm{p}'_{1}\bm{p}'_{2}}^{(f, \mathrm{e})} \hat{{S}}_{\bm{p}_{1}}\hat{{S}}_{\bm{p}_{2}} \hat{\rho}_R^{(g)} \hat{{S}}_{\bm{p}'_{1}}\hat{{S}}_{\bm{p}'_{2}} ,\nonumber \\
&&\hat{\rho}_{\rm oo, oo}^{\rm (e)}   = \sum_{\rm oo,oo}   (-1)^{\Theta^{\rm oo, oo}_{\tilde{\bm{p}},\tilde{\bm{p}}'} } \rho_{\tilde{\bm{p}}_{1}\tilde{\bm{p}}_{2}, \tilde{\bm{p}}'_{1}\tilde{\bm{p}}'_{2}}^{(f, \mathrm{e})} \hat{{S}}_{\bm{p}_{1}}\hat{{S}}_{\bm{p}_{2}}\hat{S}^\partial_{R^{(1)}R^{(2)}} \hat{\rho}_R^{(g)} \hat{S}^\partial_{R^{(1)}R^{(2)}} \hat{{S}}_{\bm{p}'_{1}}\hat{{S}}_{\bm{p}'_{2}},
\nonumber \\
&&  \hat{\rho}_{\rm oo, ee}^{\rm (e)}  = \sum_{\rm oo, ee}(-1)^{\Theta^{\rm oo, ee}_{\tilde{\bm{p}},\tilde{\bm{p}}'} }
 \rho_{\tilde{\bm{p}}_{1}\tilde{\bm{p}}_{2}, \bm{p}'_{1}\bm{p}'_{2}}^{(f, \mathrm{e})} \hat{{S}}_{\bm{p}_{1}}\hat{{S}}_{\bm{p}_{2}} \hat{S}^\partial_{R^{(1)}R^{(2)}}\hat{\rho}_R^{(g)} \hat{{S}}_{\bm{p}'_{1}}\hat{{S}}_{\bm{p}'_{2}},  \nonumber \\
 && \hat{\rho}_{\rm ee, oo}^{\rm (e)}  =\sum_{\rm ee, oo} (-1)^{\Theta^{\rm ee, oo}_{\tilde{\bm{p}},\tilde{\bm{p}}'}  }\rho_{\bm{p}_{1}\bm{p}_{2}, \tilde{\bm{p}}'_{1}\tilde{\bm{p}}'_{2}}^{(f, \mathrm{e})} \hat{{S}}_{\bm{p}_{1}}\hat{{S}}_{\bm{p}_{2}} \hat{\rho}_R^{(g)}\hat{S}^\partial_{R^{(1)}R^{(2)}}  \hat{{S}}_{\bm{p}'_{1}}\hat{{S}}_{\bm{p}'_{2}},
 \end{eqnarray}
 and 
  \begin{eqnarray}
&& \hat{\rho}_{\rm eo, eo}^{\rm (o)} = \sum_{\rm eo,eo}(-1)^{\Theta^{\rm eo, eo}_{\tilde{\bm{p}},\tilde{\bm{p}}'} }
 \rho_{\bm{p}_{1}\tilde{\bm{p}}_{2}, \bm{p}'_{1} \tilde{\bm{p}}'_{2}}^{(f, \mathrm{o})} \hat{{S}}_{\bm{p}_{1}}\hat{{S}}_{\bm{p}_{2}} \hat{S}_{R^{(2)} \bar{R}}^{\partial, (2)} \hat{\rho}_R^{(g)} \hat{S}_{R^{(2)} \bar{R}}^{\partial, (2)} \hat{{S}}_{\bm{p}'_{1}}\hat{{S}}_{\bm{p}'_{2}}, \nonumber\\
&& \hat{\rho}_{\rm eo, oe}^{\rm (o)} =  \sum_{\rm eo, oe}(-1)^{\Theta^{\rm eo, oe}_{\tilde{\bm{p}},\tilde{\bm{p}}'} }
 \rho_{{\bm{p}}_{1}\tilde{\bm{p}}_{2}, \tilde{\bm{p}}'_{1}\tilde{\bm{p}}'_{2}}^{(f, \mathrm{o})} \hat{{S}}_{\bm{p}_{1}}\hat{{S}}_{\bm{p}_{2}}  \hat{S}_{R^{(2)} \bar{R}}^{\partial, (2)}\hat{\rho}_R^{(g)}  \hat{S}^\partial_{R^{(1)}R^{(2)}} \hat{S}_{R^{(2)} \bar{R}}^{\partial, (2)} \hat{{S}}_{\bm{p}'_{1}}\hat{{S}}_{\bm{p}'_{2}},\nonumber  \\ 
&&\hat{\rho}_{\rm oe, eo}^{\rm (o)} = \sum_{\rm oe, eo} (-1)^{\Theta^{\rm oe, eo}_{\tilde{\bm{p}},\tilde{\bm{p}}'} }
\rho_{\tilde{\bm{p}}_{1}\tilde{\bm{p}}_{2}, {\bm{p}}'_{1}\tilde{\bm{p}}'_{2}}^{(f, \mathrm{o})} \hat{{S}}_{\bm{p}_{1}}\hat{{S}}_{\bm{p}_{2}} \hat{S}^{\partial}_{R^{(1)}R^{(2)}} \hat{S}_{R^{(2)} \bar{R}}^{\partial, (2)} \hat{\rho}_R^{(g)} \hat{S}_{R^{(2)} \bar{R}}^{\partial, (2)} \hat{{S}}_{\bm{p}'_{1}}\hat{{S}}_{\bm{p}'_{2}} , \nonumber \\
 && \hat{\rho}_{\rm oe, oe}^{\rm (o)}= \sum_{\rm oe, oe} (-1)^{\Theta^{\rm oe, oe}_{\tilde{\bm{p}},\tilde{\bm{p}}'} }
 \rho_{\tilde{\bm{p}}_{1}\tilde{\bm{p}}_{2}, \tilde{\bm{p}}'_{1}\tilde{\bm{p}}'_{2}}^{(f, \mathrm{o})} \hat{{S}}_{\bm{p}_{1}}\hat{{S}}_{\bm{p}_{2}} \hat{S}^\partial_{R^{(1)}R^{(2)}}  \hat{S}_{R^{(2)} \bar{R}}^{\partial, (2)}  \hat{\rho}_R^{(g)}\hat{S}^\partial_{R^{(1)}R^{(2)}} \hat{S}_{R^{(2)} \bar{R}}^{\partial, (2)} \hat{{S}}_{\bm{p}'_{1}}\hat{{S}}_{\bm{p}'_{2}}  ,
\end{eqnarray} 
with phase factors
\begin{align}
\Theta^{\rm oo, oo}_{\tilde{\bm{p}} \tilde{\bm{p}}'}
&= n_{1}(\tilde{\bm{p}}) + n_{2L}(\tilde{\bm{p}})
 + n_{1}(\tilde{\bm{p}}')+ n_{2L}(\tilde{\bm{p}}'),
&
\Theta^{\rm eo, eo}_{\tilde{\bm{p}} \tilde{\bm{p}}'}
&= n_{2R}(\tilde{\bm{p}}) + n_{2R}(\tilde{\bm{p}}'),
\nonumber \\
\Theta^{\rm oo, ee}_{\tilde{\bm{p}} \tilde{\bm{p}}'}
&= n_{1}(\tilde{\bm{p}}) + n_{2L}(\tilde{\bm{p}}),
&
\Theta^{\rm eo, oe}_{\tilde{\bm{p}} \tilde{\bm{p}}'}
&= n_{2R}(\tilde{\bm{p}}) + n_{1}(\tilde{\bm{p}}')
 + n_{2L}(\tilde{\bm{p}}')+ n_{2R}(\tilde{\bm{p}}'),
\nonumber \\
\Theta^{\rm ee, oo}_{\tilde{\bm{p}} \tilde{\bm{p}}'}
&= n_{1}(\tilde{\bm{p}}')+ n_{2L}(\tilde{\bm{p}}'),
&
\Theta^{\rm oe, eo}_{\tilde{\bm{p}} \tilde{\bm{p}}'}
&= n_{1}(\tilde{\bm{p}}) + n_{2L}(\tilde{\bm{p}})
 + n_{2R}(\tilde{\bm{p}}) + n_{2R}(\tilde{\bm{p}}'), \nonumber \\
 \Theta^{\rm oe, oe}_{\tilde{\bm{p}} \tilde{\bm{p}}'} &=n_{1}(\tilde{\bm{p}}) + n_{2L}(\tilde{\bm{p}}) + n_{2R}(\tilde{\bm{p}})  + n_{1}(\tilde{\bm{p}}')+ n_{2L}(\tilde{\bm{p}}')+ n_{2R}(\tilde{\bm{p}}'). 
 \label{eq:Thethas}
\end{align}
The occupations $n_{\bar{R}}(\tilde{\bm{p}})$ and $n_{\bar{R}}(\tilde{\bm{p}}')$ do not appear in Eq.~\eqref{eq:Thethas}, as they cancel between $\tilde{\bm{p}}$ and $\tilde{\bm{p}}'$ due to the matching condition in $\bar{R}$ enforced by tracing out $\bar{R}$, as in Sec.~\ref{sec:redspin}.
The partial transpose with respect to $R^{(2)}$ is then
\begin{eqnarray}
\left[ \hat{\rho}_{\rm ee, ee}^{\rm (e)} \right]^{T_2}= && \sum_{\rm ee, ee}\rho_{\bm{p}_{1}\bm{p}_{2}, \bm{p}'_{1}\bm{p}'_{2}}^{(f, \mathrm{e})} \hat{{S}}_{\bm{p}_{1}}\hat{{S}}_{\bm{p}'_{2}} \left[ \hat{\rho}_R^{(g)}\right]^{T_2} \hat{{S}}_{\bm{p}'_{1}}\hat{{S}}_{\bm{p}_{2}}, \nonumber \\
\left[ \hat{\rho}_{\rm ee, oo}^{\rm (e)} \right]^{T_2} &&= \sum_{\rm ee, oo} (-1)^{\Theta^{\rm ee, oo}_{\tilde{\bm{p}},\tilde{\bm{p}}'} }
 \rho_{\bm{p}_{1}\bm{p}_{2}, \tilde{\bm{p}}'_{1}\tilde{\bm{p}}'_{2}}^{(f, \mathrm{e})} \hat{{S}}_{\bm{p}_{1}}\hat{{S}}_{\bm{p}'_{2}} \hat{S}^{\partial, (2)}_{R^{(1)}R^{(2)}} \left[ \hat{\rho}_R^{(g)}\right]^{T_2} \hat{S}^{\partial, (1)}_{R^{(1)}R^{(2)}}   \hat{{S}}_{\bm{p}'_{1}}\hat{{S}}_{\bm{p}_{2}},\nonumber \\
\left[ \hat{\rho}_{\rm oo, ee}^{\rm (e)} \right]^{T_2} && = \sum_{\rm oo, ee}  (-1)^{\Theta^{\rm oo, ee}_{\tilde{\bm{p}},\tilde{\bm{p}}'} }
\rho_{\tilde{\bm{p}}_{1}\tilde{\bm{p}}_{2}, \bm{p}'_{1}\bm{p}'_{2}}^{(f, \mathrm{e})} \hat{{S}}_{\bm{p}_{1}}\hat{{S}}_{\bm{p}'_{2}} \hat{S}^{\partial, (1)}_{R^{(1)}R^{(2)}} \left[ \hat{\rho}_R^{(g)}\right]^{T_2} \hat{{S}}_{\bm{p}'_{1}}\hat{S}^{\partial, (2)}_{R^{(1)}R^{(2)}} \hat{{S}}_{\bm{p}_{2}}, \nonumber \\ 
 \left[ \hat{\rho}_{\rm oo, oo}^{\rm (e)} \right]^{T_2}&& =\sum_{\rm oo, oo}  (-1)^{\Theta^{\rm oo, oo}_{\tilde{\bm{p}},\tilde{\bm{p}}'} }
\rho_{\tilde{\bm{p}}_{1}\tilde{\bm{p}}_{2}, \tilde{\bm{p}}'_{1}\tilde{\bm{p}}'_{2}}^{(f, \mathrm{e})} \hat{{S}}_{\bm{p}_{1}}\hat{{S}}_{\bm{p}'_{2}}\hat{S}^\partial_{R^{(1)}R^{(2)}} \left[ \hat{\rho}_R^{(g)}\right]^{T_2} \hat{S}^\partial_{R^{(1)}R^{(2)}} \hat{{S}}_{\bm{p}'_{1}}\hat{{S}}_{\bm{p}_{2}}
 \end{eqnarray}
and, for the odd components,
 \begin{eqnarray}
\left[ \hat{\rho}_{\rm eo, eo}^{\rm (o)} \right]^{T_2}= && \sum_{\rm eo,eo}  (-1)^{\Theta^{\rm eo, eo}_{\tilde{\bm{p}},\tilde{\bm{p}}'} }
 \rho_{\bm{p}_{1}\tilde{\bm{p}}_{2}, \bm{p}'_{1} \tilde{\bm{p}}'_{2}}^{(f, \mathrm{o})} \hat{{S}}_{\bm{p}_{1}}\hat{{S}}_{\bm{p}'_{2}}  \hat{S}_{R^{(2)} \bar{R}}^{\partial, (2)} \left[ \hat{\rho}_R^{(g)}\right]^{T_2}  \hat{S}_{R^{(2)} \bar{R}}^{\partial, (2)}  \hat{{S}}_{\bm{p}'_{1}}\hat{{S}}_{\bm{p}_{2}} ,\nonumber\\
\left[ \hat{\rho}_{\rm eo, oe}^{\rm (o)} \right]^{T_2} && = \sum_{\rm eo, oe}  (-1)^{\Theta^{\rm eo, oe}_{\tilde{\bm{p}},\tilde{\bm{p}}'} }
 \rho_{{\bm{p}}_{1}\tilde{\bm{p}}_{2}, \tilde{\bm{p}}'_{1}\tilde{\bm{p}}'_{2}}^{(f, \mathrm{o})} \hat{{S}}_{\bm{p}_{1}}\hat{{S}}_{\bm{p}'_{2}} \hat{S}_{R^{(2)} \bar{R}}^{\partial, (2)} \hat{S}^{\partial, (2)}_{R^{(1)}R^{(2)}} \left[ \hat{\rho}_R^{(g)}\right]^{T_2}  \hat{S}^{\partial, (1)}_{R^{(1)}R^{(2)}} \hat{S}_{R^{(2)} \bar{R}}^{\partial, (2)} \hat{{S}}_{\bm{p}'_{1}}\hat{{S}}_{\bm{p}_{2}},\nonumber  \\ 
\left[ \hat{\rho}_{\rm oe, eo}^{\rm (o)} \right]^{T_2} && = \sum_{\rm oe, eo}  (-1)^{\Theta^{\rm oe, eo}_{\tilde{\bm{p}},\tilde{\bm{p}}'} }
\rho_{\tilde{\bm{p}}_{1}\tilde{\bm{p}}_{2}, {\bm{p}}'_{1}\tilde{\bm{p}}'_{2}}^{(f, \mathrm{o})} \hat{{S}}_{\bm{p}_{1}}\hat{{S}}_{\bm{p}'_{2}} \hat{S}^{\partial, (1)}_{R^{(1)}R^{(2)}} \hat{S}_{R^{(2)} \bar{R}}^{\partial, (2)} \left[ \hat{\rho}_R^{(g)}\right]^{T_2}\hat{S}^{\partial, (2)}_{R^{(1)}R^{(2)}}  \hat{S}_{R^{(2)} \bar{R}}^{\partial, (2)}  \hat{{S}}_{\bm{p}'_{1}}\hat{{S}}_{\bm{p}_{2}} ,\nonumber \\
\left[ \hat{\rho}_{\rm oe, oe}^{\rm (o)} \right]^{T_2} && = \sum_{\rm oe, oe} (-1)^{\Theta^{\rm oe, oe}_{\tilde{\bm{p}},\tilde{\bm{p}}'} }
 \rho_{\tilde{\bm{p}}_{1}\tilde{\bm{p}}_{2}, \tilde{\bm{p}}'_{1}\tilde{\bm{p}}'_{2}}^{(f, \mathrm{o})} \hat{{S}}_{\bm{p}_{1}}\hat{{S}}_{\bm{p}'_{2}} \hat{S}^\partial_{R^{(1)}R^{(2)}} \hat{S}_{R^{(2)} \bar{R}}^{\partial, (2)} \left[ \hat{\rho}_R^{(g)}\right]^{T_2}\hat{S}^\partial_{R^{(1)}R^{(2)}} \hat{S}_{R^{(2)} \bar{R}}^{\partial, (2)} \hat{{S}}_{\bm{p}'_{1}}\hat{{S}}_{\bm{p}_{2}}  .
\end{eqnarray} 
We notice that $\left[\hat{\rho}_R\right]^{T_2}$ has a different block structure compared to its fermionic counterpart of Eq.~\eqref{eq:rhot2fermion}, when considering the fermion parity in each sub-region. Indeed, differently from the purely fermionic system, the partial transpose does not exchange blocks such as $\rho^{\rm (e)}_{\rm ee, oo}$ and $\rho^{\rm (o)}_{\rm eo, oe}$ between the even and odd parity sectors, but rather creates new blocks, because the $ZX$ strings to the side of $\left[\hat{\rho}_R^{(g)}\right]^{T_2}$ carry different ``legs'' $\hat{S}^\partial_{RR'}$ (or their boundary truncations) extending to the boundary. The result is that products of the form $[\rho^{(\rm o)}_{\rm eo, eo}]^{T_2}[\rho^{\rm (e)}_{\rm ee, oo}]^{T_2}$ vanish, differently from the corresponding products $[\rho^{(f,\mathrm{o})}_{\rm eo, eo}]^{T_2}[\rho^{(f,\mathrm{e})}_{\rm ee, oo}]^{T_2}$ for fermions. We illustrate this by showing the explicit calculation of one of such terms. 
Consider the product 
\begin{multline}
[\rho^{\rm (o)}_{\rm eo, eo}]^{T_2} [\rho^{\rm (e)}_{\rm ee, oo}]^{T_2} = \sum_{\mathrm{eo,eo} }^{\bm{p}}
\sum_{ \mathrm{ee,oo} }^{\bm{q}}
 (-1)^{\Theta^{\rm eo, eo}_{\tilde{\bm{p}},\tilde{\bm{p}}'} }
  (-1)^{\Theta^{\rm ee, oo}_{\tilde{\bm{q}},\tilde{\bm{q}}'} }
   \rho_{\bm{p}_{1}\tilde{\bm{p}}_{2}, \bm{p}'_{1} \tilde{\bm{p}}'_{2}}^{(f, \mathrm{o})} 
    \rho_{\bm{q}_{1}\bm{q}_{2}, \tilde{\bm{q}}'_{1}\tilde{\bm{q}}'_{2}}^{(f, \mathrm{e})}
\hat{{S}}_{\bm{p}_{1}}\hat{{S}}_{\bm{p}'_{2}}  \hat{S}_{R^{(2)} \bar{R}}^{\partial, (2)} \left[ \hat{\rho}_R^{(g)}\right]^{T_2}\\
\cdot \hat{S}_{R^{(2)} \bar{R}}^{\partial, (2)}  \hat{{S}}_{\bm{p}'_{1}}\hat{{S}}_{\bm{p}_{2}} 
  \hat{{S}}_{\bm{q}_{1}}\hat{{S}}_{\bm{q}'_{2}} \hat{S}^{\partial, (2)}_{R^{(1)}R^{(2)}} \left[ \hat{\rho}_R^{(g)}\right]^{T_2} \hat{S}^{\partial, (1)}_{R^{(1)}R^{(2)}}   \hat{{S}}_{\bm{q}'_{1}}\hat{{S}}_{\bm{q}_{2}}  .
  \end{multline}
  Expanding $[\hat{\rho}_R^{(g)}]^{T_2}$ as in Eq.~\eqref{eq:rhoAgtransposew}, the latter expression becomes
  \begin{align}
  [\rho^{\rm (o)}_{\rm eo, eo}]^{T_2} [\rho^{\rm (e)}_{\rm ee, oo}]^{T_2}=& \frac{1}{2^{2(3L_y-1)}} \sum_{
\substack{w,\bar{w}, v, \bar{v}=\pm,\\ : w \bar{w}=v \bar{v}} }   \sum_{
\substack{\alpha,\bar{\alpha}, \beta, \bar{\beta}=\pm,\\ : \alpha \bar{\alpha}= \beta \bar{\beta}} } \sum_{\bm{\lambda}, \bm{\mu}}  
\sum_{\mathrm{eo,eo} }^{\bm{p}}
\sum_{ \mathrm{ee,oo} }^{\bm{q}}
 (-1)^{\Theta^{\rm eo, eo}_{\tilde{\bm{p}},\tilde{\bm{p}}'} }
  (-1)^{\Theta^{\rm ee, oo}_{\tilde{\bm{q}},\tilde{\bm{q}}'} }
   \rho_{\bm{p}_{1}\tilde{\bm{p}}_{2}, \bm{p}'_{1} \tilde{\bm{p}}'_{2}}^{(f, \mathrm{o})}   \rho_{\bm{q}_{1}\bm{q}_{2}, \tilde{\bm{q}}'_{1}\tilde{\bm{q}}'_{2}}^{(f, \mathrm{e})} \nonumber \\
 &  \qquad \qquad  %
    \hat{{S}}_{\bm{p}_{1}}\hat{{S}}_{\bm{p}'_{2}}  \hat{S}_{R^{(2)} \bar{R}}^{\partial, (2)} \ket{w_{\lambda_1, \lambda_2}^{(1)}\bar{v}_{\lambda_4, \lambda_3}^{(2)}}\bra{\beta_{\mu_1\mu_4}^{(1)}\bar{\alpha}_{\mu_2\mu_3}^{(2)}} 
  \hat{S}^{\partial, (1)}_{R^{(1)}R^{(2)}}   \hat{{S}}_{\bm{q}'_{1}}\hat{{S}}_{\bm{q}_{2}}\nonumber  \\
  &\qquad \qquad \cdot  \bra{v_{\lambda_1\lambda_4}^{(1)}\bar{w}_{\lambda_2\lambda_3}^{(2)}}  \hat{S}_{R^{(2)} \bar{R}}^{\partial, (2)}  \hat{{S}}_{\bm{p}'_{1}}\hat{{S}}_{\bm{p}_{2}} 
  \hat{{S}}_{\bm{q}_{1}}\hat{{S}}_{\bm{q}'_{2}} \hat{S}^{\partial, (2)}_{R^{(1)}R^{(2)}} 
  \ket{\alpha_{\mu_1, \mu_2}^{(1)}\bar{\beta}_{\mu_4, \mu_3}^{(2)}} .
  \label{eq:onestep}
\end{align}
We note that $\hat{S}^{\partial, (2)}_{R^{(1)}R^{(2)}} \hat{S}_{R^{(2)} \bar{R}}^{\partial, (2)} = \hat{S}_*^{(2)} \hat W_{x}^{ZX, (2)}$ up to a sign, where $\hat{S}^{(2)}_*$ is a valid fermion configuration in $R^{(2)}$ with a fermion at the top left and bottom right corners, as depicted in Fig.~\ref{fig:Sstar}. 
Leveraging the fact that the states $\ket{\bar{\beta}^{(2)}_{\mu\mu'}}$ have been chosen to be eigenstates of $W_{x}^{ZX, (2)}$ (see Sec.~\ref{sec:logneggs}), the matrix element in Eq.~\eqref{eq:onestep} can be evaluated as   
\begin{align}
& \bra{v_{\lambda_1\lambda_4}^{(1)}\bar{w}_{\lambda_2\lambda_3}^{(2)}}  \hat{{S}}_{\bm{p}'_{1}}\hat{{S}}_{\bm{p}_{2}} 
  \hat{{S}}_{\bm{q}_{1}}\hat{{S}}_{\bm{q}'_{2}}  \hat{S}^{(2)}_*  \hat W_{x}^{ZX, (2)} 
  \ket{\alpha_{\mu_1, \mu_2}^{(1)}\bar{\beta}_{\mu_4, \mu_3}^{(2)}} \nonumber \\
 &\qquad \qquad  \qquad \qquad  =\bar{\beta} \bra{v_{\lambda_1\lambda_4}^{(1)}\bar{w}_{\lambda_2\lambda_3}^{(2)}}  \hat{{S}}_{\bm{p}'_{1}}\hat{{S}}_{\bm{p}_{2}} 
  \hat{{S}}_{\bm{q}_{1}}\hat{{S}}_{\bm{q}'_{2}} \hat{S}^{(2)}_* 
  \ket{\alpha_{\mu_1, \mu_2}^{(1)}\bar{\beta}_{\mu_4, \mu_3}^{(2)}},\nonumber  \\
&\qquad \qquad  \qquad \qquad  = \bar{\beta} \delta_{v\alpha}\delta_{\bar{w}\bar{\beta}}\delta_{\lambda_1\mu_1}\delta_{\lambda_2\mu_4}\delta_{\lambda_3\mu_3}\delta_{\lambda_4\mu_2} \delta_{\bm{p}_1'\bm{q}_1}\delta_{\bm{p}_2^* \bm{q}_2'},
\end{align}
where $\bm{p}_{2}^{*}$ is the fermion configuration such that $\hat{S}_{\bm{p}_2^{ *}} = \hat{S}_{\bm{p}_2} \hat{S}^{(2)}_* $. Inserting into Eq.~\eqref{eq:onestep}, we obtain
\begin{align}
  [\rho^{\rm (o)}_{\rm eo, eo}]^{T_2} [\rho^{\rm (e)}_{\rm ee, oo}]^{T_2} =& \frac{1}{2^{2(3L_y-1)}}
\sum_{\mathrm{eo,eo} }^{\bm{p}}
\sum_{ \mathrm{ee,oo} }^{\bm{q}} 
 (-1)^{\Theta^{\rm eo, eo}_{\tilde{\bm{p}},\tilde{\bm{p}}'} }
  (-1)^{\Theta^{\rm ee, oo}_{\tilde{\bm{q}},\tilde{\bm{q}}'} }
 \rho_{\bm{p}_{1}\tilde{\bm{p}}_{2}, \bm{p}'_{1} \tilde{\bm{p}}'_{2}}^{(f, \mathrm{o})} 
     \rho_{\bm{p}_{1}'\bm{q}_{2}, \tilde{\bm{q}}'_{1}\tilde{\bm{p}}^{*}_{2}}^{(f, \mathrm{e})} \hat{{S}}_{\bm{p}_{1}}\hat{{S}}_{\bm{p}'_{2}}  \hat{S}_{R^{(2)} \bar{R}}^{\partial, (2)} \nonumber \\
&    %
  \left( \sum_{
   \substack{
   w,\bar{w}, v, \bar{v}, \bar{\alpha}, \beta=\pm, \\
    : w \bar{w}=v \bar{v} ,\\ 
    v  \bar{\alpha}=\beta \bar{w} }
     }
 \sum_{\bm{\lambda}}
 \bar{w} \ket{w_{\lambda_1, \lambda_2}^{(1)} \bar{v}_{\lambda_4, \lambda_3}^{(2)}}
    \bra{\beta_{\lambda_1\lambda_2}^{(1)}\bar{\alpha}_{\lambda_4\lambda_3}^{(2)}} \right)
  \hat{S}^{\partial, (1)}_{R^{(1)} R^{(2)}}   \hat{{S}}_{\bm{q}'_{1}}\hat{{S}}_{\bm{q}_{2}} .
  \label{eq:onemore}
\end{align}
Considering the term in brackets and noting that the kets and bras do not depend directly on $\bar{w}$ and $v$, the sums over the $W_x^{ZX, (2)}$ eigenvalues can be rearranged as follows,
\begin{equation}
       \left(\sum_{\bar{w}=\pm} \bar{w}  \sum_{
\substack{w, \bar{v}, \bar{\alpha} , \beta=\pm,\\ : w \bar{v} = \beta \bar{\alpha} } }  \sum_{\bm{\lambda}}
 \ket{w_{\lambda_1, \lambda_2}^{(1)}\bar{v}_{\lambda_4, \lambda_3}^{(2)}}
    \bra{\beta_{\lambda_1\lambda_2}^{(1)}\bar{\alpha}_{\lambda_4\lambda_3}^{(2)}} \right),
\end{equation}
which makes it clear that this term vanishes, since $\sum_{\bar{w}=\pm}\bar{w}=0$, and hence $[\rho^{\rm (o)}_{\rm eo, eo}]^{T_2} [\rho^{\rm (e)}_{\rm ee, oo}]^{T_2}=0$.

To proceed in computing the negativity for the spin state, we next analyze the sequence of even moments of the partial transpose. Due to the orthogonality relations just discussed between terms carrying the ``${}^{\rm (e)}$'' and ``${}^{\rm (o)}$'' label (i.e., between terms corresponding to different $\bar{R}$ parities), we first note that the square $[\hat{\rho}_R^{T_2}]^2$ has a block-diagonal structure with diagonal blocks
\begin{align}
(\hat{\rho}_R^{T_2})^{2} = & \mathrm{diag}\left\{ \left[\left(\rho^{(\mathrm{e})}_{\rm ee,ee} \right)^{T_2}\right]^2, \left[\left(\rho^{(\mathrm{e})}_{\rm oo,oo} \right)^{T_2}\right]^2 , \left(\rho^{(\mathrm{e})}_{\rm ee,oo} \right)^{T_2} \left(\rho^{(\mathrm{e})}_{\rm oo,ee} \right)^{T_2},\left(\rho^{(\mathrm{e})}_{\rm oo,ee} \right)^{T_2} \left(\rho^{(\mathrm{e})}_{\rm ee,oo} \right)^{T_2}, \right.\nonumber \\
&\quad \left. \left[\left(\rho^{(\mathrm{o})}_{\rm eo,eo} \right)^{T_2}\right]^2, \left[\left(\rho^{(\mathrm{o})}_{\rm oe, oe} \right)^{T_2}\right]^2, \left(\rho^{(\mathrm{o})}_{\rm eo,oe} \right)^{T_2} \left(\rho^{(\mathrm{o})}_{\rm oe,eo} \right)^{T_2},  \left(\rho^{(\mathrm{o})}_{\rm oe,eo} \right)^{T_2} \left(\rho^{(\mathrm{o})}_{\rm eo,oe} \right)^{T_2}\right\}.
\end{align}
%
Hence, the $2n$th power has the structure
\begin{eqnarray}
(\hat{\rho}_R^{T_2})^{2n} = & \mathrm{diag}\left\{ \left[\left(\rho^{(\mathrm{e})}_{\rm ee,ee} \right)^{T_2}\right]^{2n}, \left[\left(\rho^{(\mathrm{e})}_{\rm oo,oo} \right)^{T_2}\right]^{2n} , \left[\left(\rho^{(\mathrm{e})}_{\rm ee,oo} \right)^{T_2} \left(\rho^{(\mathrm{e})}_{\rm oo,ee} \right)^{T_2}\right]^{n}, \left[\left(\rho^{(\mathrm{e})}_{\rm oo,ee} \right)^{T_2} \left(\rho^{(\mathrm{e})}_{\rm ee,oo} \right)^{T_2}\right]^n, \right. \nonumber \\
&\quad \left. \left[\left(\rho^{(\mathrm{o})}_{\rm eo,eo} \right)^{T_2}\right]^{2n}, \left[\left(\rho^{(\mathrm{o})}_{\rm oe, oe} \right)^{T_2}\right]^{2n}, \left[\left(\rho^{(\mathrm{o})}_{\rm eo,oe} \right)^{T_2} \left(\rho^{(\mathrm{o})}_{\rm oe,eo} \right)^{T_2} \right]^n,  \left[ \left(\rho^{(\mathrm{o})}_{\rm oe,eo} \right)^{T_2} \left(\rho^{(\mathrm{o})}_{\rm eo,oe} \right)^{T_2} \right]^n \right\}.
\label{eq:2nrhoa1}
\end{eqnarray}
When computing the trace to determine the $2n$-th moment, $\mathrm{tr}(\hat{\rho}_R^{T_2})^{2n}$, one finds that the latter factorizes between the moment of the transposed ground-state density matrix $[\hat{\rho}_R^{(g)}]^{T_2}$ and a factor depending on the fermionic density matrix $\hat{\rho}^{(f)}_R$ only. We illustrate this for one term:
\begin{eqnarray}
 \mathrm{tr}\left[\left(\rho^{(\mathrm{e})}_{\rm ee,oo} \right)^{T_2} \right. && \left. \left(\rho^{(\mathrm{e})}_{\rm oo,ee} \right)^{T_2}\right]^{n} = \sum_{\rm ee, oo}^{\bm{p}^{[1]}} \sum_{\rm oo, ee}^{\bm{q}^{[1]}} \dots  \sum_{\rm ee, oo}^{\bm{p}^{[n]}} \sum_{\rm oo, ee}^{\bm{q}^{[n]}}  (-1)^{\Theta^{\rm ee, oo}_{\tilde{\bm{p}}^{[1]},\tilde{\bm{p}}^{\prime [1]}} }
  (-1)^{\Theta^{\rm oo, ee}_{\tilde{\bm{q}}^{[1]},\tilde{\bm{q}}^{\prime [1]}} }\dots (-1)^{\Theta^{\rm ee, oo}_{\tilde{\bm{p}}^{[n]},\tilde{\bm{p}}^{\prime [n]}} }
  (-1)^{\Theta^{\rm oo, ee}_{\tilde{\bm{q}}^{[n]},\tilde{\bm{q}}^{\prime [n]}} } \ \nonumber \\
&&  \rho_{\bm{p}^{[1]}_{1}\bm{p}^{[1]}_{2}, \tilde{\bm{p}}^{\prime [1]}_{1}\tilde{\bm{p}}^{\prime [1]}_{2}}^{(f, \mathrm{e})} \rho_{\tilde{\bm{q}}^{[1]}_{1}\tilde{\bm{q}}^{[1]}_{2}, \bm{q}^{\prime [1]}_{1}\bm{q}^{\prime [1]}_{2}}^{(f, \mathrm{e})} \dots  \rho_{\bm{p}^{[n]}_{1}\bm{p}^{[n]}_{2}, \tilde{\bm{p}}^{\prime [n]}_{1}\tilde{\bm{p}}^{\prime [n]}_{2}}^{(f, \mathrm{e})} \rho_{\tilde{\bm{q}}^{[n]}_{1}\tilde{\bm{q}}^{[n]}_{2}, \bm{q}^{\prime [n]}_{1}\bm{q}^{\prime [n]}_{2}}^{(f, \mathrm{e})} \nonumber \\
&& \cdot \mathrm{tr}\left[  \hat{{S}}_{\bm{p}^{[1]}_{1}}\hat{{S}}_{\bm{p}^{\prime [1]}_{2}} \hat{S}^{\partial, (2)}_{R^{(1)}R^{(2)}} \left( \hat{\rho}_R^{(g)}\right)^{T_2} \hat{S}^{\partial, (1)}_{R^{(1)}R^{(2)}}   \hat{{S}}_{\bm{p}^{\prime [1]}_{1}}\hat{{S}}_{\bm{p}^{[1]}_{2}} \hat{{S}}_{\bm{q}^{[1]}_{1}}\hat{{S}}_{\bm{q}^{\prime [1]}_{2}} \hat{S}^{\partial, (1)}_{R^{(1)}R^{(2)}} \left( \hat{\rho}_R^{(g)}\right)^{T_2} \hat{{S}}_{\bm{q}^{\prime [1]}_{1}}\hat{S}^{\partial, (2)}_{R^{(1)}R^{(2)}} \hat{{S}}_{\bm{q}^{[1]}_{2}} \dots \right. \nonumber \\
&& \left. \hat{{S}}_{\bm{p}^{[n]}_{1}}\hat{{S}}_{\bm{p}^{\prime [n]}_{2}} \hat{S}^{\partial, (2)}_{R^{(1)}R^{(2)}} \left( \hat{\rho}_R^{(g)}\right)^{T_2} \hat{S}^{\partial, (1)}_{R^{(1)}R^{(2)}}   \hat{{S}}_{\bm{p}^{\prime [n]}_{1}}\hat{{S}}_{\bm{p}^{[n]}_{2}} \hat{{S}}_{\bm{q}^{[n]}_{1}}\hat{{S}}_{\bm{q}^{\prime [n]}_{2}} \hat{S}^{\partial, (1)}_{R^{(1)}R^{(2)}} \left( \hat{\rho}_R^{(g)}\right)^{T_2} \hat{{S}}_{\bm{q}^{\prime [n]}_{1}}\hat{S}^{\partial, (2)}_{R^{(1)}R^{(2)}} \hat{{S}}_{\bm{q}^{[n]}_{2}} \right] ,
 \label{eq:mome1}
\end{eqnarray}
where the superscript $[j]$ labels the configurations summed over in the $j$th pair of factors $(\rho^{(\mathrm{e})}_{\rm ee,oo})^{T_2}(\rho^{(\mathrm{e})}_{\rm oo,ee})^{T_2}$, with the cyclic identification $[n+1]\equiv[1]$, while the subscripts $1$ and $2$ denote, as before, the parts of a configuration in $R^{(1)}$ and $R^{(2)}$, respectively.
Inserting the decomposition \eqref{eq:rhoAgtransposeeta} for $\left( \hat{\rho}_R^{(g)}\right)^{T_2}$, the trace in Eq.~\eqref{eq:mome1} reduces to the product of matrix elements of the form 
\begin{eqnarray}
\langle s^{(1)}_{\lambda_{1} \lambda_{4}}, s^{(2)}_{\lambda_{2} \lambda_{3}} | \hat{S}^{\partial, (1)}_{R^{(1)}R^{(2)}}   \hat{{S}}_{\bm{p}^{\prime [j]}_{1}}\hat{{S}}_{\bm{p}^{[j]}_{2}} \hat{{S}}_{\bm{q}^{[j]}_{1}}\hat{{S}}_{\bm{q}^{\prime [j]}_{2}} \hat{S}^{\partial, (1)}_{R^{(1)}R^{(2)}} \ket{ \bar{s}^{(1)}_{\mu_{1} \mu_{2}}, \bar{s}^{ (2)}_{\mu_{4} \mu_{3}}} = \nonumber \\
  \delta_{s\bar{s}} \delta_{\lambda_{1}\mu_{1}}\delta_{\lambda_{4}\mu_{2}}    \delta_{\lambda_{2}\mu_{4}}\delta_{\lambda_{3}\mu_{3}} \delta_{\tilde{\bm{p}}^{\prime [j]}_{1} \tilde{\bm{q}}^{[j]}_{1}}  \delta_{\bm{p}^{[j]}_{2} \bm{q}^{\prime [j]}_{2}}
\end{eqnarray}
and
\begin{eqnarray}
\langle s^{ (1)}_{\lambda_{1} \lambda_{4}}, s^{(2)}_{\lambda_{2} \lambda_{3}} | \hat{S}^{\partial, (2)}_{R^{(1)}R^{(2)}}   \hat{{S}}_{\bm{q}^{\prime [j]}_{1}}\hat{{S}}_{\bm{q}^{[j]}_{2}} \hat{{S}}_{\bm{p}^{[j+1]}_{1}}\hat{{S}}_{\bm{p}^{\prime [j+1]}_{2}} \hat{S}^{\partial, (2)}_{R^{(1)}R^{(2)}} \ket{ \bar{s}^{(1)}_{\mu_{1} \mu_{2}}, \bar{s}^{ (2)}_{\mu_{4} \mu_{3}}} =\nonumber \\
  \delta_{s\bar{s}} \delta_{\lambda_{1}\mu_{1}}\delta_{\lambda_{4}\mu_{2}}    \delta_{\lambda_{2}\mu_{4}}\delta_{\lambda_{3}\mu_{3}}
 \delta_{\bm{q}^{\prime [j]}_{1} \bm{p}^{[j+1]}_{1}}  \delta_{\tilde{\bm{q}}^{[j]}_{2} \tilde{\bm{p}}^{\prime [j+1]}_{2}} .
\end{eqnarray}
One, thus, obtains
\begin{eqnarray}
\mathrm{tr}\left[\left(\rho^{(\mathrm{e})}_{\rm ee,oo} \right)^{T_2} \left(\rho^{(\mathrm{e})}_{\rm oo,ee} \right)^{T_2}\right]^{n} = &&\sum_{\rm ee, oo}^{\bm{p}^{[1]}} \sum_{\rm oo, ee}^{\bm{q}^{[1]}} \dots  \sum_{\rm ee, oo}^{\bm{p}^{[n]}} \sum_{\rm oo, ee}^{\bm{q}^{[n]}}   \rho_{\bm{p}^{[1]}_{1}\bm{p}^{[1]}_{2}, \tilde{\bm{p}}^{\prime [1]}_{1}\tilde{\bm{p}}^{\prime [1]}_{2}}^{(f, \mathrm{e})} \rho_{\tilde{\bm{q}}^{[1]}_{1}\tilde{\bm{q}}^{[1]}_{2}, \bm{q}^{\prime [1]}_{1}\bm{q}^{\prime [1]}_{2}}^{(f, \mathrm{e})} \dots  \rho_{\bm{p}^{[n]}_{1}\bm{p}^{[n]}_{2}, \tilde{\bm{p}}^{\prime [n]}_{1}\tilde{\bm{p}}^{\prime [n]}_{2}}^{(f, \mathrm{e})} \rho_{\tilde{\bm{q}}^{[n]}_{1}\tilde{\bm{q}}^{[n]}_{2}, \bm{q}^{\prime [n]}_{1}\bm{q}^{\prime [n]}_{2}}^{(f, \mathrm{e})} \nonumber \\
&& \mathrm{tr}\left[\left( \hat{\rho}_R^{(g)}\right)^{T_2}\right]^{2n} \prod_{j}\delta_{\tilde{\bm{p}}^{\prime [j]}_{1} \tilde{\bm{q}}^{[j]}_{1}}  \delta_{\bm{p}^{[j]}_{2} \bm{q}^{\prime [j]}_{2}}
  \delta_{\bm{q}^{\prime [j]}_{1} \bm{p}^{[j+1]}_{1}}  \delta_{\tilde{\bm{q}}^{[j]}_{2} \tilde{\bm{p}}^{\prime [j+1]}_{2}}, \nonumber \\
  = && \mathrm{tr}\left[\left( \hat{\rho}_R^{(g)}\right)^{T_2}\right]^{2n}\mathrm{tr}\left[\left(\rho^{(f, \mathrm{e})}_{\rm ee,oo} \right)^{T_2} \left(\rho^{(f,\mathrm{e})}_{\rm oo,ee} \right)^{T_2}\right]^{n} .
 \label{eq:mome2}
\end{eqnarray}
The phase factors $\Theta$ in Eq.~\eqref{eq:mome1} cancel, since $n_1$ and $n_{2L}$ depend only on the $R^{(1)}$ and $R^{(2)}$ parts of a configuration, respectively, and the Kronecker deltas give $n_{1}(\tilde{\bm{p}}^{\prime [j]}) = n_{1}(\tilde{\bm{q}}^{[j]})$ and $n_{2L}(\tilde{\bm{q}}^{[j]}) = n_{2L}(\tilde{\bm{p}}^{\prime [j+1]})$. The same cancellation occurs analogously for all terms in Eq.~\eqref{eq:2nrhoa1}. 
%
Considering all the terms in Eq.~\eqref{eq:2nrhoa1} with a decomposition as in Eq.~\eqref{eq:mome2}, the fermionic contribution can be written compactly by decomposing $[\hat{\rho}_R^{(f)}]^{T_2}$ as
\begin{equation}
(\hat{\rho}_R^{(f)})^{T_2} = \hat{\Sigma}_+ + \hat{\Sigma}_-, 
\end{equation}
with $\hat{\Sigma}_+$ and $\hat{\Sigma}_-$ denoting, respectively, terms for which the partial transpose does not alter the parity in $R^{(2)}$ (i.e., of the form $\hat{\rho}^{(f)}_{ij, ij}$ for $i,j\in \{\mathrm{e,o}\}$) and terms for which it does ($\hat{\rho}^{(f)}_{ij, kl}$ with $j\ne l$), namely,
\begin{eqnarray}
&& \hat \Sigma_\pm = \frac{1}{2} \left[(\hat{\rho}_R^{(f)})^{T_2} \pm \hat{\mathcal{P}}_2 (\hat{\rho}_R^{(f)})^{T_2} \hat{\mathcal{P}}_2  \right], 
\label{eq:sigmadx}
\end{eqnarray}
where $\hat{\mathcal{P}}_2=(-1)^{\sum_{p\in R^{(2)}}\hat{f}_p^\dag \hat{f}_p}$ is the fermion parity in $R^{(2)}$. The $2n$th moment of partial transpose for the spin system can then be decomposed as
\begin{equation}
\mathrm{tr}\left[\rho_R^{T_2}\right]^{2n} =  \mathrm{tr}\left[(\rho_R^{(g)})^{T_2}\right]^{2n} \left[\mathrm{tr}\hat{\Sigma}_+^{2n} + \mathrm{tr}\hat{\Sigma}_-^{2n}  \right].
\end{equation}
Through the replica limit [Eq.~\eqref{eq:replicanegativity}] we thus obtain the following decomposition of the logarithmic negativity,
\begin{eqnarray}
\mathcal{E} =&& \mathcal{E}^{(g)} + \widetilde{\mathcal{E}}^{(f)},\nonumber  \\
&& = (L_y-1)\log(2) + \widetilde{\mathcal{E}}^{(f)},
\end{eqnarray}
where we used Eq.~\eqref{eq:negativitygs} for the ground state contribution $\mathcal{E}^{(g)}$ and where we defined the fermionic contribution
\begin{equation}
 \widetilde{\mathcal{E}}^{(f)} = \log\left( \Vert  \hat{\Sigma}_+ \Vert_1+ \Vert \hat{\Sigma}_- \Vert_1\right).
\end{equation}
In Sec.~\ref{sec:boundneg}, we will discuss this fermionic contribution for the long-time state of the system.

\section{Floquet dynamics and high-frequency analysis}

In this section, we discuss the analytical high-frequency approximation of the dynamics presented in the main text. 

\subsection{The Floquet problem}
\label{sec:Floquetproblem}

For homogeneous driving at frequency $\omega=2\pi/\tau$ throughout the lattice and in the absence of disorder and vortices ($e$ particles), the time-periodic Hamiltonian in fermion and in quasimomentum representation reads as 
\begin{equation}\label{eq:Hkt}
\hat H(t) = \hat{H}(t+\tau)= \frac{1}{2}\sum_{\bm{k}}(\hat{f}_{\bm{k}}^\dag, \hat{f}_{-\bm{k}}){\mathcal{H}}_{\bm{k}}(t)\begin{pmatrix}
\hat{f}_{\bm{k}} \\
\hat{f}_{-\bm{k}}^\dag
\end{pmatrix},
\end{equation}
with 
\begin{equation}
\label{eq:mathcalH}
{\mathcal{H}}_{\bm{k}}(t) =\begin{pmatrix}
 \varepsilon_{\bm{k}}(t) & -i \Delta_{\bm{k}}(t)
\\
i \Delta_{\bm{k}}^*(t) &  -\varepsilon_{-\bm{k}}(t)
\end{pmatrix}
\end{equation}
and
\begin{eqnarray}
& \varepsilon_{\bm{k}}(t) = \mu_f - 2 \Omega(t) [\cos(k_x) + \cos(k_y)],\nonumber  \\
& \Delta_{\bm{k}}(t) = - 2\Omega (t)[\sin(k_x) - \sin(k_y)].
\end{eqnarray}
In the following, we will indicate with $K_0$ the set of $\bm{k}$ values for which $\Delta_{\bm{k}}(t)$ vanishes, that is 
\begin{equation}
\label{eq:ksinK0}
k_y = \pi-k_x \ (\mathrm{mod}\ 2\pi), \qquad k_y = k_x \ (\mathrm{mod}\ 2\pi).
\end{equation}
The drive cannot alter the occupation of these quasimomenta. The exact cardinality $|K_0|$ depends on the parity properties of $L_x$ and $L_y$, but it can generally be bounded by
\begin{equation}
\label{eq:K0bound}
|K_0| \le 2 \min(L_x, L_y). 
\end{equation}
Indeed, for each $k_x\in(0, 2\pi]$, there are at most two possible values of allowed $k_y$ according to Eq.~\eqref{eq:ksinK0} $(\mathrm{mod}\ 2\pi)$ and likewise for each $k_y$. However, consider the case in which $L_x=L_y=L$ and $L$ is a multiple of four. Now the quasimomenta $k_x=\pi/2$ or $k_x=3\pi/2$ are allowed and the corresponding two values of $k_y$ mentioned above coincide, resulting in $|K_0|=2(L-1)$.
 
The Hamiltonian generates the end-of-period time evolution operator
\begin{equation}
 \hat{U}(\tau) = \mathcal{T}\exp \left(-i \int_0^\tau \!\! dt\ \hat{H}(t)  \right),
\end{equation}
with $\mathcal{T}$ the time-ordering operator.
As the initial state is the vacuum $\ket{0}$ of the operators $\hat{f}_{\bm{k}}$, the time-evolved state $\ket{\psi(t)}=\hat{U}(t) \ket{0}$ at time $t$ is the vacuum of the operators $\hat{f}_{\bm{k}}(t) =  \hat{U}(t) \hat{f}_{\bm{k}} \hat{U}^\dag (t) $,
\begin{equation}
\label{eq:annihilft}
\hat{f}_{\bm{k}}(t) \ket{\psi(t)} = 0.
 \end{equation}
 The operators $\big(\hat{f}_{\bm{k}}(t),\hat{f}^\dag_{-\bm{k}}(t)\big)^T$ then satisfy
  \begin{eqnarray}
 \frac{d}{dt} \hat{f}_{\bm{k}}(t) && = -i [\hat{H}(t), \hat{f}_{\bm{k}}(t)], \nonumber \\
 \frac{d}{dt} \hat{f}_{-\bm{k}}^\dag(t) && = -i [\hat{H}(t), \hat{f}_{-\bm{k}}^\dag(t)].
 \label{eq:comm1BdG}
\end{eqnarray}
Denoting with $\mathcal{A}_{\bm{k}}(t)$ the unitary transformation relating the operators $\big(\hat{f}_{\bm{k}}(t),\hat{f}^\dag_{-\bm{k}}(t)\big)^T$ and $\big(\hat{f}_{\bm{k}},\hat{f}^\dag_{-\bm{k}}\big)^T$ according to
 \begin{equation}
 \label{eq:eqforfm1}
\begin{pmatrix}
\hat{f}_{\bm{k}}(t)\\
\hat{f}^\dag_{-\bm{k}}(t)
\end{pmatrix} =  \mathcal{A}^\dag_{\bm{k}}(t) 
\begin{pmatrix}
\hat{f}_{\bm{k}}\\
\hat{f}^\dag_{-\bm{k}}
\end{pmatrix},
 \end{equation}
Eq.~\eqref{eq:comm1BdG} can be written in terms of the Bogoliubov-de Gennes Hamiltonian as
 \begin{equation}
 \label{eq:eqforf0}
 \frac{d}{dt} \begin{pmatrix}
\hat{f}_{\bm{k}}(t)\\
\hat{f}^\dag_{-\bm{k}}(t)
\end{pmatrix} = i \mathcal{A}_{\bm{k}}^\dag(t) {\mathcal{H}}_{\bm{k}}(t) \mathcal{A}_{\bm{k}}(t)  
\begin{pmatrix}
\hat{f}_{\bm{k}}(t)\\
\hat{f}^\dag_{-\bm{k}}(t)
\end{pmatrix}.
 \end{equation}
 Comparing Eq.~\eqref{eq:eqforf0}~ with Eq.~\eqref{eq:eqforfm1} and its time derivative reveals that $\mathcal{A}_{\bm{k}}(t)$ fulfills 
 \begin{equation}
 \partial_t \mathcal{A}_{\bm{k}}^\dag = i \mathcal{A}_{\bm{k}}^\dag (t)\mathcal{H}_{\bm{k}}(t).
 \end{equation} 
 Hence, $\mathcal{A}_{\bm{k}}$ is given by the evolution operator generated by the Bogoliubov-de Gennes Hamiltonian, i.e.
 \begin{equation}
 \mathcal{A}_{\bm{k}}(t) = \mathcal{U}_{\bm{k}}(t),
 \end{equation}
since it fulfills the same Schr\"odinger equation,
  \begin{equation}
  \label{eq:eqforf}
   i \partial_t {\mathcal{U}}_{\bm{k}}(t) = {\mathcal{H}}_{\bm{k}}(t){\mathcal{U}}_{\bm{k}}(t),
   \end{equation}
 with the initial condition ${\mathcal{U}}_{\bm{k}}(0)=\mathds{1}_{\bm{k}}$. The operator ${\mathcal{U}}_{\bm{k}}(t)$ can be represented as 
\begin{equation}
\label{eq:Ukuv}
{ \mathcal{U}}_{\bm{k}}(t) = \begin{pmatrix}
 u_{\bm{k}}(t) & v_{\bm{k}}(t) \\
 v_{-\bm{k}}^*(t) & u_{-\bm{k}}^*(t)
\end{pmatrix},
\end{equation}
 for appropriate functions $u_{\bm{k}}(t)$ and $v_{\bm{k}}(t)$. 
 
Using Eqs.~\eqref{eq:eqforfm1}, \eqref{eq:Ukuv} and \eqref{eq:annihilft}, the correlation matrix $\Lambda_{\bm{k}}(t)$ for quasimomentum $\bm{k}$, using the notation $\langle \cdot \rangle(t) = \bra{\psi(t)} \cdot \ket{\psi(t)}$, is given by
\begin{align}
\Lambda_{\bm{k}} (t) 
&=  \begin{bmatrix}
\langle \hat f_{\bm{k}}^\dag \hat f_{\bm{k}}\rangle(t)  & \langle \hat f_{\bm{k}}^\dag \hat f_{-\bm{k}}^\dag \rangle (t) \\
\langle \hat f_{-\bm{k}} \hat f_{\bm{k}} \rangle (t) & \langle \hat f_{-\bm{k}}  \hat f_{-\bm{k}}^\dag \rangle (t) 
\end{bmatrix},\nonumber \\
& = \begin{bmatrix}
|v_{\bm{k}}(t)|^2  & v_{\bm{k}}^*(t) u^*_{-\bm{k}}(t)\\
v_{\bm{k}}(t) u_{-\bm{k}}(t) &1-|v_{\bm{k}}(t)|^2  
\end{bmatrix}.
\label{eq:Mkn}
\end{align}
%
For a time-periodic Hamiltonian ${\mathcal{H}}_{\bm{k}}(t)={\mathcal{H}}_{\bm{k}}(t+\tau)$ as considered here, Floquet's theorem applies to Eq.~\eqref{eq:eqforf}, such that the evolution operator ${\mathcal{U}}_{\bm{k}}(t)=\mathcal{T}\exp\left(-i\int_0^t\!\!dt'\ \mathcal{H}_{\bm{k}}(t')\right)$ can be decomposed as
 \begin{equation}
 {\mathcal{U}}_{\bm{k}}(t) =  {\mathcal{V}}_{\bm{k}}(t) e^{-i \mathcal{H}_{\bm{k}}^{(F)} t},
 \end{equation}
 with $ {\mathcal{V}}_{\bm{k}}(t)={\mathcal{V}}_{\bm{k}}(t+\tau)$ time-periodic and where the Floquet Hamiltonian $\mathcal{H}_{\bm{k}}^{(F)}$ is time-independent~\cite{Goldman2014, Bukov2015, Eckardt2017}. 
 The eigenvectors of the one-cycle evolution operator ${\mathcal{U}}_{\bm{k}}(\tau)$ define single-particle Floquet modes.  
 If ${\mathcal{D}}_{\bm{k}}$ is the matrix that diagonalizes ${\mathcal{U}}_{\bm{k}}(\tau)$, namely
 \begin{equation}
 {\mathcal{D}}_{\bm{k}} ^\dag {\mathcal{U}}_{\bm{k}}(\tau) {\mathcal{D}}_{\bm{k}} = \begin{pmatrix}
 e^{-i q_{\bm{k}} \tau} & 0 \\
 0 & e^{i q_{-\bm{k}}\tau}
 \end{pmatrix},
 \end{equation}
 with quasienergies $q_{\bm{k}}$, then the field operators $(\hat m_{\bm{k}}, \hat m_{-\bm{k}}^\dag)$ for the single-particle Floquet modes are related to $(\hat{f}_{\bm{k}},\hat{f}_{-\bm{k}}^\dag)$ by 
\begin{equation}
\begin{bmatrix}
\hat m_{\bm{k}} \\
\hat{m}_{-\bm{k}}^\dag
\end{bmatrix} = {\mathcal{D}}_{\bm{k}}^\dag\begin{bmatrix}
\hat f_{\bm{k}} \\
\hat{f}_{-\bm{k}}^\dag
\end{bmatrix} .
\end{equation}
Determining the explicit form of the evolution operator, correlation matrix and Floquet modes in Eqs.~\eqref{eq:Ukuv} and \eqref{eq:Mkn} amounts to solving the time-dependent Schr\"odinger equation for the Floquet problem. As this is not possible analytically, we next perform a high-frequency (or rotating-wave approximation) analysis, providing an approximate solution in the weak driving regime, $\Omega\ll \omega$.

\subsection{High-frequency (weak-driving) limit}

For a weak and resonant drive $\Omega(t)=\Omega \cos(\omega t)$ with $\omega=2\mu_f$, we can perform a rotating-wave approximation (RWA) analysis, starting from the quasimomentum-space two-level Hamiltonian of Eq.~\eqref{eq:mathcalH}. After integrating out the term $\mu_f {\sigma}_z$ in Eq.~\eqref{eq:mathcalH}, the Hamiltonian reads as
\begin{align}
\mathcal{H}_{\bm{k}}(t) = &- 2 a_{\bm{k}} \cos(\omega t)\sigma_z - b_{\bm{k}} \sigma_y  - b_{\bm{k}}[-i e^{2i\omega t}\sigma_+ + i e^{-2i \omega t} \sigma_-],
\end{align}
where
\begin{align}
& a_{\bm{k}} = \Omega [\cos(k_x) + \cos(k_y)], \nonumber \\
& b_{\bm{k}} = \Omega [\sin(k_x) - \sin(k_y)].
\label{eq:Bk}
\end{align}
The Floquet Hamiltonian is then well approximated by the time-averaged (i.e., RWA) Hamiltonian
\begin{equation}
{\mathcal{H}}^{ (F)}_{\bm{k}}= B^{(F)}_{\bm{k}, y} \sigma_y,
\end{equation}
where we defined the effective magnetic field component $B_{\bm{k},y}^{(F)} = -b_{\bm{k}}$, and it provides an accurate approximation of the exact dynamics in the weak-driving regime we are considering here, $\Omega\ll \omega$~\cite{Goldman2014, Bukov2015,  Eckardt2017}. The corresponding RWA evolution operator after $n$ periods is
\begin{align}
{\mathcal{U}}^{(F)}_{\bm{k}}(n \tau) = & e^{-i n \tau B_{\bm{k},y}^{(F)} \sigma_y},\nonumber  \\
= & \begin{bmatrix}
\cos\left(B_{\bm{k},y}^{(F)} n \tau \right) & -\sin\left(B_{\bm{k},y}^{(F)} n \tau \right)\\
\sin\left( B_{\bm{k},y}^{(F)} n \tau \right) & \cos\left(B_{\bm{k},y}^{(F)} n \tau\right)
\end{bmatrix}.
\label{eq:Uk_t}
\end{align}
In the light of Eqs.~\eqref{eq:eqforfm1}, \eqref{eq:Ukuv} and \eqref{eq:Mkn}, the RWA correlation matrix is thus
\begin{align}
 \Lambda_{\bm{k}}^{(F)}(n\tau) = &
\begin{pmatrix} 
\sin^2\left(B_{\bm{k},y}^{(F)} n\tau\right) & -\frac{1}{2}\sin \left(2 B_{\bm{k},y}^{(F)} n \tau\right)\\
-\frac{1}{2}\sin\left( 2B_{\bm{k},y}^{(F)} n \tau\right) & \cos^2\left(B_{\bm{k},y}^{(F)} n \tau \right)
\end{pmatrix}.
\label{eq:corr_mat_k}
\end{align}
From Eq.~\eqref{eq:corr_mat_k}, we clearly see that the occupation of the $\bm{k}$ modes for $B_{\bm{k}, y}^{(F)}\ne 0$ oscillates perpetually and averages to $1/2$ over time. 
For the quasimomenta $\bm{k}\in K_0$ whose occupation is unaltered by the drive, the (RWA approximation of the) Floquet modes match the static ones $(f_{\bm k}, f_{-\bm{k}}^\dag)$. 
For $B_{\bm{k},y}^{(F)}> 0$, the field operators $\hat{m}_{\bm{k}}$ and $\hat{m}_{-\bm{k}}^\dag$ for single-particle Floquet modes $\ket{m_{\bm{k}}}=\hat{m}_{\bm{k}}^\dag\ket{0}$ with associated quasienergy $q_{\bm{k}}=B_{\bm{k},y}^{(F)}$, can be chosen as 
\begin{equation}
\label{eq:fmodesk}
\begin{bmatrix}
\hat m_{\bm{k}}\\
\hat m_{-\bm{k}}^\dag
\end{bmatrix} =\frac{1}{\sqrt{2}} \begin{pmatrix}
1 & -i\\
1 &i
\end{pmatrix}\begin{bmatrix}
\hat f_{\bm{k}}\\
\hat f_{-\bm{k}}^\dag
\end{bmatrix}.
\end{equation}

Having determined expressions characterizing the properties of the system during its time evolution, we next discuss the ensemble that predicts the long-time dynamics.

\subsection{Periodic Gibbs ensemble}
The long-time dynamics of an integrable Floquet system is predicted to be well captured by a periodic Gibbs ensemble (PGE)~\cite{Lazarides2014a}, or Floquet generalized Gibbs ensemble~\cite{Russomanno2012}, having the form
\begin{equation}\label{eq:PGE0}
\hat\rho^{ \rm PGE} = \prod_{\bm{k}} \frac{e^{-\zeta_{\bm{k}} \hat{m}_{\bm{k}}^\dag\hat{m}_{\bm{k}}}}{Z_{\bm{k}}}
\end{equation}
at stroboscopic times, where $\hat{m}_{\bm{k}}^\dag\hat{m}_{\bm{k}}$ are conserved quantities given by the occupation number of the single-particle Floquet mode $\ket{m_{\bm{k}}}$ and $Z_{\bm{k}}=1+\exp(-\zeta_{\bm{k}})$ is a normalization constant. The coefficients $\zeta_{\bm{k}}$ are determined by the initial values of the conserved quantities. As the initial state is the fermion vacuum $\ket{0}$, fulfilling $\hat{f}_{\bm{k}}\ket{0}=0$, they satisfy
\begin{equation}
\mathrm{tr}\left(\hat{\rho}^{\rm PGE} \hat{m}_{\bm{k}}^\dag\hat{m}_{\bm{k}}\right) = \frac{1}{e^{\zeta_{\bm{k}}}+1} = \bra{0}\hat{m}_{\bm{k}}^\dag\hat{m}_{\bm{k}} \ket{0}.
\end{equation}
From the RWA expression of the Floquet modes in Eq.~\eqref{eq:fmodesk}, the initial values are given by
\begin{equation}
\bra{0}\hat{m}_{\bm{k}}^\dag\hat{m}_{\bm{k}} \ket{0} =
 \begin{cases} 
 \frac{1}{2}, & \mbox{for}\ \bm{k} \not\in K_0,\\
0, & \mbox{for}\ \bm{k} \in K_0.
\end{cases}
\end{equation}
Note, in particular, that quasimomenta $\bm{k}\in K_0$ are unoccupied.
The PGE thus assumes the maximally-mixed form in the complement $\overline{K}_0$, defined by $\bm{k}\not\in K_0$, 
\begin{eqnarray}
\label{eq:PGEexplicit}
\hat\rho^{\rm PGE} && = \left(\prod_{\bm{k}\in K_0} \ket{0_{\bm{k}}}\!\bra{0_{\bm{k}}} \right)\left(\prod_{\bm{k}\not\in K_0}  \frac{\hat
{\mathds{1}}_{\bm{k}}}{2}\right), \nonumber \\
&& = \frac{1}{2^{|\overline{K}_0|}} \prod_{\bm{k}\in K_0} (1-\hat{n}_{\bm{k}}),
\end{eqnarray}
with $|\overline{K}_0 |$ denoting the cardinality of $\overline{K}_0$.

We stress that, although Eq.~\eqref{eq:PGEexplicit} predicts that the infinite-temperature ensemble (for $\bm{k}\not\in K_0$) describes well the behaviour of local observables at long times, this result is not a manifestation of Floquet eigenstate thermalization. The system is integrable and will thus not lose memory of its initial conditions. However, our drive is designed precisely to resonantly proliferate fermionic quasiparticles and thus yield infinite-temperature-like expectation values.

Indicating with $\langle \circ \rangle_{\rm PGE}=\mathrm{tr}(\circ \hat{\rho}^{\rm PGE})$ the PGE expectation values, Eq.~\eqref{eq:PGEexplicit} yields the PGE correlation matrix
\begin{equation}
\Lambda^{\rm PGE}_{\bm{k}\in K_0}= \begin{pmatrix}
\langle \hat{f}_{\bm{k}}^\dag \hat{f}_{\bm{k}}\rangle_{\rm PGE} & \langle \hat{f}_{\bm{k}}^\dag \hat{f}_{-\bm{k}}^\dag \rangle_{\rm PGE}\\
\langle \hat{f}_{-\bm{k}} \hat{f}_{\bm{k}}\rangle_{\rm PGE}& \langle \hat{f}_{-\bm{k}} \hat{f}_{-\bm{k}}^\dag \rangle_{\rm PGE}
\end{pmatrix} = \begin{pmatrix}
0 & 0\\
0& 1
\end{pmatrix}, \qquad \Lambda^{\rm PGE}_{\bm{k}\not\in K_0} =\frac{1}{2} \begin{pmatrix}
1 & 0 \\
0 & 1
\end{pmatrix}.
\end{equation}
To study entanglement properties, the correlations in real space are also needed. From the quasimomentum-space results, one finds that the anomalous correlators vanish, $\langle \hat{f}_p^\dag \hat{f}_{p'}^\dag \rangle_{\rm PGE} =  \langle \hat{f}_p \hat{f}_{p'} \rangle_{\rm PGE} =0$, while
\begin{eqnarray}
\langle \hat f_p^\dag \hat f_{p'}\rangle_{\rm PGE} && = \frac{1}{L_xL_y}\sum_{\bm{k}} e^{-i \bm{k}\cdot(\bm{r}_p - \bm{r}_{p'})}\langle \hat f_{\bm{k}}^\dag \hat f_{\bm{k}}\rangle_{\rm PGE},\nonumber \\
 &&= \frac{\delta_{pp'}}{2}- \frac{1}{2L_xL_y} \sum_{\bm{k} \in K_0} e^{-i \bm{k}\cdot(\bm{r}_p - \bm{r}_{p'})} .
\end{eqnarray}
Hence, the normal correlation matrix $C_{pp'}^{\rm PGE} = \langle \hat f_p^\dag \hat f_{p'}\rangle_{\rm PGE} $ has the form
\begin{equation}
C^{\rm PGE}=\frac{1}{2}(1 - \mathcal{K}_0),
\end{equation}
with 
\begin{equation}
\label{eq:mathcalK0}
\mathcal{K}_{0, pp'} = \frac{1}{L_x L_y}\sum_{\bm{k}\in K_0} e^{-i \bm{k}\cdot(\bm{r}_p- \bm{r}_{p'})}.
\end{equation}
The deviation $\mathcal{K}_0$ from the fully-mixed-state result can be evaluated explicitly for finite $L_x$ and $L_y$, but the result depends on the parity properties of $L_x$ and $L_y$, as discussed in Sec.~\ref{sec:Floquetproblem}. For example, for $L_x=L_y=L$ and $L$ a multiple of four, one obtains 
\begin{eqnarray}
\langle \hat f_p^\dag \hat f_{p'}\rangle_{\rm PGE} =&& \frac{1}{2}\delta_{pp'} - \frac{1}{2L^2}\sum_{k_x}  \left[e^{-i k_x(x_{pp'}+y_{pp'})} +(-1)^{y_{pp'}} e^{-i k_x(x_{pp'}-y_{pp'})}\right] +\frac{1}{L^2}\cos\left[\frac{\pi}{2}(x_{pp'}+y_{pp'})\right], 
\end{eqnarray}
where $x_{pp'}=x_p-x_{p'}$ and $y_{pp'}=y_p - y_{p'}$ are differences of the lattice coordinates $(x_p, y_p)$ of site $p$.

Given $C^{\rm PGE}$, we can formally determine the real-space representation of $\hat{\rho}^{\mathrm{PGE}}_{ R}$ in a sub-region $R$, since $\hat{\rho}^{\mathrm{PGE}}_{ R}$ is diagonalized by the same transformation diagonalizing $C_R^{\rm PGE}$. Given the eigenvalues $h_j$ of $\mathcal{K}_{0R}$, which lead to eigenvalues $c_j = (1- h_j)/2$ of $C_R^{\rm PGE}$, $\hat{\rho}^{\rm PGE}_R$  can be written in diagonal form as
\begin{eqnarray}
\hat{\rho}^{\rm PGE}_R && =\prod_{j} \left[\left(\frac{1-h_j}{2}\right)\hat{c}^\dag_j\hat{c}_j + \left(\frac{1+h_j}{2}\right) \hat{c}_j\hat{c}_j^\dag \right],\label{eq:rhoAPGE0} \nonumber \\
&& = \frac{1}{2^{|R|} }\prod_{j} \left[ 1+h_j(1-2\hat{c}_j^\dag \hat{c}_j)\right] , \label{eq:rhoAPGE1}
\end{eqnarray}
where the modes $(\bm{\hat{c}}, \bm{\hat{c}}^\dag)$ are related to $(\bm{\hat{f}}, \bm{\hat{f}}^\dag)$ by the unitary diagonalizing $C_R^{\rm PGE}$ (and $\mathcal{K}_{0R}$). The expression \eqref{eq:rhoAPGE1} will be instrumental to derive bounds on the logarithmic negativity in the next subsection.

\subsection{Long-time $f$-contribution to the logarithmic negativity}
\label{sec:boundneg}

Applying the results of Sec.~\ref{sec:decompnegativity}, the logarithmic negativity of the PGE between two adjacent non-contractible regions $R^{(1)}$ and $R^{(2)}$ as in Fig.~1 can be decomposed as
\begin{equation}
\mathcal{E}= \mathcal{E}^{(g)}+ \widetilde{\mathcal{E}}^{\rm PGE},
\end{equation}
where $\mathcal{E}^{(g)}=(L_y-1)\log(2)$ is the logarithmic negativity of the ground state, carrying the topological term $-\log(2)$, and
\begin{equation}
\widetilde{\mathcal{E}}^{\rm PGE}= \log(\Vert \hat{\Sigma}^{\mathrm{PGE}}_{ +} \Vert_1 + \Vert  \hat{\Sigma}^{\mathrm{PGE}}_{ -} \Vert_1)
\end{equation}
is the contribution given by $f$ quasiparticles, where $ \hat{\Sigma}^{\mathrm{PGE}}_{ \pm}$ are the components of $(\hat{\rho}^{ \mathrm{PGE}}_R)^{T_2}$ whose parity labels in $R^{(2)}$ remain unchanged ($+$) or are switched ($-$) by the partial transpose, see Eq.~\eqref{eq:sigmadx}. 

We here show that, for a sub-region $R$ of fixed width $w_R$, the $f$ contribution $\widetilde{\mathcal{E}}^{\rm PGE}$ vanishes for increasing system size. To this end, we first introduce the deviation $\hat{\delta} \equiv \hat{\rho}^{\rm PGE}_R   - \mathds{1}_R/2^{|R|}$ of the reduced PGE density matrix from the maximally-mixed state. The components $\hat{\Sigma}^{\mathrm{PGE}}_{ +} $ and $\hat{\Sigma}^{\mathrm{PGE}}_{-}$ have then the form
\begin{eqnarray}
 \hat{\Sigma}^{\mathrm{PGE}}_{ +} && = \frac{\hat{\mathds{1}}_R}{2^{|R|}} +\frac{1}{2}\left( \hat \delta^{T_2} + \hat{\mathcal{P}}_2 \hat \delta^{T_2} \hat{\mathcal{P}}_2 \right),\nonumber \\
   \hat{\Sigma}^{\mathrm{PGE}}_{ -} && = \frac{1}{2}\left( \hat \delta^{T_2} - \hat{\mathcal{P}}_2 \hat \delta^{T_2} \hat{\mathcal{P}}_2 \right),
\end{eqnarray}
with $\hat{\mathcal{P}}_2=(-1)^{\sum_{p\in R^{(2)} }\hat{f}_p^\dag \hat{f}_p}$. As $\hat{\mathcal{P}}_2$ is unitary, by the triangle inequality their trace norm is bounded as
\begin{equation}
\Vert \hat{\Sigma}^{\mathrm{PGE}}_{+}  \Vert_1 \le 1 +  \Vert \hat \delta^{T_2} \Vert_1, \qquad \Vert \hat{\Sigma}^{\mathrm{PGE}}_{-}  \Vert_1 \le   \Vert \hat \delta^{T_2} \Vert_1,
\end{equation}
such that $\widetilde{\mathcal{E}}^{\rm PGE}$, in turn, fulfills
\begin{equation}
\label{eq:wildeeps}
\widetilde{\mathcal{E}}^{\rm PGE} \le \log( 1 + 2\Vert \hat{\delta}^{T_2}\Vert_1).
\end{equation}
To obtain an explicit bound, we search for an upper bound on $\Vert \hat{\delta}^{T_2}\Vert_1$. To circumvent the necessity of computing the partial transpose, we use the fact that the trace norm is upper bounded by 
\begin{equation}
\label{eq:norm12}
\Vert \hat{\delta}^{T_2}\Vert_1 \le 2^{|R|/2} \Vert \hat{\delta}^{T_2}\Vert_2 =  2^{|R|/2} \Vert \hat{\delta}\Vert_2,
\end{equation}
 where $\Vert \hat O \Vert_2 =\sqrt{ \mathrm{tr}[\hat O^\dag \hat O]} $ is the Hilbert-Schmidt norm, which is preserved by the partial transpose, $\Vert \hat{O}^{T_2}\Vert_2 =\Vert \hat{O} \Vert_2 $. We can then estimate $\Vert \hat{\delta}\Vert_2$ as follows. By rewriting
\begin{equation}
\Vert \hat\delta \Vert_2^2 = \mathrm{tr} [(\hat{\rho}_R^{\rm PGE})^2] - \frac{1}{2^{|R|}}
\end{equation}
and noting that, as $\hat{\rho}_R^{\rm PGE}$ is Gaussian, its purity is obtained from the eigenvalues of the $C$ [introduced in Eq.~\eqref{eq:rhoAPGE0}] as
\begin{eqnarray}
 \mathrm{tr} [(\hat{\rho}_R^{\rm PGE})^2] = \prod_j [c_j^2 + (1-c_j)^2] && = \prod_j \frac{1+h_j^2}{2},\nonumber  \\
 && \le 2^{-|R|} \exp\left({\mathrm{tr}[\mathcal{K}_{0R}^2]}\right).
 \label{eq:boundka}
\end{eqnarray}
From Eq.~\eqref{eq:mathcalK0}, the trace of $\mathcal{K}_{0R}^2$ is given by
\begin{equation}
\mathrm{tr} \mathcal{K}_{0R}^2 = \frac{1}{(L_xL_y)^2}\sum_{\bm{r}_p, \bm{r}_{p'} \in R} \sum_{\bm{k}, \bm{k}'\in K_0} e^{-i(\bm{k}-\bm{k}')\cdot (\bm{r}_p-\bm{r}_{p'})}.
\end{equation}
The sums over the $y_p$ and $y_{p'}$ variables yield a factor $L_y^2 \delta_{k_{y}, k_y'}$, leaving
\begin{eqnarray}
\mathrm{tr} \mathcal{K}_{0R}^2 && = \frac{1}{L_x^2 } \sum_{\substack{\bm{k}, \bm{k}'\in K_0 \\ k_y=k'_y}} \left\vert \sum_{x_p \in R}  e^{-i(k_x-k_x')x_p} \right\vert^2 , \nonumber \\
&& \le  \frac{2 |K_0| w_R^2 }{L_x^2 }  ,
\end{eqnarray}
where we considered that the sum over $\bm{k}$ and $\bm{k}'$ can include at most $2|K_0|$ pairs. Returning to $\Vert \hat{\delta}^{T_2} \Vert_1$ via Eqs.~\eqref{eq:boundka} and \eqref{eq:norm12} and using that $|K_0|/L_x\le 2 \min (L_x, L_y)/L_x \le 2 $ according to Eq.~\eqref{eq:K0bound}, we have thus determined that
\begin{equation}
\Vert \hat{\delta}^{T_2} \Vert_1 \le \sqrt{ \exp\left(\frac{4 {w}_R^2}{L_x}\right) - 1}.
\end{equation}
Hence, $\Vert \hat{\delta}^{T_2} \Vert_1$ vanishes for large system size. By virtue of Eq.~\eqref{eq:wildeeps} we can conclude that $\widetilde{\mathcal{E}}^{\rm PGE}$ also vanishes.

\subsection{Additional details about numerical simulations}

The numerical simulation data reported in the main text have been obtained as follows:
\begin{itemize}
\item Fig.~2(a)-(b): The Floquet dynamics of Eq.~\eqref{eq:eqforf} is solved by exact numerical time propagation and determination of the Floquet modes by diagonalization of the one-period evolution operator $\mathcal{U}_{\bm{k}}(\tau)$. The real-space correlation matrix $\Lambda_{ij}=\langle \hat{F}_i^\dag \hat{F}_{j}\rangle$, for $\hat{\bm{F}}=( \hat{\bm{f}}, \hat{\bm{f}}^\dag)$ (and the Majorana covariance matrix $\Gamma_{i,j}=(i/2)\langle [\hat{\gamma}_i, \hat{\gamma}_j] \rangle$) is then determined for both the time-evolved state and the PGE (through numerical fast Fourier transform, given the solution for the evolution in quasimomentum space). The entropy in $R$ is computed from the eigenvalues $\lambda_j$ of the correlation matrix $\Lambda_R$ restricted to the modes in region $R$~\cite{peschelCalculationReducedDensity2003, peschelReducedDensityMatrices2009, Surace2021},
\begin{equation}
\mathcal{S}_R^{(f)} = - \frac{1}{2} \sum_j \left[\lambda_j \log \lambda_j +(1-\lambda_j)\log(1-\lambda_j)\right].
\end{equation}
\item Fig.~2(c): The effective exponent $\beta_{\rm eff}$ for the perimeter-law to area-law crossover of the contractible Wilson loops is computed through numerical gradient as
\begin{equation}
\beta_{\rm eff}(l, t) = \frac{\partial \log (-\log|W^Z_{\square_l}(t)|)}{\partial \log l} .
\end{equation}
By virtue of Eq.~\eqref{eq:wilsontoparity} and Wick's theorem, the quantities $-\log |W^Z_{\square_l}(t)| $ are computed as 
\begin{equation}
-\log|W^Z_{\square_l}(t)|= -\frac{1}{2}\log |\det \Gamma_{\square_l}(t)|
\end{equation}
from the Majorana covariance matrix $\Gamma_{\square_l}(t)$ at time $t$ restricted to the modes in $\square_l$, where $\log |\det(\circ)|$ is computed through the dedicated NumPy routines~\cite{Harris2020}.
For a dependence of the form 
\begin{equation}
W^Z_{\square_l} = W_0 \exp\left(-\sum_{\beta=0}^{\beta_{\rm max}} \alpha_\beta l^\beta  \right) ,
\end{equation}
as is predicted for $W^Z_{\square_l}(t)$ with $\beta_{\rm max}=2$ in the area-law regime and with $\beta_{\rm max}=1$ in the perimeter-law regime ~\cite{Fradkin1978, Kogut1979}, $\beta_{\rm eff}$ yields
\begin{equation}
\beta_{\rm eff}(l,t)
  = \frac{\displaystyle\sum_{\beta=0}^{\beta_{\rm max}} \beta\, \alpha_\beta\, l^{\beta}}
         {\displaystyle\sum_{\beta=0}^{\beta_{\rm max}} \alpha_\beta\, l^{\beta} - \log|W_0|}.
         \end{equation}
 In the large $l$ limit, 
\begin{equation}
\beta_{\rm eff} = \beta_{\rm max} + \mathcal{O}(l^{-1}) 
\end{equation}
and thus $\beta_{\rm eff}$ estimates $\beta_{\rm max}$ for large loops and away from the crossover region.
\end{itemize}


\end{document}